\documentclass{aa}  
\usepackage{amstext,amsmath,mathtools}
\usepackage{graphicx}
\usepackage{txfonts}
\usepackage{subfig}
\usepackage{url}
\usepackage{natbib}
\bibpunct{(}{)}{;}{a}{}{,}
\usepackage{fancyvrb}

\usepackage{multirow} 
\usepackage{dcolumn} 
\begin{document}
\frenchspacing
\title{Evaluation of image-shift measurement algorithms for solar
  Shack--Hartmann wavefront sensors}

\titlerunning{Evaluation of image-shift measurement algorithms for
  solar SH WFS}

\author{Mats G. L\"ofdahl\inst{1,2}}
\authorrunning{M.G. L\"ofdahl}

\offprints{M. L\"ofdahl, \email{mats@astro.su.se}}

\institute{Institute for Solar Physics of the Royal Swedish
  Academy of Sciences, AlbaNova University Center,
  SE-106\,91 Stockholm, Sweden
  \and
  Stockholm Observatory, AlbaNova University Center,
  SE-106\,91 Stockholm, Sweden}

\date{Accepted 13 September 2010}

\abstract
{Solar Shack--Hartmann wavefront sensors measure differential
  wavefront tilts as the relative shift between images from different
  subapertures. There are several methods in use for measuring these
  shifts.}
{We evaluate the inherent accuracy of the methods and the effects of
  various sources of error, such as noise, bias mismatch, and
  blurring. We investigate whether Z-tilts or G-tilts are measured.}
{We test the algorithms on two kinds of artificial data sets, one
  corresponding to images with known shifts and one corresponding to
  seeing with different $r_0$.}
{Our results show that the best methods for shift measurements are
  based on the square difference function and the absolute difference
  function squared, with subpixel accuracy accomplished by use of
  two-dimensional quadratic interpolation. These methods measure
  Z-tilts rather than G-tilts.}
{}


\maketitle
%


\section{Introduction}
\label{sec:introduction}

High-resolution observational astronomy with telescopes operated on
the ground relies on methods for combating the effects from turbulence
in the Earth's atmosphere. Turbulence varies with time scales on the
order of ms, mixing air that varies in temperature and therefore in
refractive index. The wavefronts of the light from astronomical
objects, which are flat when they enter the atmosphere, are distorted
by the optically active atmosphere, which causes image motion and
blurring in the images collected through the telescope. Measurements
of the wavefronts allow us to correct for these effects, both in real
time and in post processing.

There are several different methods for measuring in real time the
variation of the wavefront phase of the light over the telescope
pupil. For solar telescopes, the one exclusively used is the
Shack--Hartmann \citep[SH;][]{shack71production} wavefront sensor
(WFS). This is true for the adaptive optics (AO) systems of present
1-meter class, high-resolution solar telescopes
\citep{1998SPIE.3353...72R, scharmer00workstation, scharmer03adaptive,
  2003SPIE.4853..187V, 2004SPIE.5171..179R}. SH wide-field WFSs are
used for characterizing the atmospheric turbulence above existing and
future solar telescopes
\citep{waldmann07untersuchung,2008SPIE.7015E.154W,scharmer10s-dimm+}.

An SH WFS estimates the wavefront from measurements of the local
wavefront tilt in several regions of the pupil called subpupils. The
local tilts manifest themselves as image motion in the focal plane
associated with each subpupil. For night-time telescopes, this image
motion is (comparably) easily measured by tracking the peak of the
image of a natural or artificial point source. As no point sources are
available on the sun, one instead has to resort to measuring relative
image motion by matching solar features in the different subpupil
images. This can be done by finding the minimum of the degree of
mismatch of the images as a function of image shift. This procedure
requires solving two distinct subproblems: 1) computing the mismatch
function as a function of image shift in steps of whole pixels and 2)
finding the minimum of the mismatch function with subpixel precision.
The latter involves interpolation between the whole-pixel grid points.
This procedure for solar correlation tracking was apparently pioneered
by \citet{1977lock.reptR....S} and \citet{1983A&A...119...85V}.

The SH WFSs developed for the AO systems currently in use at different
solar telescopes use a variety of shift measurement methods, so do the
sensors used for turbulence characterization. This suggests that the
choices of methods were based on personal preferences and
technological momentum from other applications rather than on a
thorough evaluation of the performance of these methods on relevant
data.

Several algorithms for measuring the mismatch and performing the
interpolation were investigated by \citet{yi92software}. However,
because the purpose of their work was different from ours (image
remapping in post-processing), they used much larger fields of view
(FOVs) than is currently practical for real-time solar SH WFS.

As part of a project on characterization of the atmospheric turbulence
using SH WFS, \citet{waldmann07untersuchung} and
\citet{2008SPIE.7015E.154W} address these problems for setups more
similar to ours. The recent MSc thesis of
\citet{johansson10cross-correlation} has expanded on Waldmann's work
in this area. She implemented a number of different methods for
image-shift measurements and tested them on synthetic data relevant to
solar SH WFSs. Comparisons with their methods and results will appear
throughout this paper.

\begin{table*}[t]
  \caption{Correlation algorithms (CAs).}  
  \label{tab:shift_algorithms}
  \centering
  \begin{tabular}{lll}
    \hline\hline\noalign{\smallskip}
    Acronym & Name & Mismatch, $c_{i,j}$, for shift $i,j$ \\
    \hline\noalign{\smallskip}
    SDF & Square Difference Function & $ \displaystyle \sum_{x,y}
    \bigl( g(x,y) - g_\text{ref}(x+i,y+j) \bigr)^2$\\ 
    CFI & Covariance Function, image domain  &
    $\displaystyle -\sum_{x,y}
    \protect\overbracket[.5pt][1pt]{g}(x,y)\cdot \protect\overbracket[.5pt][1pt]{g_\text{ref}}(x+i,y+j)$\\ 
    CFF& Covariance Function, Fourier domain  &$\displaystyle
    -\mathfrak{F}^{-1}\bigl\{ \mathfrak{F}\{ w_2\cdot\protect\overbracket[.5pt][1pt]{g}(x,y) \}
    \cdot
    \mathfrak{F}^*\{ w_2\cdot\protect\overbracket[.5pt][1pt]{g_\text{ref}}(x,y) \} 
    \bigr\}(i,j)$\\
    \noalign{\smallskip}
    ADF & Absolute Difference Function &  $\displaystyle\sum_{x,y} \bigl| g(x,y) -
    g_\text{ref}(x+i,y+j) \bigr|$\\
    ADF$^2$& Absolute Difference Function, Squared &
    $\displaystyle\biggl( \sum_{x,y} \bigl| g(x,y) -
    g_\text{ref}(x+i,y+j) 
    \bigr| \biggr)^2$\\
    \hline
  \end{tabular}
  \tablefoot{$\mathfrak{F}$ denotes the Fourier transform and an asterisk
    as a superscript denotes complex conjugation. $w_2$ is a 2D
    Hamming window. The notation $\protect\overbracket[.5pt][1pt]{g}$ is used for a version
    of $g$, where a fitted plane has been subtracted.}
\end{table*}

In this paper we investigate, by use of artificial data relevant to
present solar SH WFSs, a number of different algorithms for measuring
whole-pixel image shifts and interpolating to get subpixel accuracy.
The algorithms are defined in Sect.~\ref{sec:algorithms}. We
investigate the inherent accuracy of the algorithms in
Sect.~\ref{sec:algorithm-accuracy}, by testing them on identical
images with known shifts, as well as the influence of noise and
variations in intensity level. In Sect.~\ref{sec:image-shift-versus}
we use the best methods on images formed through artificial seeing and
evaluate the performance for different seeing conditions. We discuss
our results in Sect.~\ref{sec:conclusions}.

\section{Algorithms}
\label{sec:algorithms}

\subsection{Correlation algorithms}
\label{sec:corr-algor}

In Table~\ref{tab:shift_algorithms} we define five different
correlation algorithms (CAs), which we use to calculate the image
mismatch on a grid of integer pixel shifts $i,j$. The names and
acronyms of the CAs are also given in the table. These mismatch values
make a matrix $\mathbf c$ with elements $c_{i,j}$. A coarse estimate
of the image shift, $(\delta x,\delta y)$, is then given by the
indices corresponding to the grid position with the minimum mismatch
value, $(i_\text{min},j_\text{min})$. This shift should be sought
within a maximum range in order to reduce the number of false matches
to other parts of the granulation pattern. The algorithms in
Sect.\@~\ref{sec:subpixel-interp} are then used to refine this
estimate to subpixel accuracy.

Perhaps most straight-forward is the SDF algorithm, in which one
calculates the mismatch in a Least Squares (LS) sense.

Subtracting the intensity mean, expanding the square in the SDF
algorithm, and retaining only the cross term gives twice the
covariance (with negative sign), which is the basis of the following
two algorithms. The CFI algorithm calculates the covariance in the
image domain. It is the one being used for the Dunn Solar Telescope
system \citep{1998SPIE.3353...72R}.
The correlation coefficient differs from the covariance only by
division with the standard deviations of the two images, so methods
based on the former \citep[e.g., ][]{2008SPIE.7015E.154W} should give
results similar to those of CFI.

The covariance can also be calculated in the Fourier domain, taking
advantage of the Fast Fourier Transform (FFT). The CFF algorithm was
developed by \citet{1983A&A...119...85V} for an image stabilization
system and is used today in the KAOS AO implementation used at the
Vacuum Tower Telescope \citep{2003SPIE.4853..187V}. For small images,
such as those involved in SH WFS calculations, one can expect errors
from wrap-around effects. That is, because of the assumption of
periodicity implicit in finite-size Fourier transforms, for large
shifts, structures in one image are not matched with structures at the
shifted location but with structures shifted in from the opposite side
of the image. 

\citet{knutsson05extended} derived a method based on the Fourier
spectrum of the correlation function (correlation spectrum phase). It
is supposed to give results similar to the CFF method but with some
accuracy sacrificed for speed. We have not investigated this method.

The CFF method requires apodization of the images, i.e., the
multiplication of a window function that reduces ringing effects from
the discontinuities caused by the Fourier wrap-around. When not
explicitly stated otherwise, we use a Hamming window written as
\begin{equation}
  w_2(x,y) = w_1(x)\cdot w_1(y),
\end{equation}
where $w_1$ is the 1-dimensional Hamming window
\citep{enochson68programming},
\begin{equation}
  w_1(x) =  a+(a-1)\cos \Bigl(\frac{2\pi x}{N-1}\Bigr),
\end{equation}
where $a=0.53836$ and $N$ is the linear size of the window. See also
Sect.~\ref{sec:window-function} below.

The ADF algorithm is fast because it can be calculated very
efficiently in CPU instructions available for many architectures,
particularly for 8-bit data.  \citet{2003SPIE.4853..351K} use ADF for
the IR AO system of the McMath--Pierce solar telescope. So do
\citet{2009OptRv..16..558M} in their recently presented AO system used
for the domeless solar telescope at Hida Observatory.

Because the shape of the ADF minimum does not match the assumption of
a parabolic shape implicit in the subpixel algorithms, squaring the
ADF correlation values leads to an improvement.  This adds a
completely negligible computational cost to that of ADF, as squaring
is done after summing. In fact, because it does not move
$(i_\text{min},j_\text{min})$, only the at most nine grid points used
for subpixel interpolation (see Sect.~\ref{sec:subpixel-interp})
have to be squared.  This ADF$^2$ method was developed by
G.\@~Scharmer in 1993 for use in the correlation tracker systems of
the former Swedish Vacuum Solar Telescope \citep{shand95latency} and
is in use in the AO and tracker systems of the Swedish 1-meter Solar
Telescope \citep[SST; ][]{scharmer00workstation,scharmer03adaptive}.

\citet{1977lock.reptR....S} showed that a linear trend in intensity
shifts the covariance peak from the correct position. Therefore one
should subtract a fitted plane from both $g$ and $g_\text{ref}$ before
applying the CAs. However, the granulation data used in our
simulations have negligible trends and for the difference based
algorithms (ADF, ADF$^2$, and SDF), a consistent bias in the intensity
level cancels automatically. We therefore saved computing time in our
tests by limiting the pre-processing of the data to only subtracting
the mean values, and only when calculating CFI and CFF.

\begin{table*}[t]
  \caption{Interpolation algorithms (IA) or subpixel minimum finding
    algorithms.}
  \label{tab:subpix_algorithms}
  \centering
  \begin{tabular}{lll}
    \hline\hline\noalign{\smallskip}
    Acronym & Name & Location of minimum \\
    \hline\noalign{\smallskip}
    2LS &  2D Least Squares & $(x_2,y_2)$\quad where\quad
    $
    \begin{cases}
      a_2 = \bigl(\langle s_{1,j}\rangle_j - \langle s_{-1,j}\rangle_j\bigr)/2 \\
      a_3 = \bigl(\langle s_{1,j}\rangle_j - 2\langle s_{0,j}\rangle_j + \langle s_{-1,j}\rangle_j\bigr)/2 \\
      a_4 = \bigl(\langle s_{i,1}\rangle_i - \langle s_{i,-1}\rangle_i\bigr)/2 \\
      a_5 = \bigl(\langle s_{i,1}\rangle_i - 2\langle s_{i,0}\rangle_i + \langle s_{i,-1}\rangle_i\bigr)/2\\
      a_6 = (s_{1,1}-s_{-1,1}-s_{1,-1}+s_{-1,-1})/4 
    \end{cases}
    $\\
    \noalign{\smallskip}
    2QI &2D Quadratic Interpolation &  $(x_2,y_2)$\quad where\quad
    $
    \begin{cases}
      a_2 = (s_{1,0}-s_{-1,0})/2\\
      a_3 = (s_{1,0}-2s_{0,0}+s_{-1,0})/2\\
      a_4 = (s_{0,1}-s_{0,-1})/2\\
      a_5 = (s_{0,1}-2s_{0,0}+s_{0,-1})/2\\
      a_6 = (s_{1,1}-s_{-1,1}-s_{1,-1}+s_{-1,-1})/4 
    \end{cases}
    $\\
    \noalign{\smallskip}
    1LS &1D Least Squares &  $(x_1,y_1)$\quad  using\quad  $a_2, a_3, a_4, a_5$
    from 2LS above\\
    \noalign{\smallskip}
    1QI &1D Quadratic Interpolation &  $(x_1,y_1)$\quad  using\quad  $a_2, a_3, a_4, a_5$
    from 2QI above\\
    \hline
  \end{tabular}
  \tablefoot{Notation in the table refers to the elements,
    $s_{i,j}$, of the submatrix $\mathbf s$ of the correlation matrix
    centered on its sample minimum, $(i_\text{min},j_\text{min})$, the
    expression for a second order polynomial of two variables, $f(x,y)
    = a_1 + a_2x + a_3x^2 + a_4y + a_5y^2 + a_6xy$, and the location
    of its 1D minimum, $x_1 = i_\text{min} - a_2/2a_3$; $y_1 =
    j_\text{min} - a_4/2a_5$ and its 2D minimum, $x_2 = i_\text{min} +
    (2a_2a_5-a_4a_6)/(a_6^2-4a_3a_5)$; $y_2 =j_\text{min} +
    (2a_3a_4-a_2a_6)/(a_6^2-4a_3a_5)$. The notation
    $\langle\cdot\rangle_i$ denotes averaging over index
    $i\in\{-1,0,1\}$.}
\end{table*}

\subsection{Subpixel interpolation}
\label{sec:subpixel-interp}

The methods in Sect.~\ref{sec:corr-algor} are responsible for making a
coarsely sampled 2D correlation function with a reasonable shape. The
interpolation algorithms (IAs) in this section then have to find the
minimum with better accuracy than given by the sampling grid.

\citet{1988ApJ...333..427N} found that interpolation methods should be
based on polynomials of degree 2. Methods based on polynomials of
higher degree systematically underestimate the shift for small
displacements, while first degree polynomials give a systematic
overestimation. The algorithms we consider can all be described as
fitting a conic section,
\begin{equation}
  f(x,y) = a_1 + a_2 x + a_3 x^2 + a_4 y + a_5 y^2 + a_6 xy,
\end{equation}
to the 3$\times$3-element submatrix $\mathbf s$ of $\mathbf c$
centered on the sample minimum of $\mathbf c$ with elements
\begin{equation}
  \label{eq:1}
  s_{i,j}=c_{i+i_\text{min},j+j_\text{min}}; \qquad i,j=-1,0,1.
\end{equation}
The interpolated shift vector, $(\delta x,\delta y)$, is the position
of the minimum of the fitted function, $(x_\text{min},y_\text{min})$.
The algorithms differ in the number of points used and whether the
fitting is done in 2D or in each dimension separately. The
definitions, names, and acronyms of the different methods are given in
Table~\ref{tab:subpix_algorithms}.

The 1QI method is based on numerical 1D derivatives using the center
row (column) of $\mathbf s$ \citep{1986ApOpt..25..392N}. It is
equivalent to a LS estimate using only the center row (column) of
$\mathbf s$. It does not use the corner elements of $\mathbf s$.

The 1LS algorithm consists of, separately for the $x$ and $y$
directions, fitting a 1D polynomial of degree 2 to all the elements in
$\mathbf s$. This is equivalent to the procedure of
\citet{waldmann07untersuchung}, projection of the data onto the axes
(i.e., summing the rows (columns)), and doing LS fitting on the
result. Waldmann uses Lagrange interpolation but this is
  mathematically equivalent to a LS fit.

The 2QI algorithm was derived by \citet[their Eq.\@~(9)]{yi92software}
as an extension of the 1QI algorithm.

The 2LS algorithm is based on expressions for the conic section
coefficients derived by \citet{waldmann07untersuchung}.
\citet{johansson10cross-correlation} found that Waldmann's expression
for one of the 2LS coefficients is missing a factor 2. We are using
the corrected expression.

\citet{waldmann07untersuchung} compared four different subpixel
methods: 1LS, 2LS, and a 2D fit to a Gaussian. He found that the
Gaussian fit gave the best results and the polynomial method worked
almost as well. Because the Gaussian fit is more computationally
heavy, he adopted the latter. \citet{johansson10cross-correlation}
tested the Gaussian fit and got results comparable to Waldmann's but
much better results for the polynomial fits.  Based on Johansson's
results, we have not evaluated Gaussian fits as a method for subpixel
interpolation.
   
\citet{2009OptRv..16..558M} use a method for subpixel interpolation,
where they find the centroid of a spot generated by inverting and then
clipping the mismatch measured with ADF. This method is not considered
here.

\begin{figure*}[t]
  \centering
  \def\figwidth{17cm}
  \hbox{\centering
  \subfloat[\label{fig:GI}]{\includegraphics[width=5.8cm]{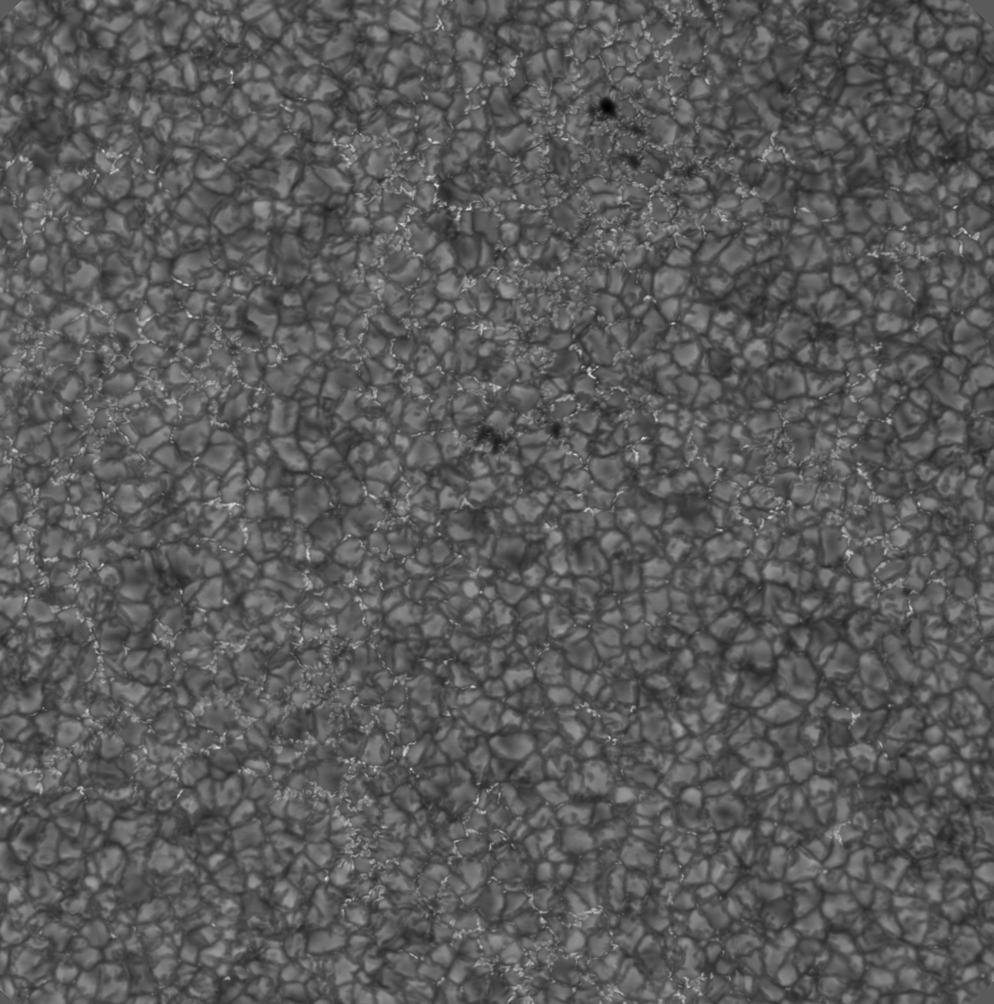}}
  \quad
  \subfloat[\label{fig:degraded}]{\includegraphics[width=5.8cm]{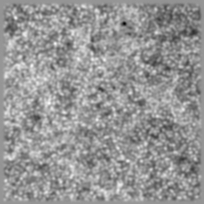}}
  }
  \caption{Granulation image used for simulation experiments.
    \textbf{(a)}~430.5~nm (G-band) SST diffraction limited granulation
    image (GI). 2000$\times$2000 pixels, image scale 0\farcs041/pixel.
    \textbf{(b)}~Degraded to resolution of a 9.8 cm subpupil, shifted
    by a whole number of pixels, and re-sampled to 200$\times$200
    0\farcs41-pixels.}
\end{figure*}

\section{Algorithm accuracy}
\label{sec:algorithm-accuracy}

The goal of this experiment with artificial data is to establish the
accuracy of the algorithms themselves, for granulation images (GI),
which are identical except for a known shift and detector
imperfections such as noise and bias. In reality, the images will be
different because the atmospheric turbulence not only shifts the
images but also smears them differently. Those effects will be
addressed in Sect.\@~\ref{sec:image-shift-versus} below.

\subsection{Artificial data}
\label{sec:perfect-data-recipe}

For this experiment, the images should be identical except for a known
shift, defined to subpixel precision without introducing errors that
are caused by the re-sampling needed for subpixel image shifting. We
therefore start with a high-resolution image, degrade the resolution,
shift it by a known number of whole high-resolution pixels, and then
down-sample it to the SH image scale.

Specifically, we use a 2000$\times$2000-pixel, high-quality SST G-band
image with an image scale of 0\farcs041/pixel, see Fig.~\ref{fig:GI}.
The image was recorded on 25 May 2003 by Mats Carlsson et al.\@ from
ITA in Oslo and corrected for atmospheric turbulence effects by use of
Multi Frame Blind Deconvolution \citep{lofdahl02multi-frame}. We
degraded it to 9.8~cm hexagonal (edge to edge) subpupil resolution at
500~nm. This degraded image was shifted by integer steps from 0 to 20
times the high-resolution pixels, as well as in steps of 10 pixels
from 30 to 60. The so degraded and shifted GIs were box-car compressed
by a factor 10 to 200$\times$200 pixels of size 0\farcs41. This
procedure gives images with known subpixel shifts, $\delta x$ and
$\delta y$, without any re-sampling, except for the
compression. Figure~\ref{fig:degraded} shows a sample compressed
image.

The data with 0--20 high-resolution pixel shifts were made to test
subpixel accuracy, while the 30--60-pixel shifts are for testing
linearity with larger shifts.

The diffraction limited resolution,
$\lambda/D_\text{sub}\approx1\arcsec$ at 500~nm, corresponds to
$>2$~pixels. This means subpixel accuracy corresponds to
super-resolution accuracy.

The resulting images had more contrast than the real data from our SH
WFS, so some bias was added to change the RMS contrast to $\sim$3\% of
the mean intensity. The resulting images were stored in two versions,
with and without Gaussian noise with a standard deviation of
0.5\%\footnote{This noise level corresponds to photon noise from a CCD
  with 40 ke$^-$ full well. For a WFS set up to be running all day,
  the noise level would be larger when the sun is at low elevation and
  the image therefore darker. However, it is likely that the
  performance is then limited by other effects than noise, such as
  image warping from anisoplanatism.}. The digitization noise of a
12-bit camera is insignificant compared to the Gaussian noise but may
be significant for an 8-bit camera. We do not include the effects of
digitization in our simulations.

\subsection{Processing}
\label{sec:perfect-processing}

The 200$\times$200-pixel FOV is much larger than the FOV of the SH
WFS, which allows the use of many different subfields in order to get
better statistics. Centered on each of 17$\times$17 grid positions,
subimages, $g$, of size 16$\times$16 or 24$\times$24 pixels, were
defined. The subimages defined in the unshifted reference image,
$g_\text{ref}$, were larger in order to accommodate a shift range
limited to $\pm8$~pixels along each axis direction, except for CFF,
which uses two images of equal size. Note also that for CFF, the size
of the correlation matrix is limited by the subimage size. For
16$\times$16-pixel subfields, this limits the range to $\pm6$ pixels
(in reality to even less).

The different sizes of $g$, 16$\times$16 and 24$\times$24 pixels, have
two purposes: 1) We want to see how a change in size affects some of
the methods and 2) we will compare CFF using 24$\times$24 pixels with
the other methods using 16$\times$16 pixels. If the image geometry on
the detector accommodates a 24$\times$24-pixel $g_\text{ref}$, then it
can also accommodate 24$\times$24-pixel $g$ subfields if no oversize
reference image is needed.

We measured the shifts with each combination of CAs in
Sect.\@~\ref{sec:corr-algor} and IAs in
Sect.\@~\ref{sec:subpixel-interp}. We do this with and without noise
and with and without multiplying the reference image by 1.01, giving
an approximate 1\% bias mismatch. The bias mismatch sensitivity is
investigated because it is known to be a problem with the ADF and
ADF$^2$ methods.

\begin{figure*}[!t]
  \centering
  \def\figwidth{0.48\textwidth}
    \includegraphics[bb=18 8 494 343,clip,width=\figwidth]{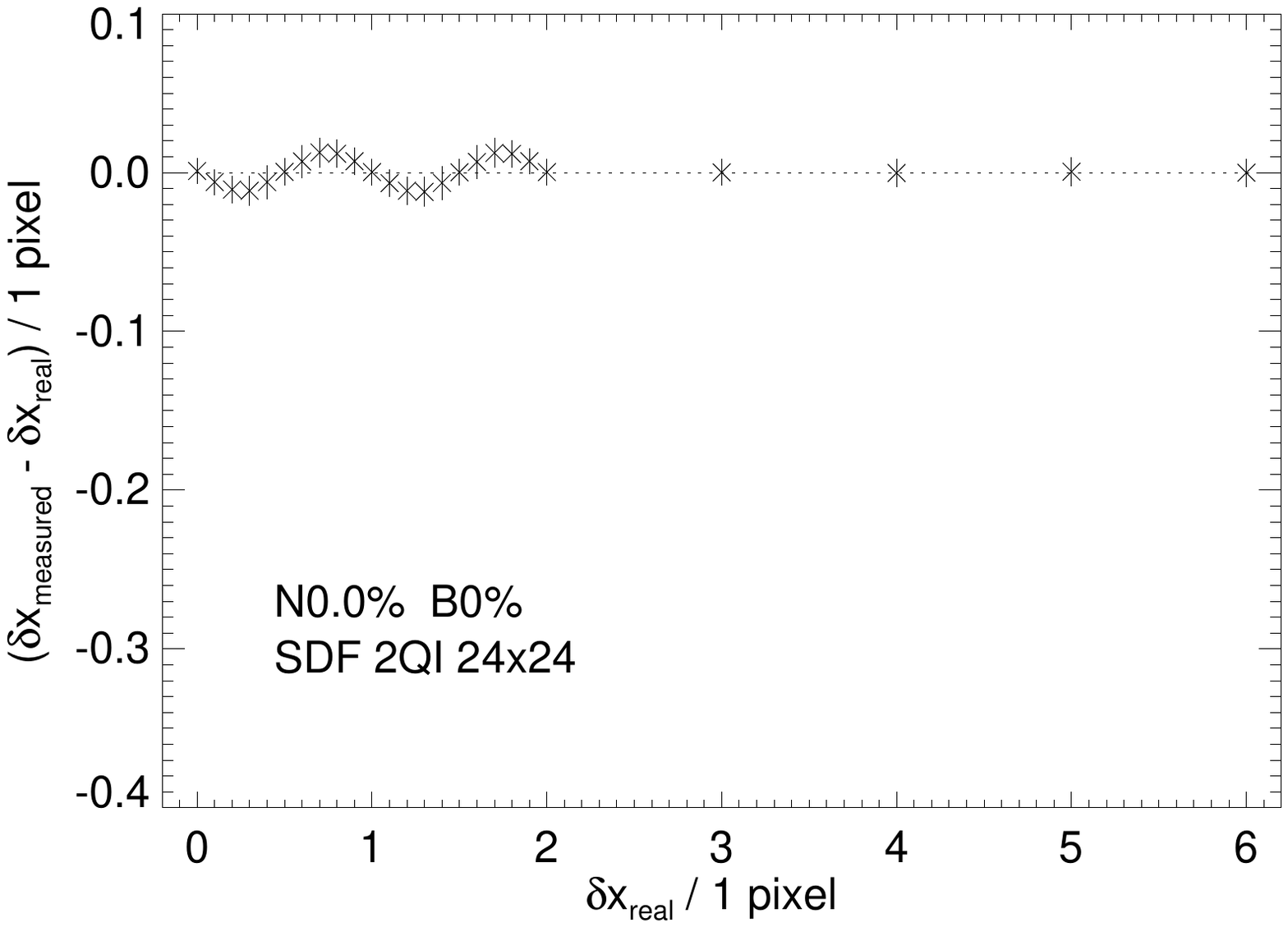}\quad
    \includegraphics[bb=18 8 494 343,clip,width=\figwidth]{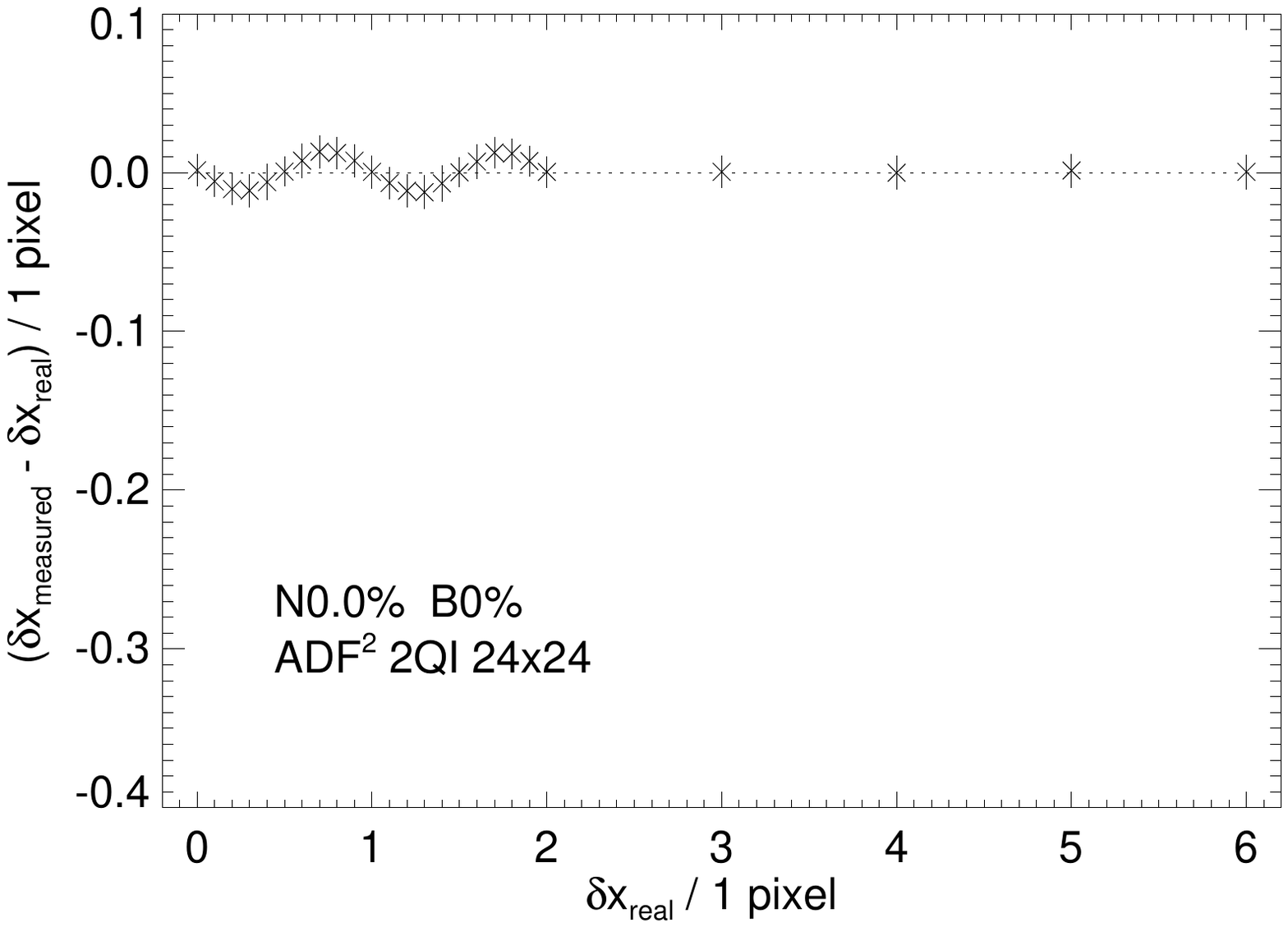}\\
    \includegraphics[bb=18 8 494 343,clip,width=\figwidth]{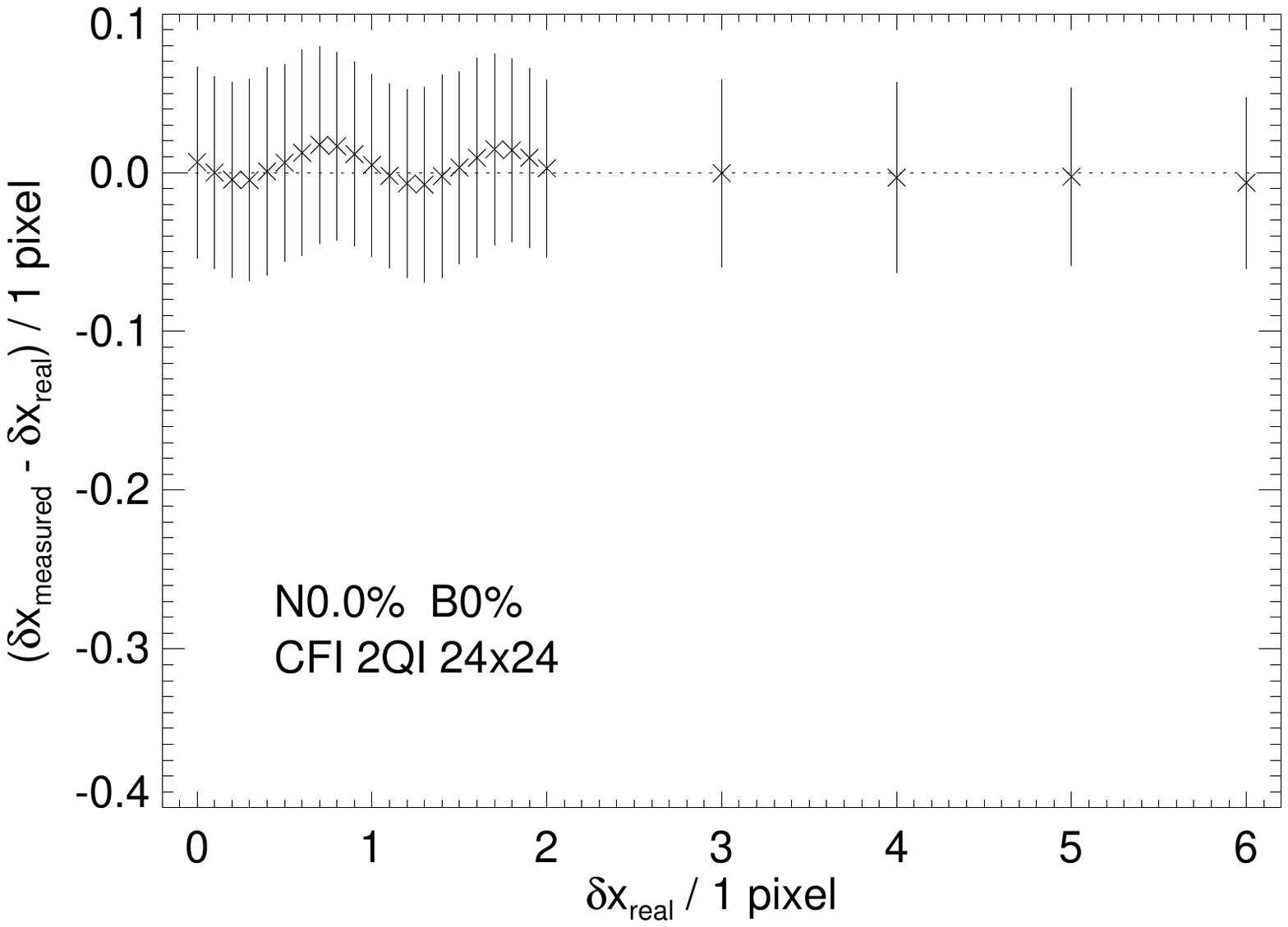}\quad
    \includegraphics[bb=18 8 494 343,clip,width=\figwidth]{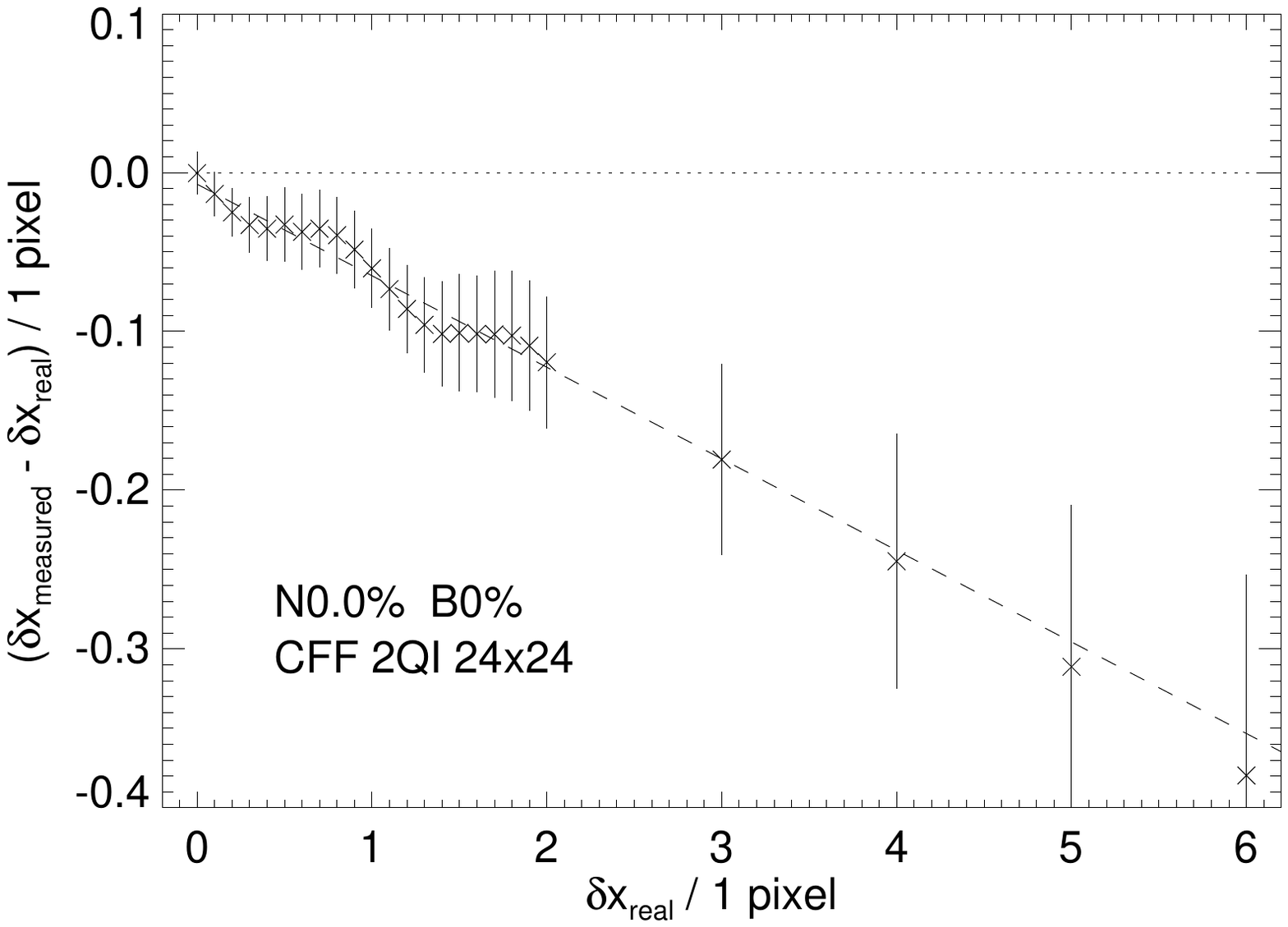}
  \caption{Errors in measured shifts vs.\@ real shifts for four
    different CFs, zero noise $N$, and zero bias mismatch $B$. Mean
    values and $\pm1\sigma$ error bars calculated with robust
    statistics. The dashed line in the CFF panel corresponds to
      $p_1-1$, where $p_1$ is fitted to Eq.~(\ref{eq:linesineone}).}
  \label{fig:perfect-scatter-four-methods}
\end{figure*}

For each real shift, $\delta x_\text{real}$ (and $\delta
y_\text{real}$, we will simplify the notation by referring to both
axis directions with $x$ when possible), we calculate the mean and
standard deviation, $\sigma$, of the measured shifts, $\delta
x_\text{measured}$, by use of the outlier-robust statistics based on
Tukey's Biweight as implemented in the IDL Astronomy User's Library
\citep{1993ASPC...52..246L}.

After removing $4\sigma$ outliers, we fit the data to the relationship
\begin{equation}
  \label{eq:linesineone}
  \delta x_\text{measured} 
  = p_0 + p_1\delta x_\text{real} + p_2\sin(a\, \delta x_\text{real}),
\end{equation}
where $a=2\pi/1$~pixel, by use of the robust nonlinear LS curve
fitting procedure MPFIT \citep{more78levenberg,2009ASPC..411..251M}.
The linear coefficient, $p_1$, of these fits is listed in the tables
below.

\subsection{Results}
\label{sec:perfect-results}

In Fig.\@~\ref{fig:perfect-scatter-four-methods}, the SDF, ADF$^2$,
CFI, and CFF CAs are compared for the case of no noise and no bias
mismatch, using the 2QI IA and 24$\times$24-pixel subfields. The
errors are a mix of systematic and random errors. The SDF and ADF$^2$
algorithms give a tighter correlation between the true and measured
image shifts than the CFI and CFF algorithms.
\citet{1988ApJ...333..427N} found similar undulating effect of small
errors near whole and half pixel shifts calculated by CFF and larger
errors in between. \citet{ballesteros96two-dimensional} make a similar
observation with ADF.

The ADF and CFF methods produce many outliers, which is the main
reason we need the robust statistics mentioned in
Sect.~\ref{sec:perfect-processing}. A \emph{very} small fraction of
these outliers, on the order $10^{-5}$ of all cases for CFF, zero for
all other CAs, are caused by the correlation matrix minimum falling on
an outer row or column. The rest are caused by false minima in the
correlation matrix, which happen to be deeper than the real minimum.
This can happen occasionally if a secondary minimum is located on or
near a grid point, while the real minimum is between grid points.  The
effect is more severe for ADF because of the more pointy shape of its
minimum. For CFF, minima corresponding to large shifts are attenuated
by two effects: apodization lowers the intensity away from the center
of the subimage and the digital Fourier transform wraps in
mismatching structure from the other side of the subimage.  This can
lead to detection of false minima corresponding to small shifts.

The dashed line in the CFF panel represents the linear part of the fit
to Eq.~(\ref{eq:linesineone}); its slope is $p_1-1$. The CFF method
systematically underestimates large shifts. Comparison between the
fitted line and the mean values shows a slight nonlinearity in the
CFF method, making the underestimation worse for larger shifts. 

The deviation of $p_1$ from unity and the undulations are systematic
errors. The former can be calibrated while the latter mix with the
random errors represented by the error bars.

\begin{figure*}[!t]
  \centering
  \def\figwidth{0.48\textwidth}
    \includegraphics[bb=12 8 494 343,clip,width=\figwidth]{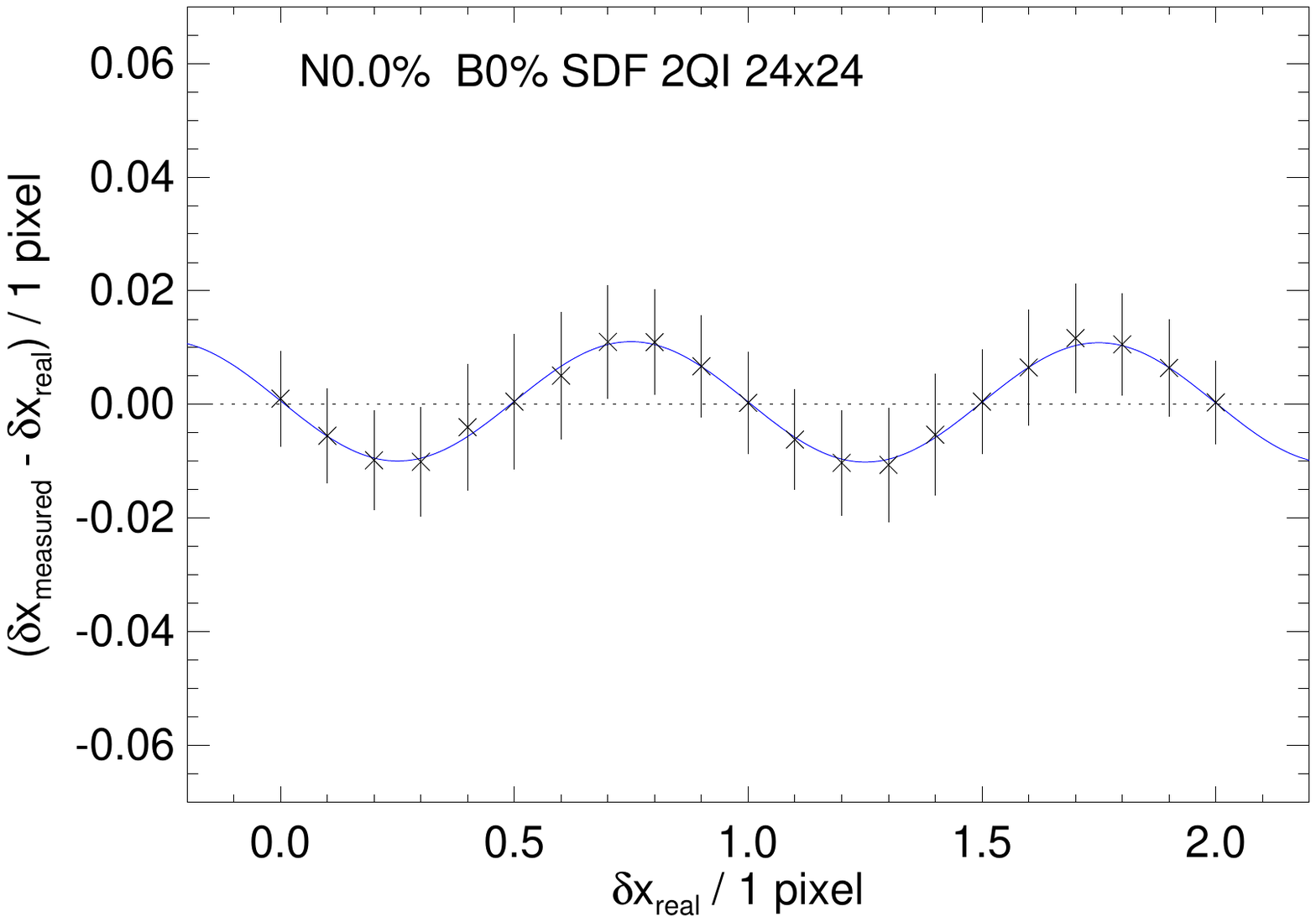}\quad
    \includegraphics[bb=12 8 494 343,clip,width=\figwidth]{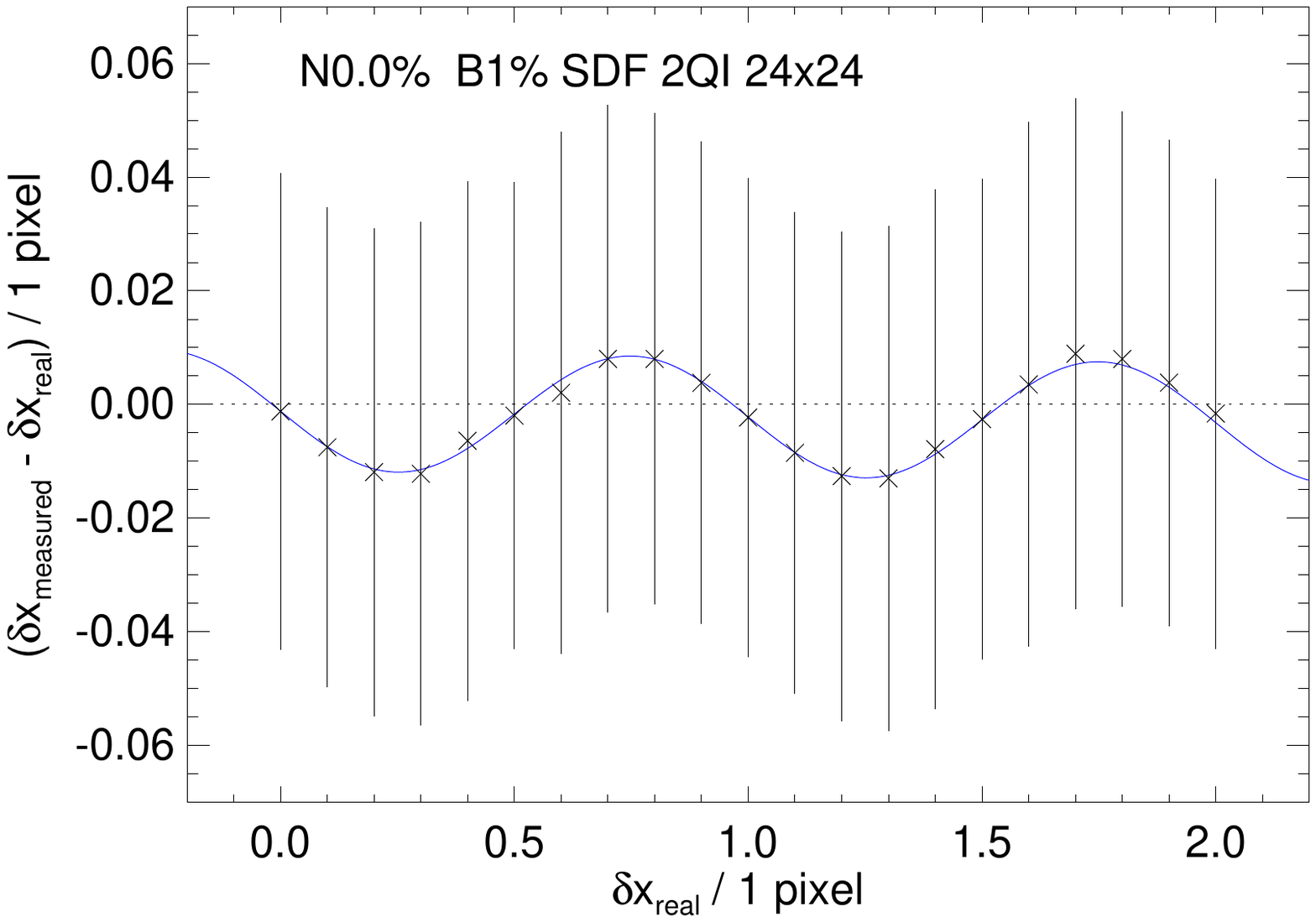}\\
    \includegraphics[bb=12 8 494 343,clip,width=\figwidth]{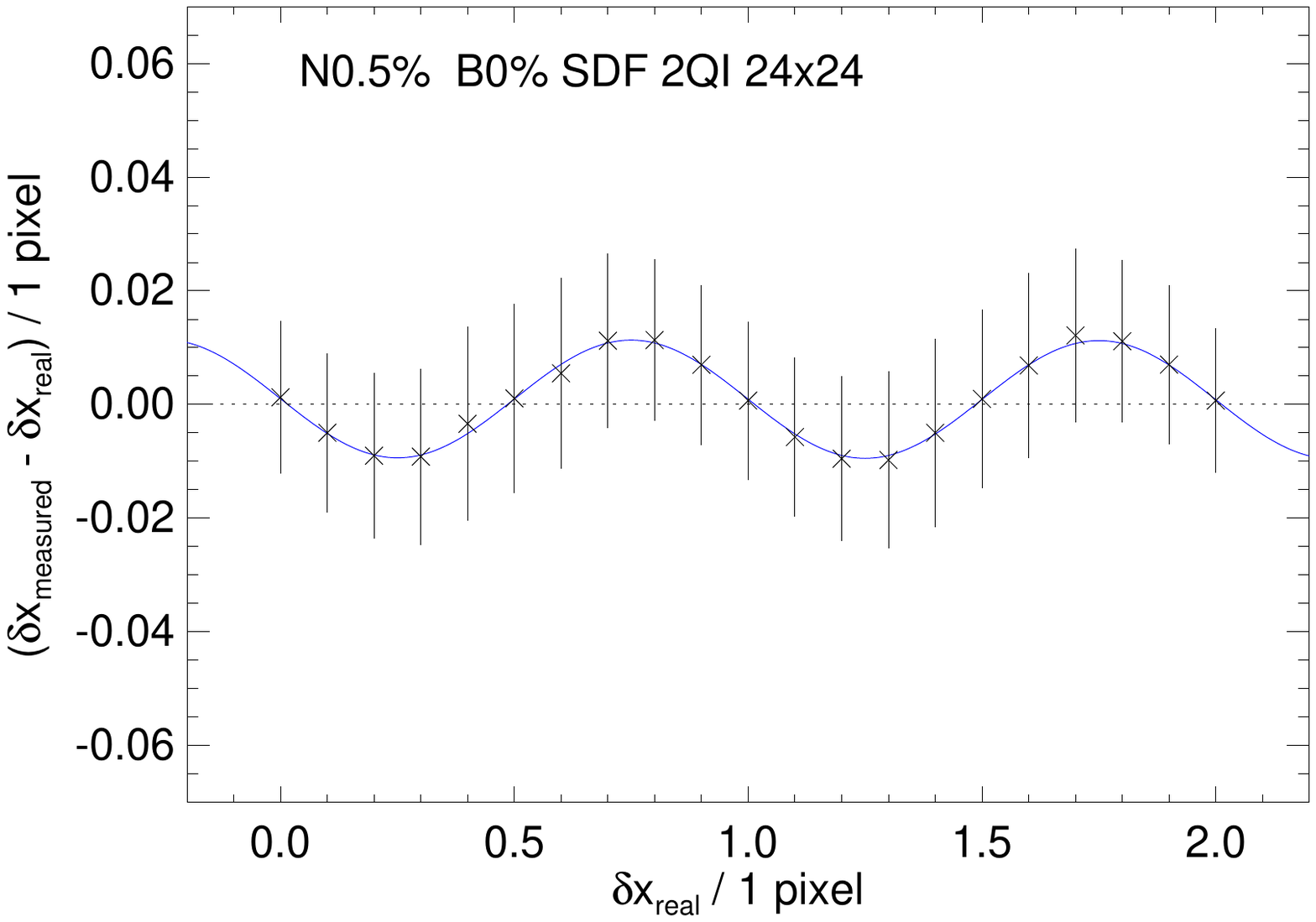}\quad
    \includegraphics[bb=12 8 494 343,clip,width=\figwidth]{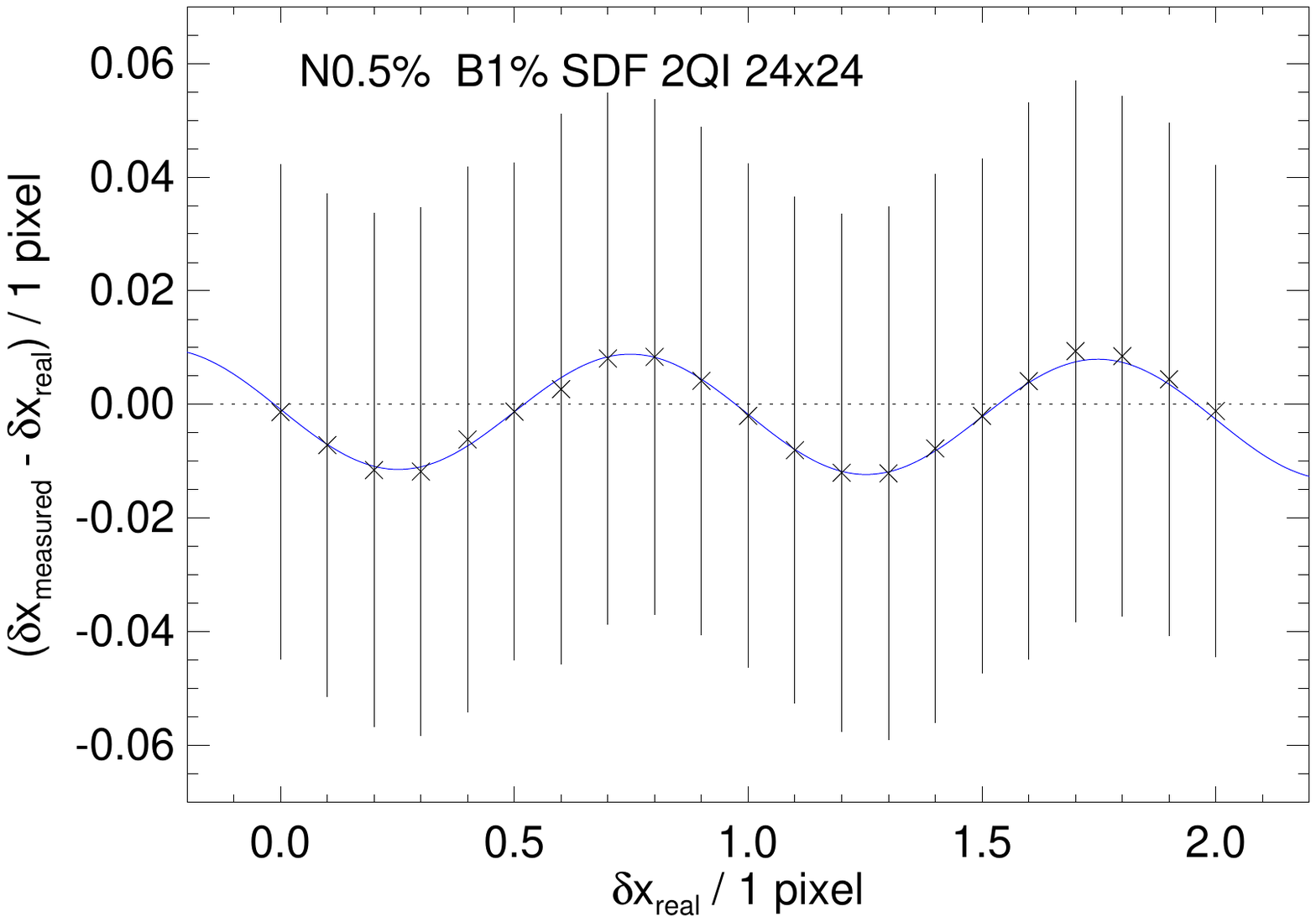}
  \caption{Results using The SDF/2QI methods with 24$\times$24-pixel
    subfields. Small shifts, $\delta r\protect\la2$~pixels, and different
    combinations of noise, $N$, and bias mismatch, $B$, as indicated
    in the labels of each tile. Error bars are $\pm 1\sigma$. The blue
    line is a fit to Eq.~(\ref{eq:linesineone}).}
  \label{fig:perfect-scatter-SDF-different-versions}
\end{figure*}

Figure~\ref{fig:perfect-scatter-SDF-different-versions} illustrates
the effects of noise and of bias mismatch for SDF. Here we limit the
plots to $\delta r=(\delta x_\text{real}^2+\delta
y_\text{real}^2)^{1/2}\le 2$~pixels. The blue line corresponds to the
fit of Eq.~(\ref{eq:linesineone}). The amplitudes of the undulations,
$p_2$, is $\sim$0.01~pixel. Noise does not appear to change the
undulating errors significantly but it does make the random errors
more dominant. Bias mismatch has a similar effect and overwhelms the
differences due to noise. It also appears to change a minute
systematic overestimation of the shifts to an equally small
underestimation.

\subsubsection{Tables}
\label{sec:tables}

The complete results of these simulations are presented in
Tables~\ref{tab:sim-n0b0}--\ref{tab:sim-n5b0-flat}.

While the standard deviations in the plots are for each real shift
individually, and therefore do not include the undulations, the
standard deviations in the tables are for intervals of real shifts and
therefore do include the undulations. The measured shifts might be
expected to have near-Gaussian distributions, which is true for
ADF$^2$ and SDF and mostly for CFI. ADF often has complicated,
multi-peak distributions. All CFF distributions are double-peaked
and/or asymmetrical.

We give the results for all shifts, $\delta r$, but also separately
for small shifts ($<1$ pixels and $<2$ pixels, resp.), medium shifts
(shifts between 3 and 5 pixels in length), and large shifts ($>5$
pixels). The small shifts are relevant to AO performance in closed
loop. The large shift results tell us something about performance in
open loop, which is relevant for wavefront sensor calibration, site
testing, image restoration, and when trying to close an AO loop.

\subsubsection{Identical images}
\label{sec:perfect-data}

In Table\@~\ref{tab:sim-n0b0}, we show the performance of the
different methods with noise free data.

We begin by noting that in many cases the errors for large shifts are
smaller than those for small shifts. This may seem counterintuitive
but it is simply a consequence of using only whole pixel shifts for
the larger shifts. The IAs have smaller systematic errors on the grid
points, see the undulations in
Fig.\@~\ref{fig:perfect-scatter-four-methods}.

The ADF CA produces many $4\sigma$ outliers. It is not surprising that
it gives much worse results than ADF$^2$, because the two CAs by
definition share the location of the whole pixel minimum but the ADF
minimum does not have the parabolic shape assumed by the IAs.

SDF and ADF$^2$ are clearly better than the CFI and CFF methods. CFF
can compete for the very smallest shifts ($<1$~pixel) if we use
24$\times$24 pixels for CFF and 16$\times$16 pixels for the other CAs
(which is the most fair comparison, since the optics need to
accommodate 24$\times$24 pixels in order to get the oversize reference
image for the non-CFF CAs).

Except for CFF, the errors appear more or less independent of the
magnitude of the shifts. CFF deteriorates significantly in three ways
at larger shifts: The number of outliers increases, the random error
increases, and the shifts are systematically underestimated, as shown
by the nonunity slopes. The former two effects can be explained by the
assumption of periodicity of the digital Fourier transform. The signal
for large shifts is diluted by mismatching granulation shifted in from
the opposite end of the FOV.

For SDF and ADF$^2$, the 2D IAs are clearly better than the 1D ones.
2QI is marginally better than 2LS. The best results with SDF and
ADF$^2$ are an error RMS of less than 0.02 pixels, corresponding to
0\farcs008. Increasing the subfield size from 16 to 24 pixels squared
reduces the error by approximately 30\%.

\subsubsection{Noisy images}
\label{sec:noisy-data}

Adding 0.5\% noise to the images give the results shown in
Table\@~\ref{tab:sim-n5b0}. The errors grow but the behavior is
similar to the zero-noise case.

The best results with SDF and ADF$^2$ 16$\times$16 are an error RMS of
less than 0.03 pixels, corresponding to 0\farcs012. CFF with
24$\times$24 gives similar results for small shifts.
As in the case of zero noise, increasing the subfield size from 16 to
24 pixels squared reduces the error by approximately 30\%.

With 24$\times$24 pixels, SDF and ADF$^2$ give results similar to the
16$\times$16-pixel zero-noise case. Surprisingly, the number of
outliers for ADF is significantly reduced by the added noise. But
$\sigma$ increases more than for ADF, so this may be an effect of
making the error distribution more Gaussian.

\subsubsection{Bias mismatch}
\label{sec:bias-mismatch}

Subtracting a fitted plane (or just the mean intensity as in our
experiment, see Sect.\@~\ref{sec:corr-algor}) from each subimage
removes mismatches in intensity bias for CFI and CFF. We did not do
this for the difference based CAs (SDF, ADF, and ADF$^2$), where a
consistent bias cancels. However, if there is a bias mismatch between
the images this cancellation is not effective. Such bias mismatch
could come from, e.g., variations in a thin cloud layer in the case
when $g_\text{ref}$ is not from the same exposure as $g$, or small
drifts in the pupil location on the SH, causing variations in the
light level from the outermost subpupils.

We re-processed all the data after multiplying the reference image by
1.01, introducing a 1\% bias mismatch with the other images.  In
Tables\@~\ref{tab:sim-n0b1} and~\ref{tab:sim-n5b1} we show the results
for the difference based CAs.  As expected, the CFI and CFF results
did not change from the ones in Tables\@~\ref{tab:sim-n0b0}
and~\ref{tab:sim-n5b0} and they are therefore not repeated.
 
For small shifts, the SDF and ADF$^2$ results are now worse than for
CFF, even when comparing the methods using the same image size.
SDF is more robust against bias mismatch than ADF$^2$. There is now no
real difference between ADF and ADF$^2$, possibly indicating that the
parabolic shape of the ADF$^2$ correlation function is destroyed by
the bias mismatch. The SDF error RMS is 0.7 pixels for 16$\times$16
and 0.4 pixels for 24$\times$24, which makes it the best method for
large shifts.

With a bias mismatch, 0.5\% noise added to the images does not
significantly change the results for the difference based CAs.
We conclude that bias mismatch should be removed in pre-processing by
subtraction of the intensity mean.

\subsubsection{Window function}
\label{sec:window-function}

\begin{figure}[tb]
  \centering
  \includegraphics[width=0.4\linewidth]{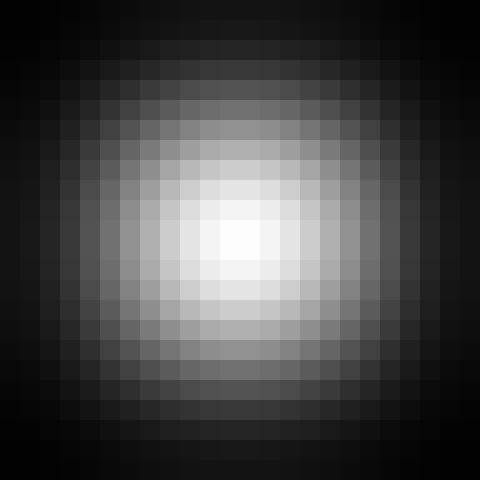}\quad 
  \includegraphics[width=0.4\linewidth]{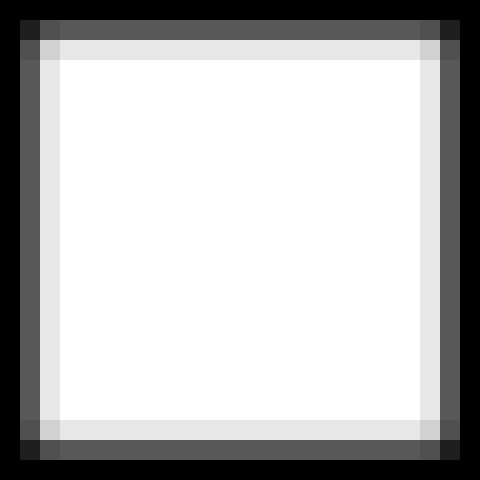}
  \caption{Window functions, 24$\times$24 pixels. \textbf{Left:}
    Hamming window. \textbf{Right:} Flat-top window.}
  \label{fig:windows}
\end{figure}

The CFF method requires apodization, i.e., the multiplication of the
subimage (after subtracting the fitted plane) by a window function.
The intention is to reduce ringing effects from the discontinuities
caused by Fourier wrap-around. For the results above, we used a 2D
Hamming window. In
Tables~\ref{tab:sim-n0b0-flat}--\ref{tab:sim-n5b0-flat} we show the
results when we instead use a window with the center $\sim$50\% of the
area flat, and a taper only in the pixels outside this area. This is
similar to the window used by \citet{waldmann07untersuchung}. See
Fig.~\ref{fig:windows} for the two types of window functions.

The results are mixed. The errors are larger but there are fewer
outliers. The results are similar with and without noise. The
systematic underestimation is reduced, resulting in slopes closer to
unity. Based on these data we cannot say which window is better, they
would have to be evaluated specifically for any new situation where
one wants to use CFF. The important result for our purposes is that
the comparison between SDF and ADF$^2$ vs CFF does not depend on the
window function used.

\section{Image shift as a measure of wavefront tilt}
\label{sec:image-shift-versus}

For wavefront sensing, the quantities that really need to be measured
are the average wavefront gradients at the positions corresponding to
the subapertures. One assumes that a shift in image position
corresponds perfectly to the average gradient of the wavefront across
the subaperture. However, in addition to local gradients, continuous
wavefront aberrations across the telescope aperture also result in
local wavefront curvature. Is the assumption valid anyway and how good
is it for different seeing conditions, as quantified with Fried's
parameter $r_0$?

Using Kolmogorov statistics for different $r_0$ without any assumption
of partial correction by an AO system makes the statistics from this
experiment relevant to open-loop WFS. I.e., systems for measuring
seeing statistics \cite[e.g., ][]{scharmer10s-dimm+} and the capture
phase for AO systems, but not necessarily AO systems in closed loop.

\subsection{Artificial data}
\label{sec:seeing-data-recipe}

\begin{figure*}[!t]
  \centering
  \def\figwidth{0.28\linewidth}
  \def\figwidth{0.3\textwidth}
  \def\figwidth{4cm}
  \centering
  \subfloat[\label{fig:hexwavefront}]{\includegraphics[bb=112 112 912 912,clip,width=\figwidth]{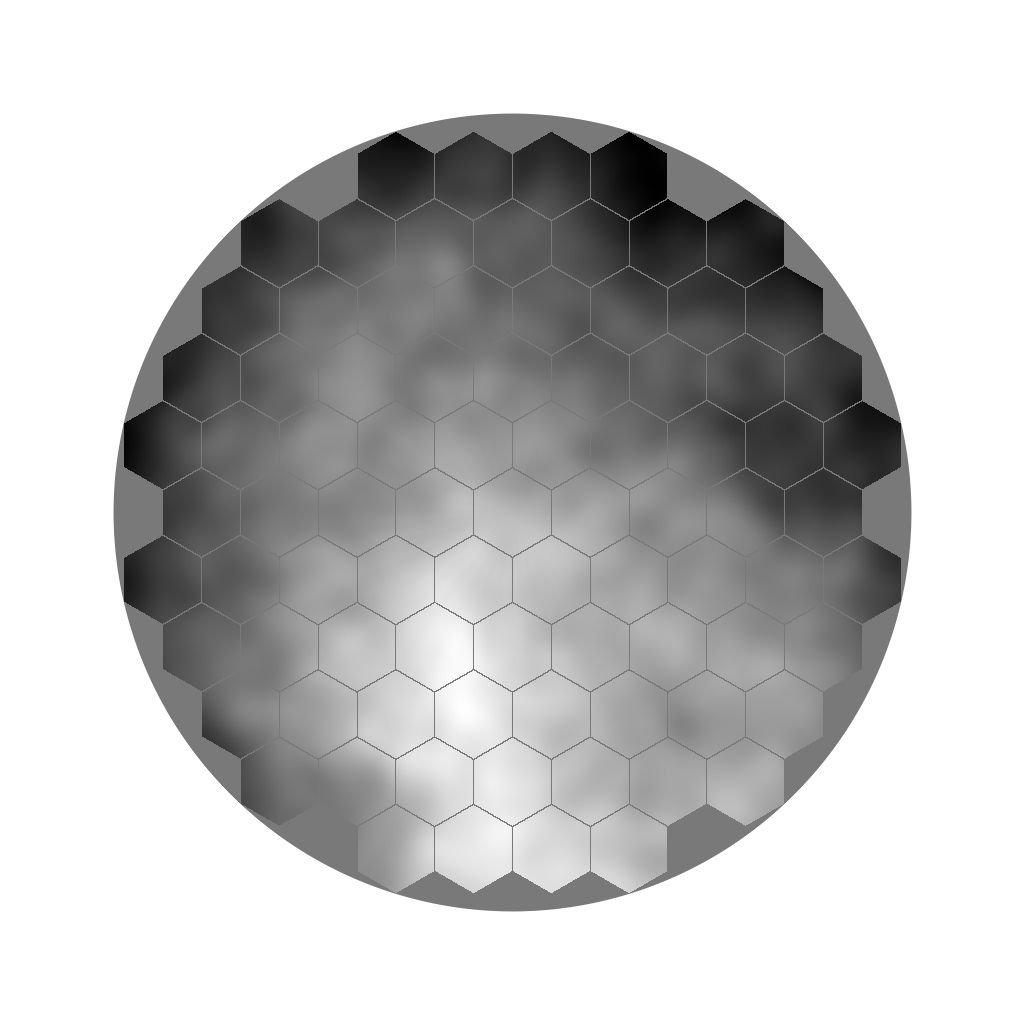}}\quad
  \subfloat[\label{fig:hexplanes}]{\includegraphics[bb=112 112 912 912,clip,width=\figwidth]{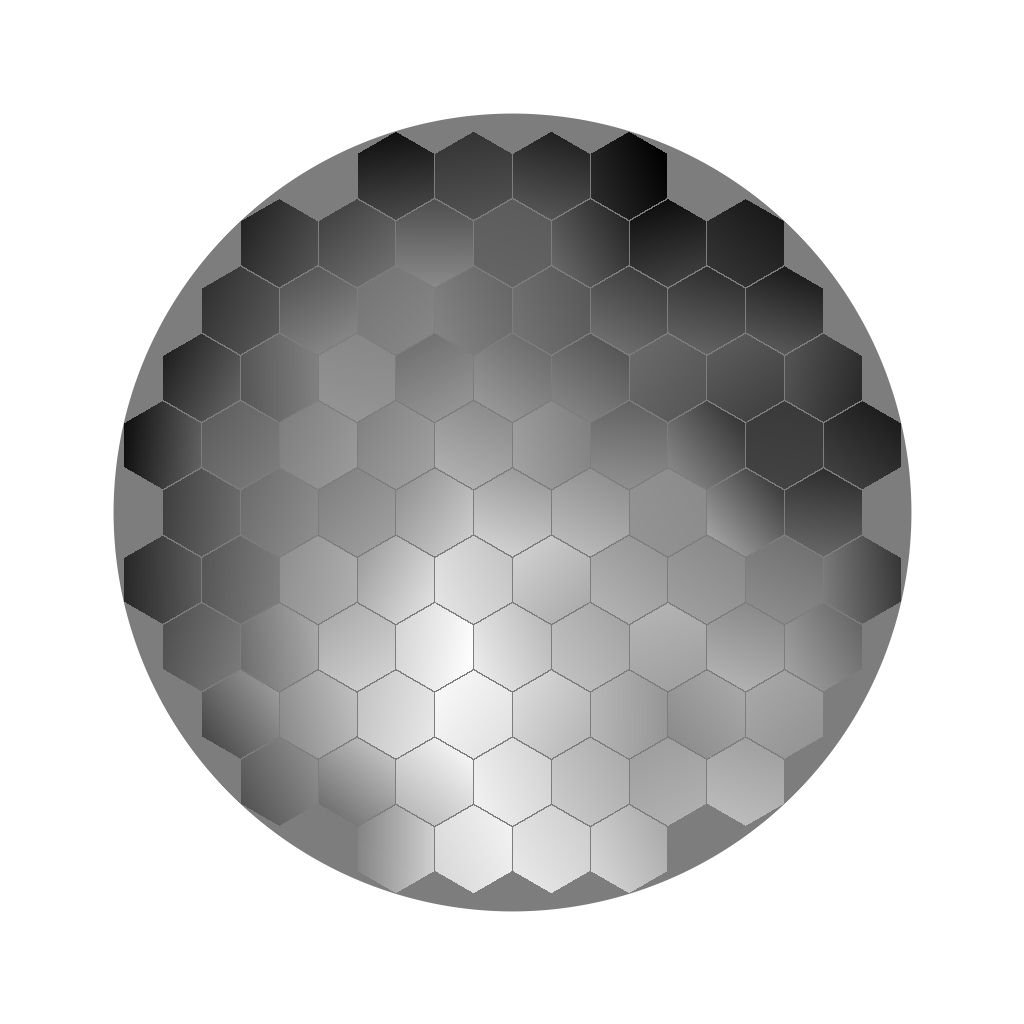}}\quad
  \subfloat[\label{fig:hextilts}]{\includegraphics[bb=112 112 912 912,clip,width=\figwidth]{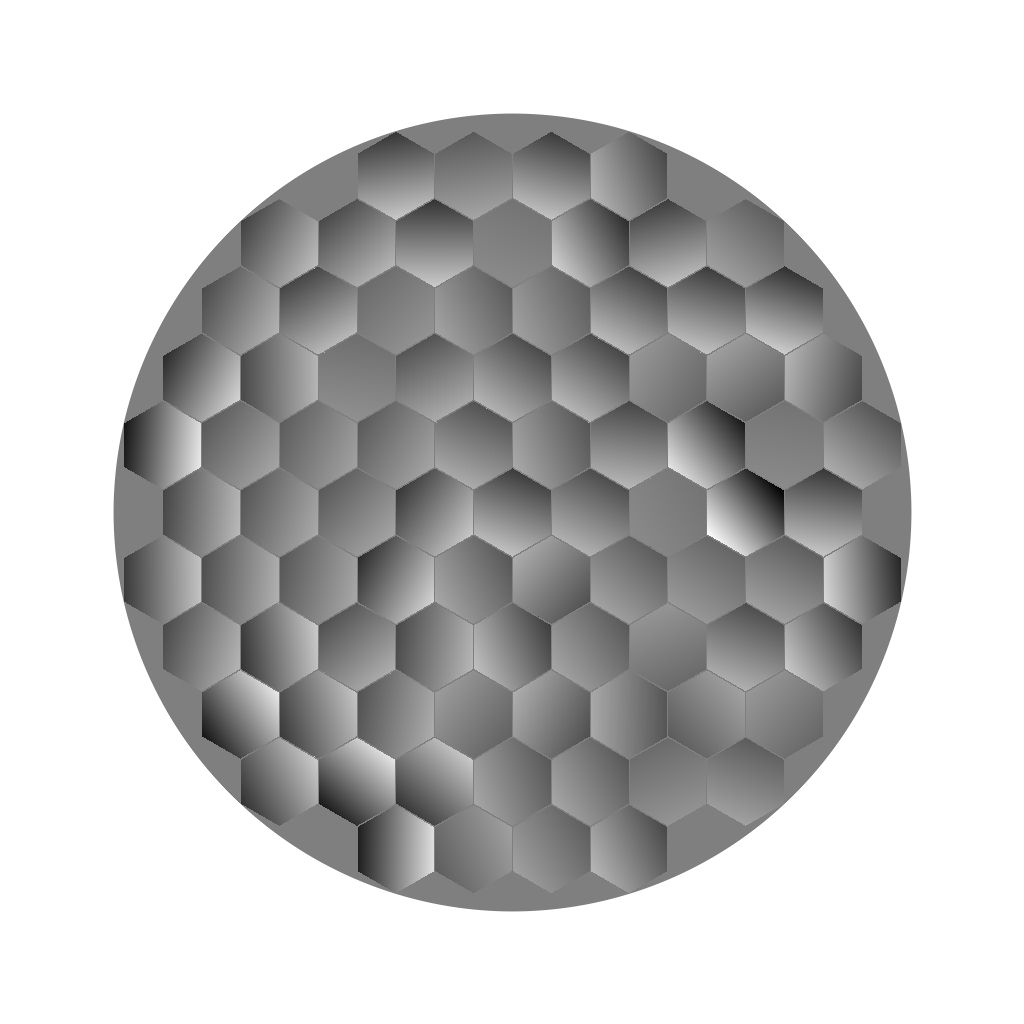}}
  \caption{Sample wavefront and the subpupil pattern of 85 hexagonal
    microlenses. The circle represents the 98~cm SST pupil and the
    gaps between adjacent microlenses have been exaggerated for
    clarity. The geometry is actually such that there is no space
    between the microlenses.
    \textbf{\protect\subref{fig:hexwavefront}}~The 1003-mode simulated
    wavefront phase. \textbf{\protect\subref{fig:hexplanes}}~Piecewise
    plane approximation. \textbf{\protect\subref{fig:hextilts}}~Local
    tilts only.}
  \label{fig:hexfig}
\end{figure*}

The setup corresponds to a filled 98-cm pupil (like the SST) with 85
subapertures, each 9.55~cm edge-to-edge. In
Fig.\@~\ref{fig:hexwavefront} we show one sample Kolmogorov wavefront
phase. The superimposed pattern shows the geometry of 85 hexagonal
microlenses circumscribed by a telescope pupil.
Figure~\ref{fig:hexplanes} demonstrates the local plane approximation
implicit in SH wavefront sensing, while Fig.\@~\ref{fig:hextilts}
shows just the tilts.

The exact geometry does not matter for the results reported here,
since our tests do not involve the step where wavefronts are
reconstructed from the shift measurements. However, the geometry
discussed is along the lines planned for the next generation SST AO
system, and in that respect motivates the particular subaperture size
and shape investigated.

We generated 100 wavefront phases, $\phi_i$, following Kolmogorov
statistics. For making each simulated phase screen, the following
procedure was used. 1003 random numbers were drawn from a standard
normal distribution, scaled with the square root of the atmospheric
variances and used as coefficients for atmospheric Karhunen--L\`oeve
(KL) functions 2--1004. We used KL functions based directly on the
theory of \citet{fried78probability}, as implemented by
\citet{dai95modal}. These modes are numbered in order of decreasing
atmospheric variance, and the exact range of indices is motivated by
KL$_1$ being piston and KL$_{1005}$ a circular mode, starting a higher
radial order. Figure~\ref{fig:hexwavefront} shows a sample wavefront
masked with the pattern of the 85 hexagonal microlens geometry.

For all simulations, we used a wavelength of 500~nm and a telescope
aperture diameter of $D_\text{tel}=98$~cm. In order to cover a range
of different seeing conditions, we scaled these wavefront phases to
different values of Fried's parameter, $r_0\in\{5,7,10,15,20\}$~cm, by
multiplying with $(D_\text{tel}/r_0)^{5/6}$. Using the same wavefront
shapes scaled differently like this, the performance for different
values of $r_0$ should be directly comparable.

For each random wavefront, separately scaled to each value of $r_0$,
and for each subpupil defined by a microlens, we generated an image
by convolving the GI in Fig.\@~\ref{fig:GI} with a PSF based on the
subpupil and the local wavefront phase.

We want to examine the effect of using different subfield sizes.
Increased subfield size can be used in different ways: Either one can
change the image scale, so the same amount of granulation fits in the
FOV but in better pixel resolution. Or one can keep the original image
scale so more granulation fits in the FOV. (Or something in between.)
Therefore we make images at three different image scales by box-car
compressing them by three different integer factors, 7, 10, and 13.
The resulting image scales are 0\farcs29/pixel, 0\farcs41/pixel, and
0\farcs53/pixel.

To summarize: the images we have generated were downgraded to the
resolution of the subpupil, shifted by the local wavefront tilt and
also somewhat blurred by the local wavefront curvature, the latter in
particular for data with small~$r_0$. A bias was then added to make
the RMS contrast of the granulation pattern approximately 3\% of the
mean intensity.

\subsection[]{Processing}
\label{sec:seeing-processing}

For each subfield size, noise, and $r_0$, relative shifts and tilts
were calculated for 490,000 randomly selected pairs of subpupil
images. We operate on image pairs corresponding to subpupils from
different random wavefronts, so results are not influenced
systematically by spatial correlations.

For the shift measurements, we use all the CF methods from
Sect.~\ref{sec:algorithms} except ADF, applied to the two images in a
pair. We use only the 2QI method for subpixel interpolation.

Shift measurements were calculated twice, with and without 1\% noise
added to the images. We increased the noise level from the 0.5\% used
in Sect.~\ref{sec:noisy-data} to make the effect of noise clearer and
thus allow better comparison of different methods. We do not
investigate bias mismatch in this experiment. The conclusion from
Sect.~\ref{sec:algorithm-accuracy} is that bias mismatch should be
compensated for before applying the shift measurement methods.

For comparison with the shift measurements, $\delta x_\text{shift}$,
we fitted Zernike tip and tilt to each wavefront, within each
hexagonal subpupil. The relative wavefront tilt for an image pair is
the difference between the Zernike tilts for the two subpupils in the
pair. These relative tip/tilt coefficients in radians, $\alpha_x$, are
converted to image shift,
\begin{equation}
  \label{eq:5}
  \delta x_\text{tilt} = 
 \frac{2\lambda}{\pi r D_\text{tel}} \alpha_x
\end{equation}
where $r$ is the image scale in rad/pixel. We calculate the robust
statistics of the resulting shifts, remove $4\sigma$ outliers, and fit
the data to the relationship
\begin{equation}
  \label{eq:linesine2}
  \delta x_\text{shift} = p_0 + p_1\delta x_\text{tilt} 
  + p_2\sin(a\, \delta x_\text{tilt}),
\end{equation}
where $a=2\pi/1$~pixel. Compare Eq.~(\ref{eq:linesineone}).

\subsection{Results}
\label{sec:seeing-results}

Table~\ref{tab:seeing-simulation} shows the results from images with
image scale 0\farcs41/pixel and different subfield sizes, $N\times N$
pixels with $N=16$, 24, and 36. In seconds of arc, this corresponds to
6\farcs56, 9\farcs84, and 14\farcs76 squared. The results using the
additional image scales 0\farcs29/pixel and 0\farcs53/pixel and a
single array size, $N=24$, are in
Table~\ref{tab:seeing-simulation-comp}.

In addition to the tabulated results, we calculated the following: The
standard deviation of $\delta x_\text{tilt}$ varies linearly with
$r_0^{-5/6}$ as expected from Kolmogorov statistics:
$\sigma_\text{tilt}=2.68$, 2.02, 1.50, 1.07, 0.84~pixels $=1\farcs10$,
0\farcs83, 0\farcs62, 0\farcs44, 0\farcs34 for $r_0=5$, 7, 10, 15,
20~cm, resp. Correlation coefficients calculated after removal of
outliers are very high: they round to 1.00 in all cases, except CFF
and CFI at $N=16$ for which they are 0.98--0.99.

\subsubsection{Failures}
\label{sec:failures}

If the errors were normal distributed, using a $4\sigma$ limit for
defining outliers should give a failure rate of 0.0063\%. With almost
$5\cdot10^5$ samples we expect the normal distribution to be well
realized but the actual failure rates are larger. All CAs show an
excess but by far it is the CFF results that suffer from the highest
number of outliers, particularly for small subfields and small
$r_0$. CFI is also slightly worse than SDF and ADF$^2$.

\begin{figure}[tb]
  \centering
  \def\tilewidth{8.5cm}
  \includegraphics[bb=15 10 495 344, width=\tilewidth]{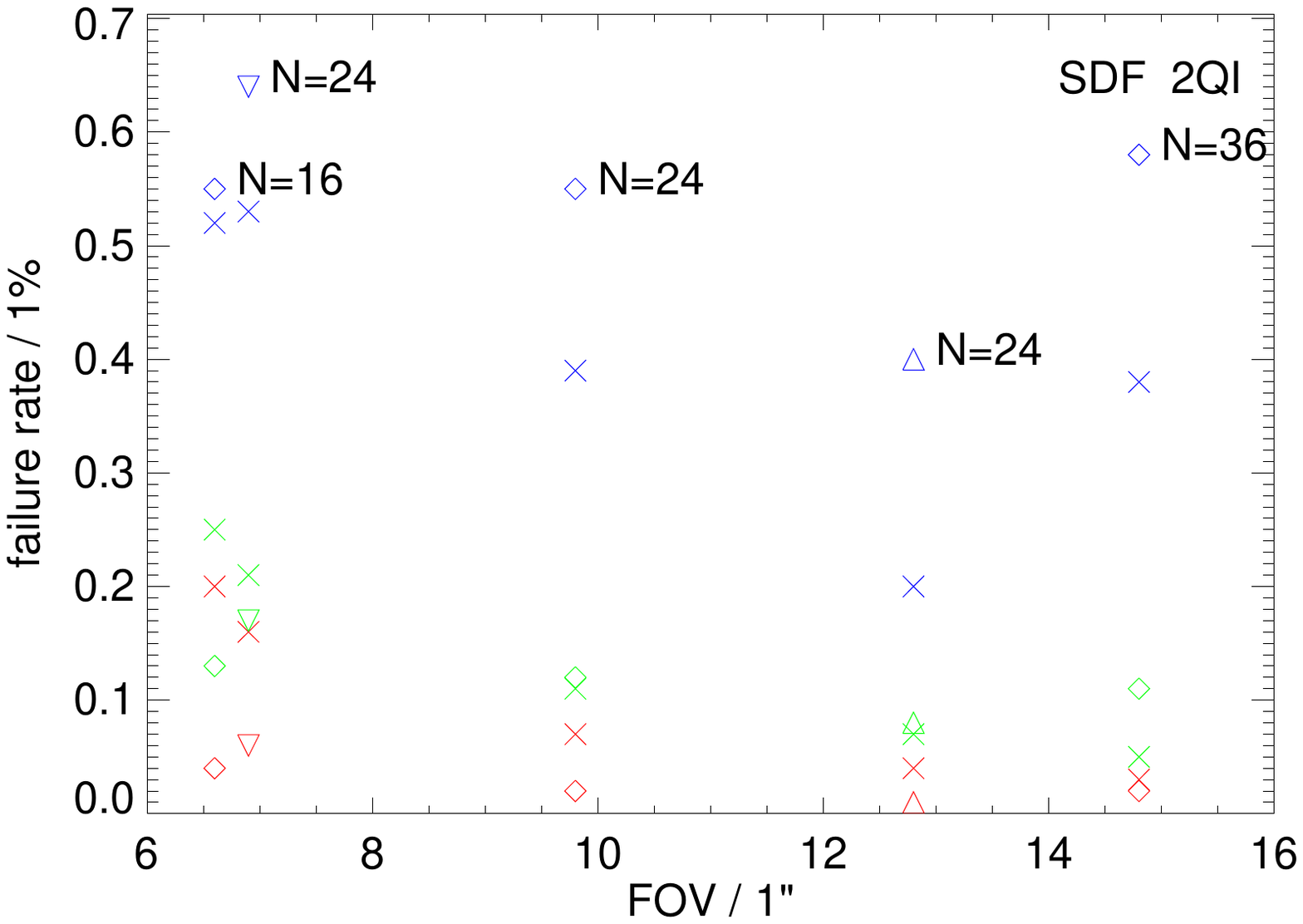}
  \includegraphics[bb=15 10 495 344, width=\tilewidth]{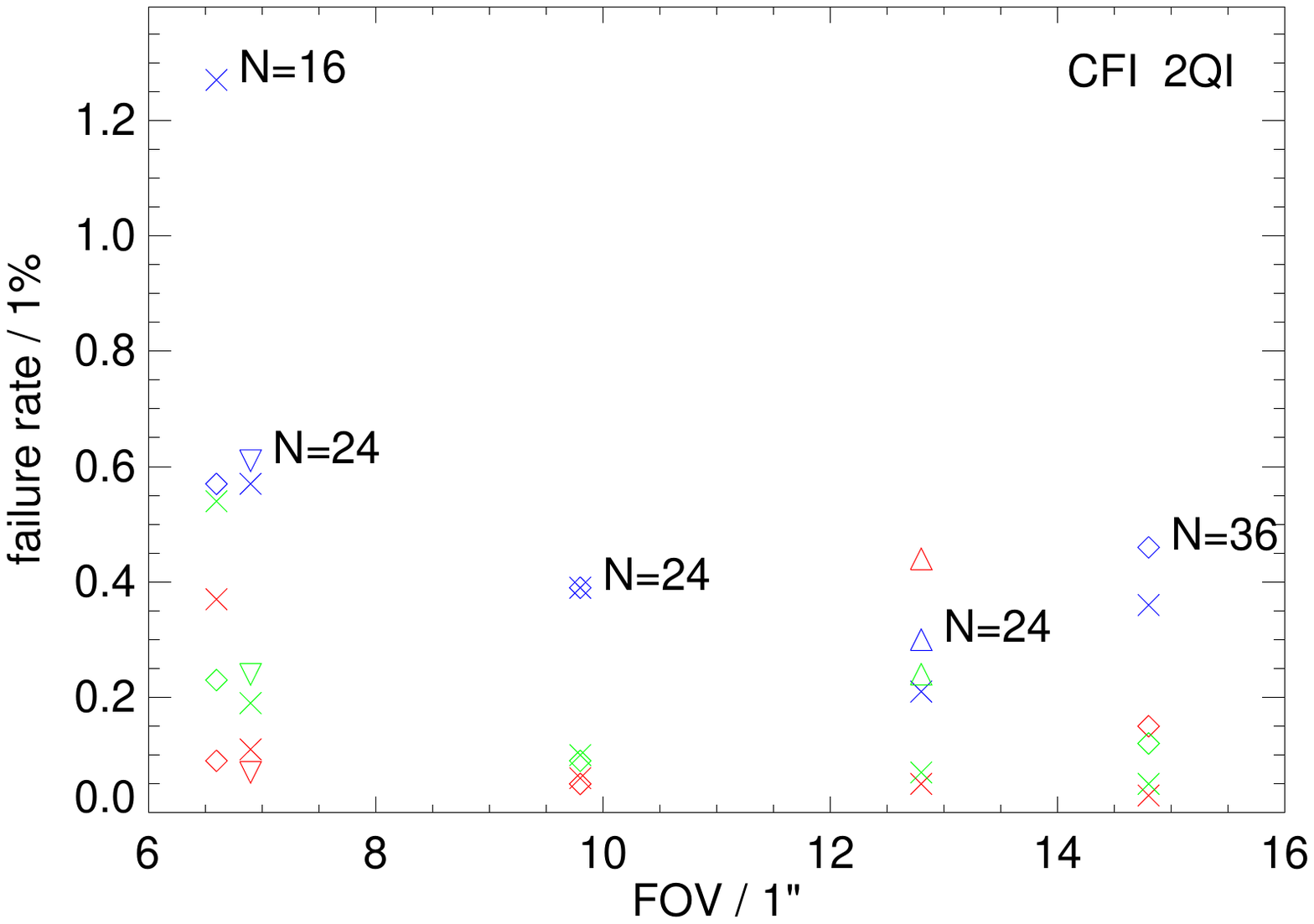}
  \includegraphics[bb=15 10 495 344, width=\tilewidth]{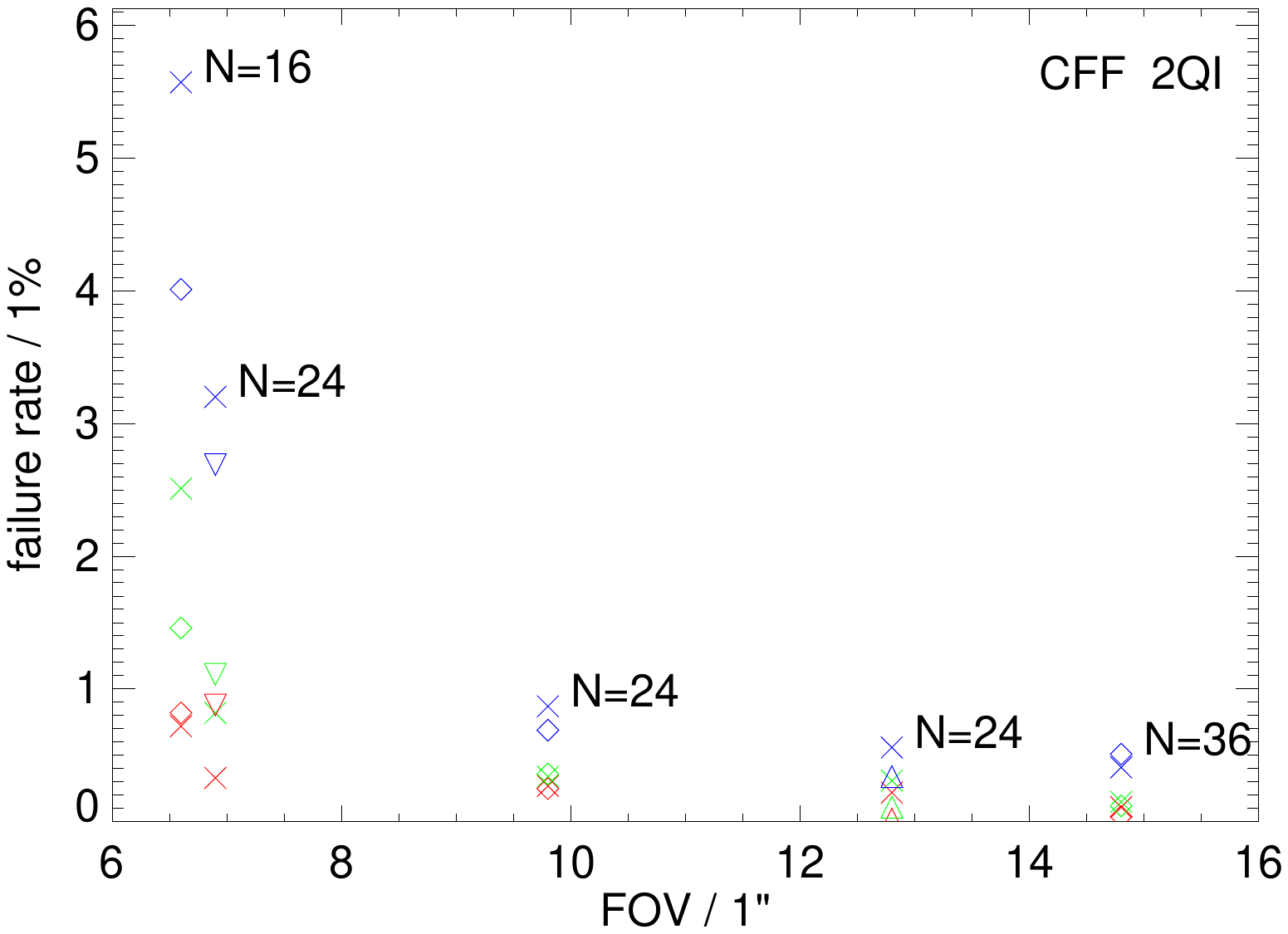}    
  \caption{Failure rates in percent for different FOV sizes and CAs.
    Red:~$r_0=20$~cm; Green:~$r_0=10$~cm; Blue:~$r_0=7$~cm. $N\times
    N$-pixel subfields. Diamonds~($\Diamond$): image scale
    0\farcs41/pixel, varying $N=\{16, 24, 36\}$ as labeled in the
    plots; Down triangles~($\triangledown$): 0\farcs29/pixel, $N=24$;
    Up triangles ($\vartriangle$): 0\farcs53/pixel, $N=24$.  All the
    above without noise. Crosses ($\times$): corresponding FOVs, with
    noise. Note different vertical scales.}
  \label{fig:fail}
\end{figure}

All failure rates are $\la10$\%. They are $>2$\% for $r_0=5$~cm with
all methods when using the smallest image scale, 0\farcs29/pixel, and
for small FOVs and $r_0=5$, 7~cm (and CFI $r_0=5$~cm, noise,
$N=16$). Figure~\ref{fig:fail} illustrates how failure rates sometimes
increase with noise, sometimes decrease. With noise, the failure rate
decreases with large FOV in arcsec (exception: $N=36$ and
$r_0=7$~cm). The variation is less systematic without noise.

\subsubsection{Shift errors}
\label{sec:shift-errors}

We now plot some of the results in pixels rather than seconds of arc
so we can compare with the results in
Sect.~\ref{sec:algorithm-accuracy}.

\begin{figure*}[!t]
  \centering
  \def\figwidth{0.3\linewidth}
  \includegraphics[bb=58 0 430 345,clip,width=\figwidth]{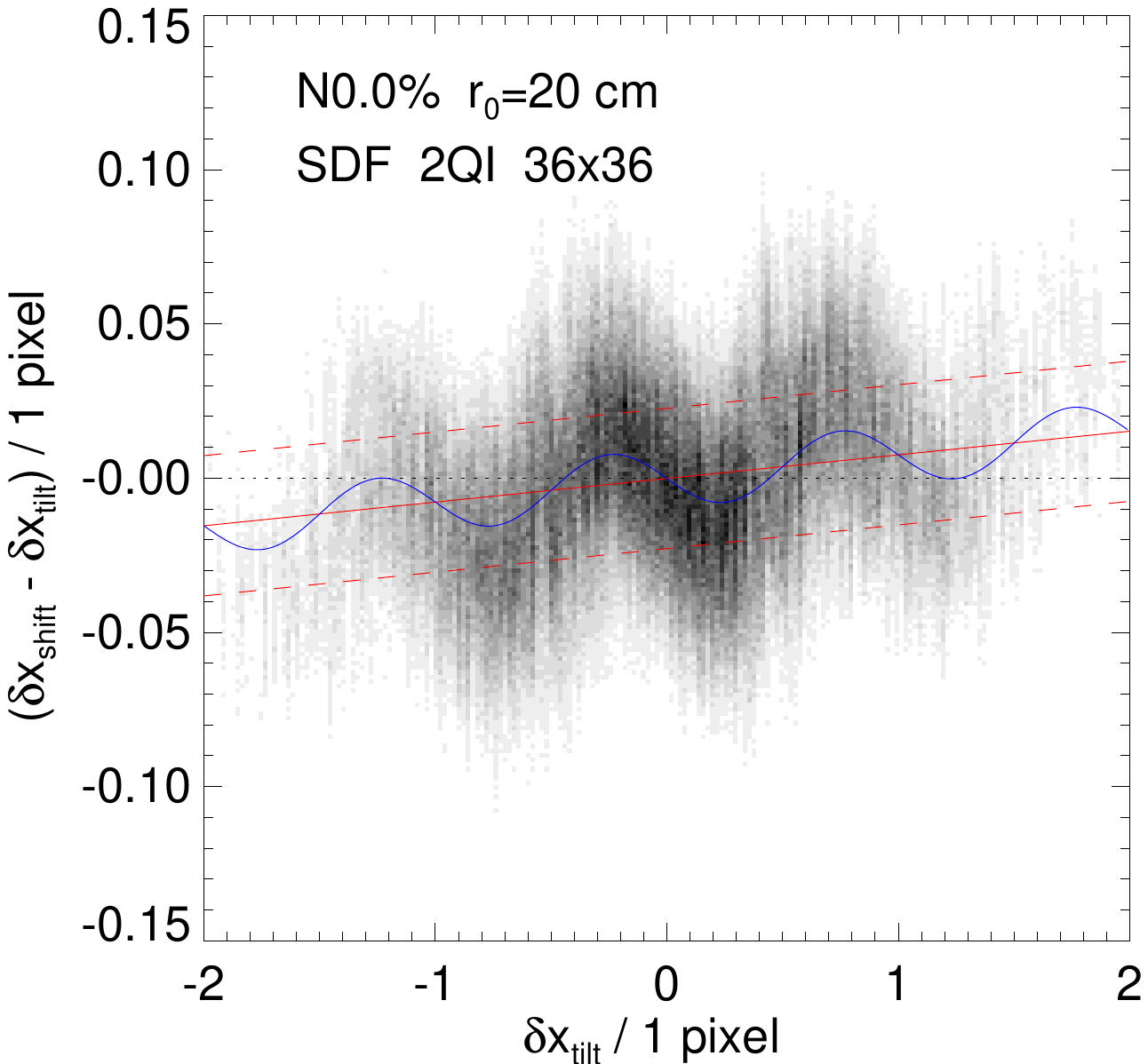}
  \quad
  \includegraphics[bb=58 0 430 345,clip,width=\figwidth]{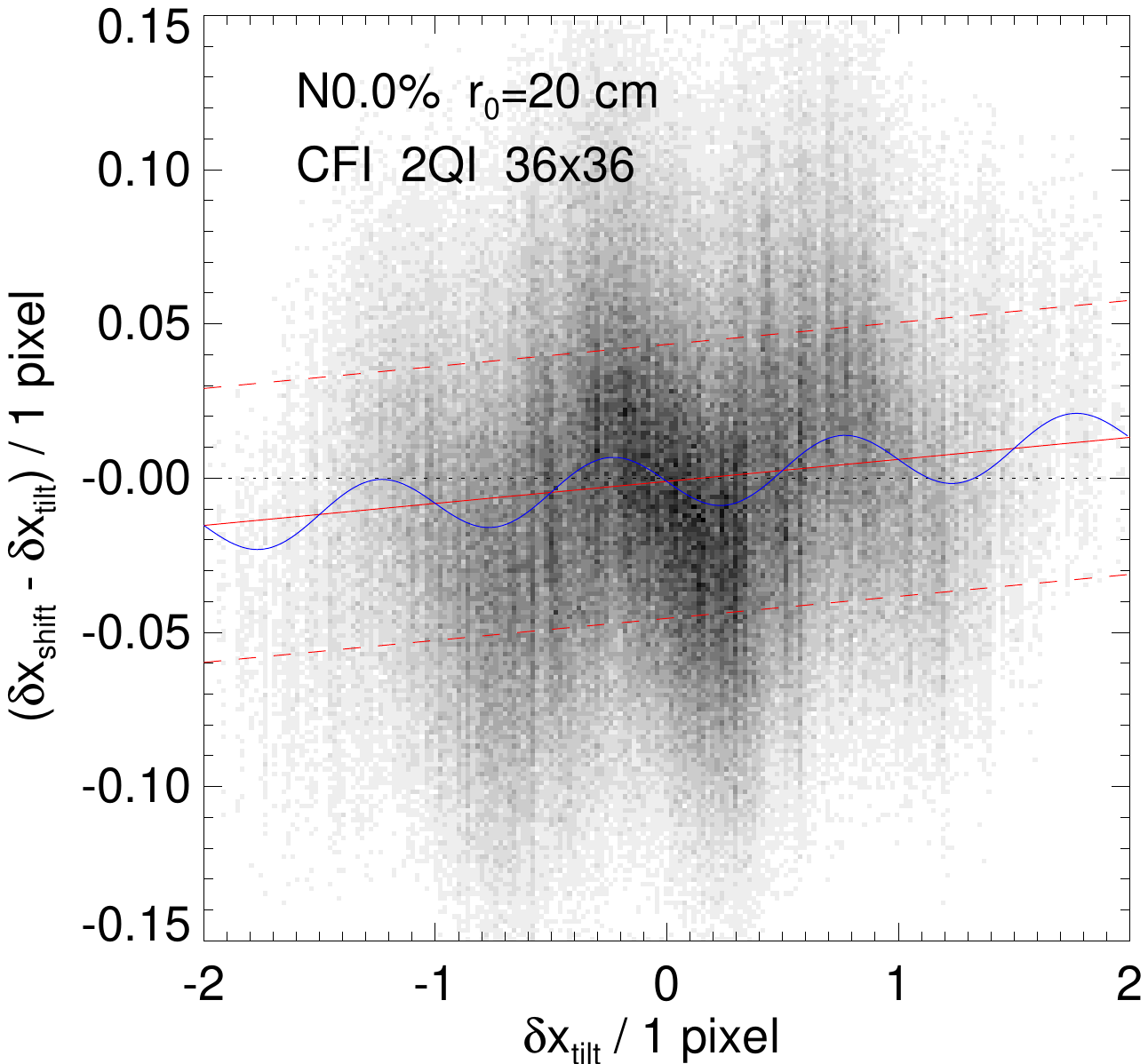}
  \quad
  \includegraphics[bb=58 0 430 345,clip,width=\figwidth]{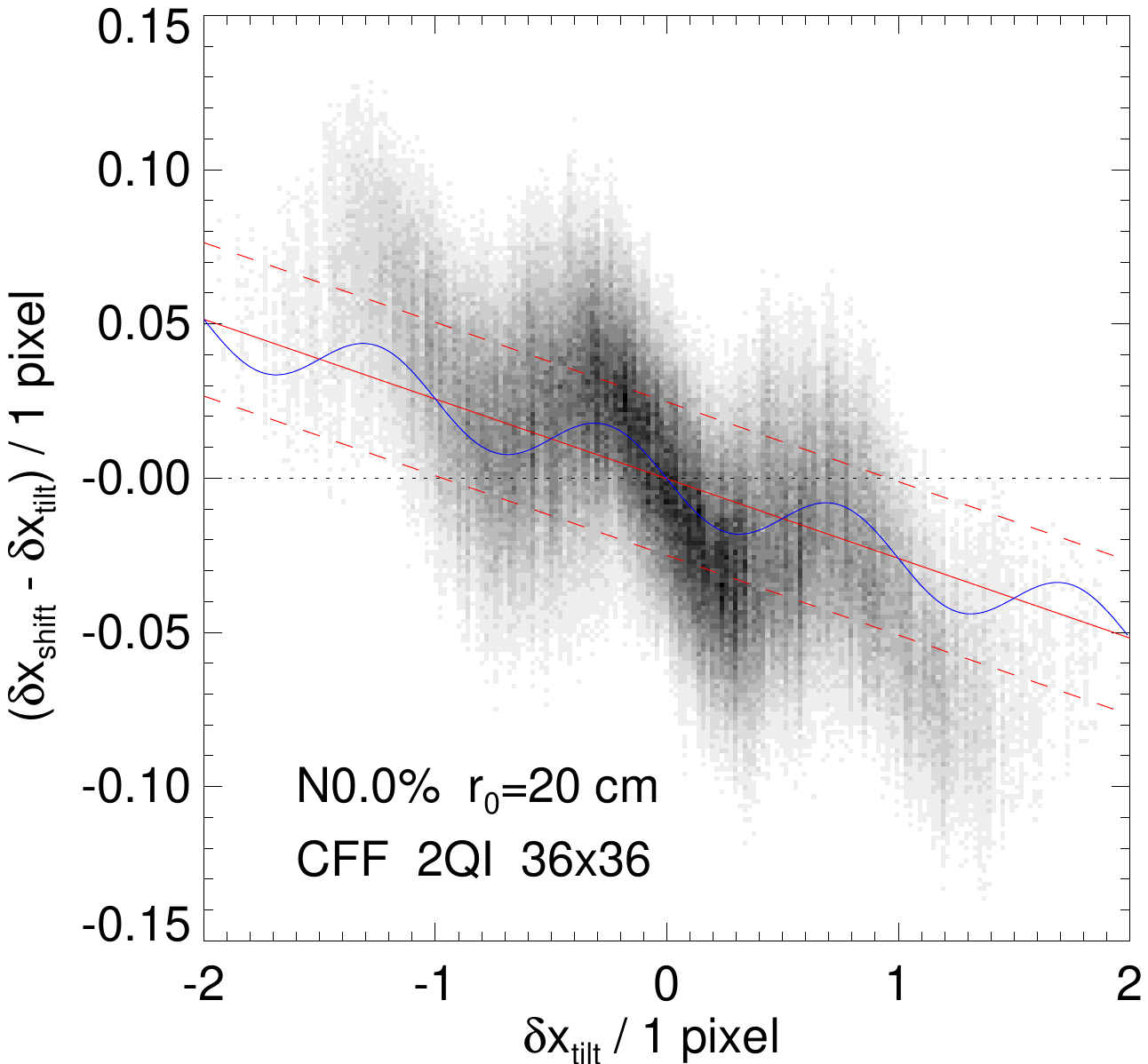}\\
  \includegraphics[bb=58 0 430 345,clip,width=\figwidth]{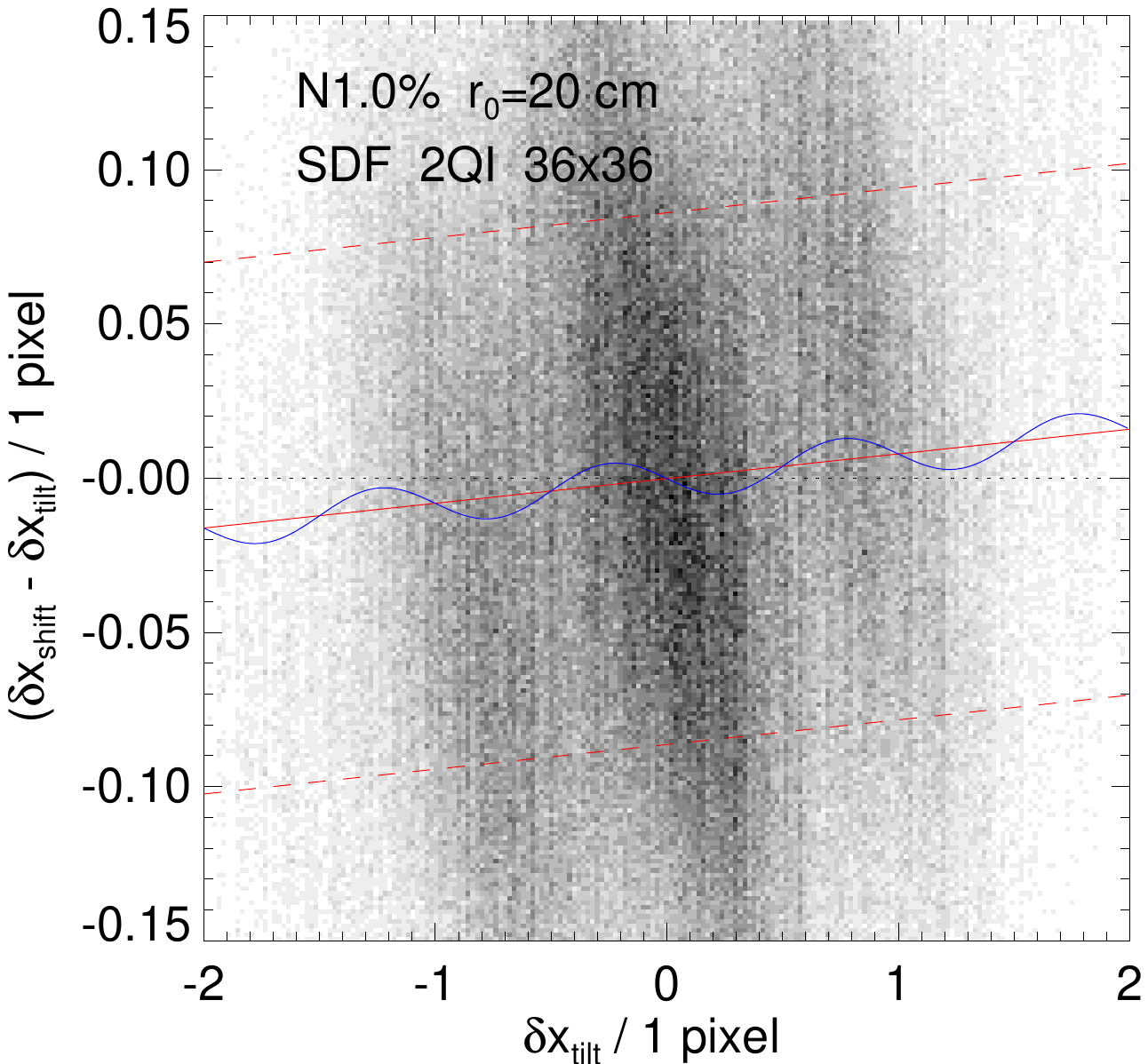}
  \quad
  \includegraphics[bb=58 0 430 345,clip,width=\figwidth]{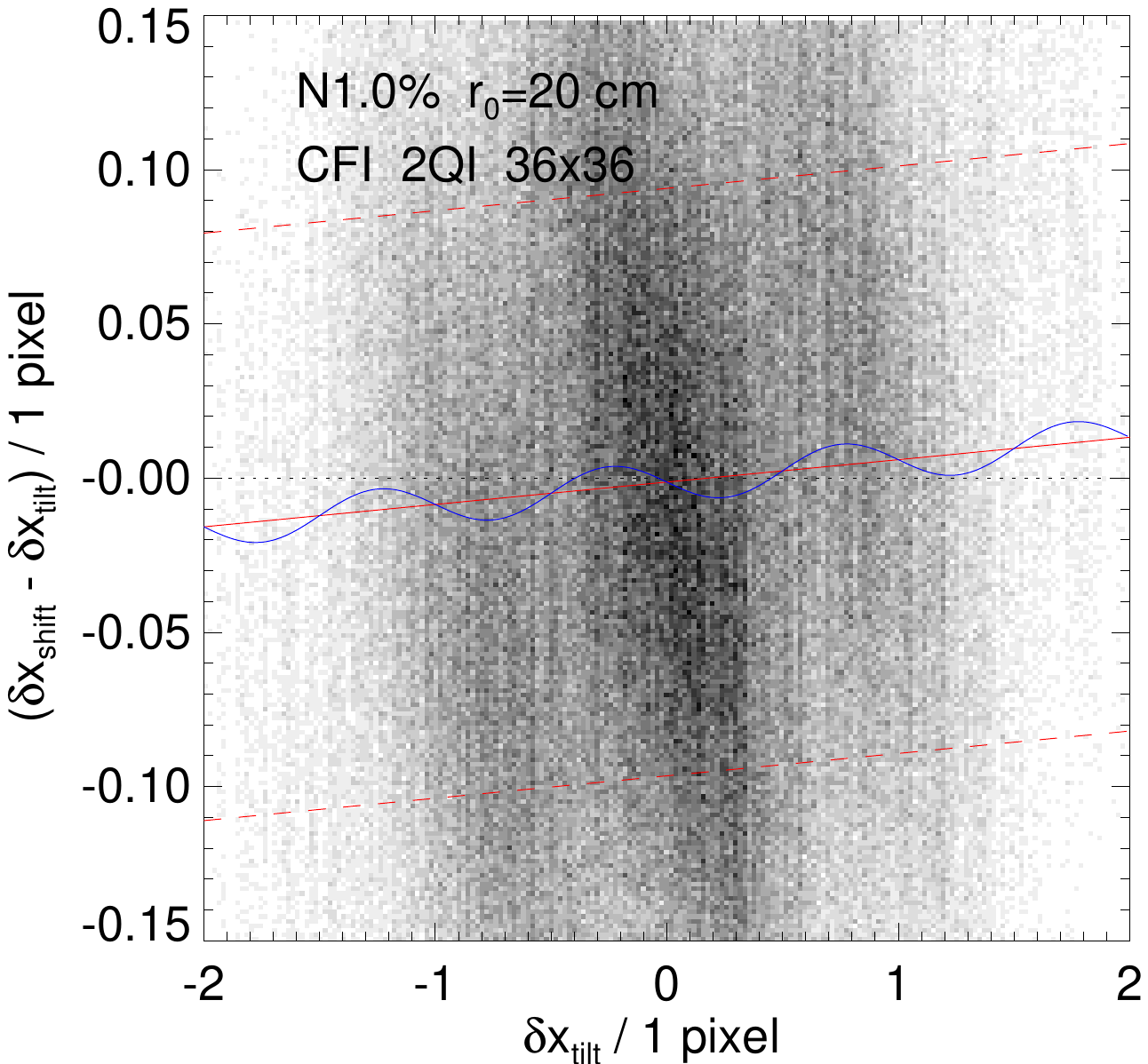}
  \quad
  \includegraphics[bb=58 0 430 345,clip,width=\figwidth]{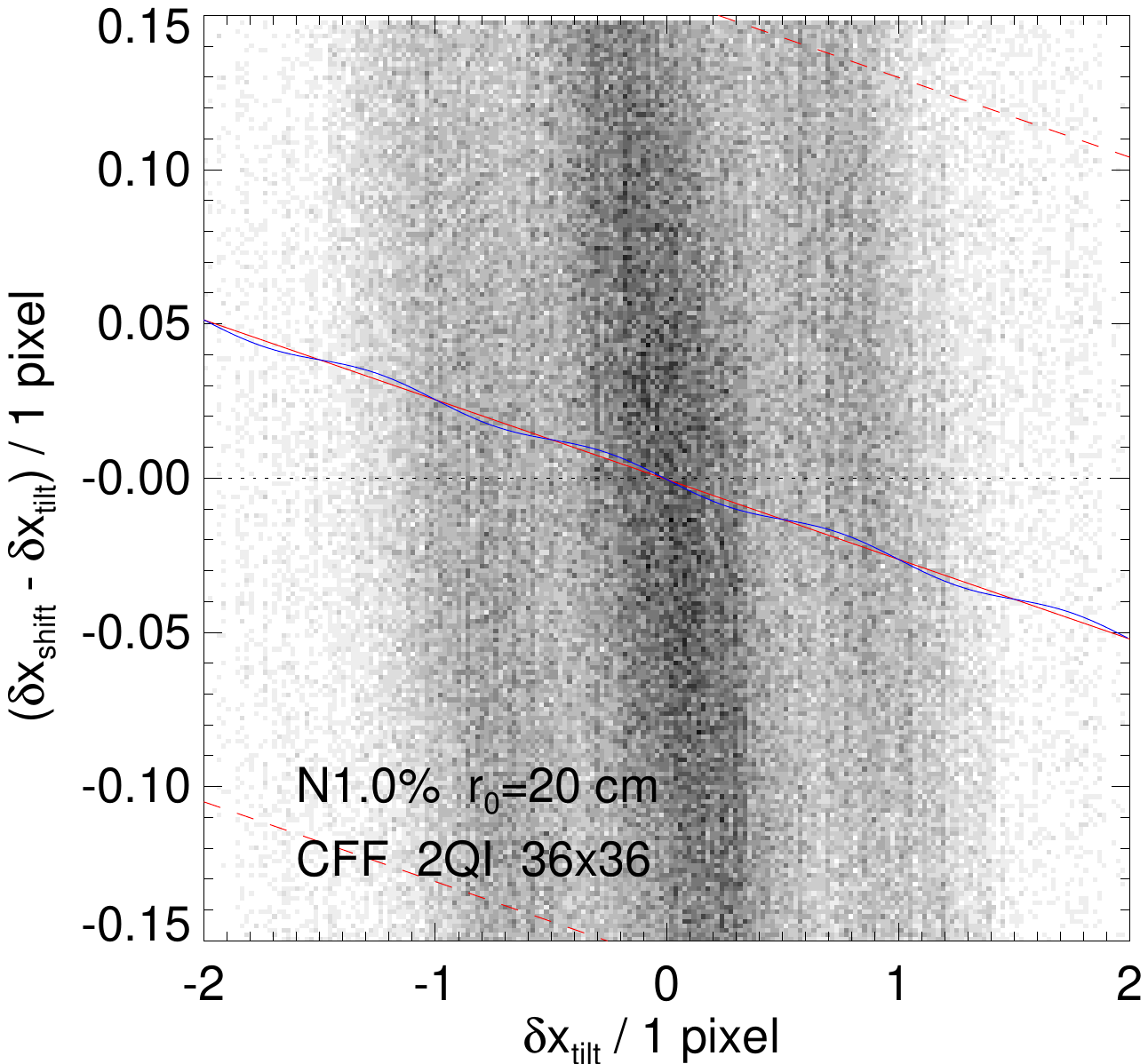}
  \caption{2D histograms of shift measurement errors for simulated
    seeing. 36$\times$36-pixel subfields, 0\farcs41/pixel,
    $r_0=20$~cm. \textbf{Top:} no noise; \textbf{Bottom:} 1\% noise.
    Blue line: fit to Eq.~(\ref{eq:linesine2}); Red lines: linear part
    of fit (dashed: $\pm1\sigma$).}
  \label{fig:scatter-seeing-36x36-r0-20cm}
\end{figure*}

Figure~\ref{fig:scatter-seeing-36x36-r0-20cm} shows the easiest case:
the largest subfields, 36$\times$36 pixels, and the best seeing,
$r_0=20$~cm, resulting in the smallest shifts. The ADF$^2$ results are
so similar to the SDF results that we only show the latter of those
two methods.

\begin{figure*}[!t]
  \centering
  \def\figwidth{0.3\linewidth}
  \includegraphics[bb=58 0 430 345,clip,width=\figwidth]{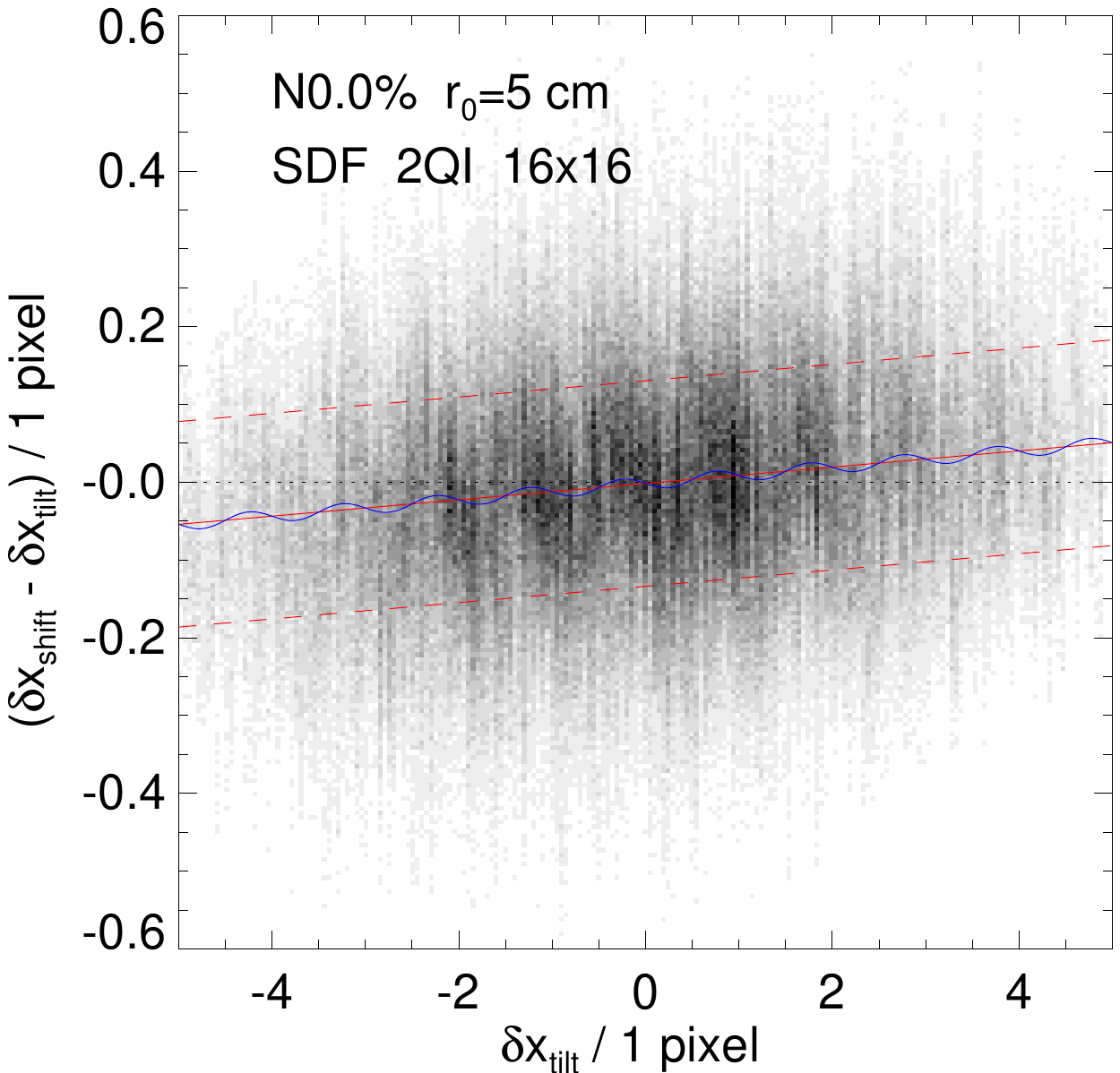}
  \quad
  \includegraphics[bb=58 0 430 345,clip,width=\figwidth]{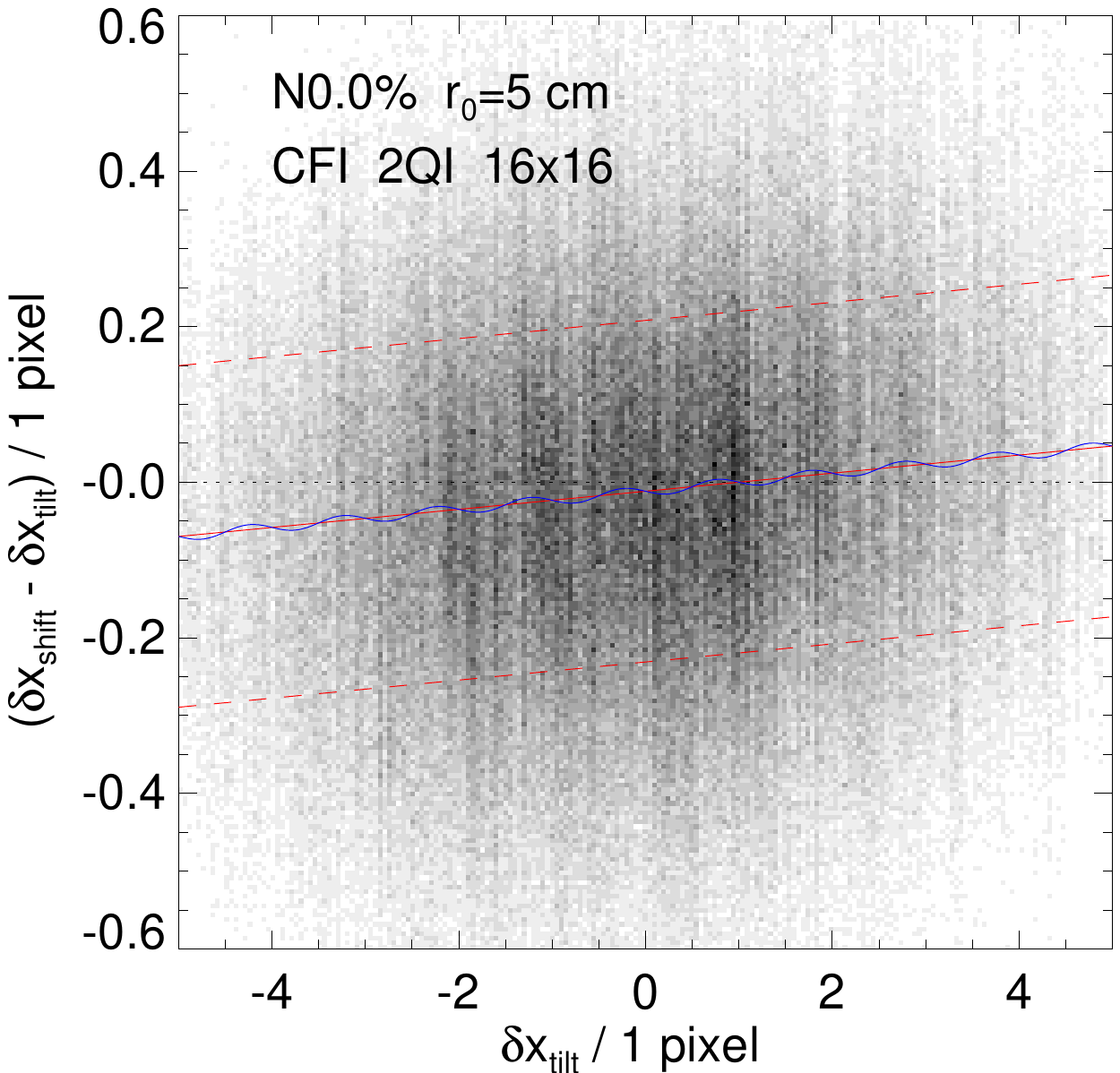}
  \quad
  \includegraphics[bb=58 0 430 345,clip,width=\figwidth]{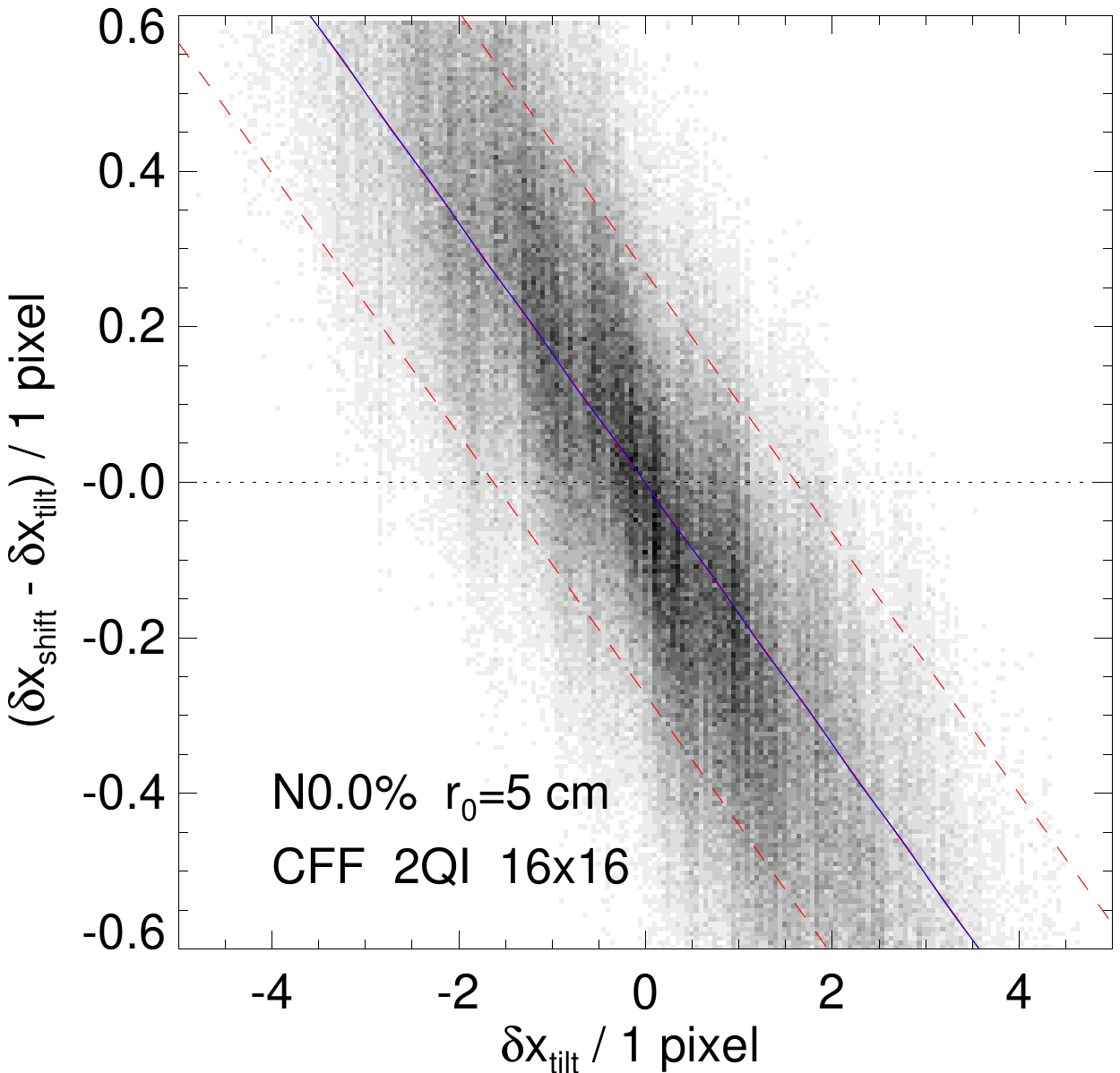}\\
  \includegraphics[bb=58 0 430 345,clip,width=\figwidth]{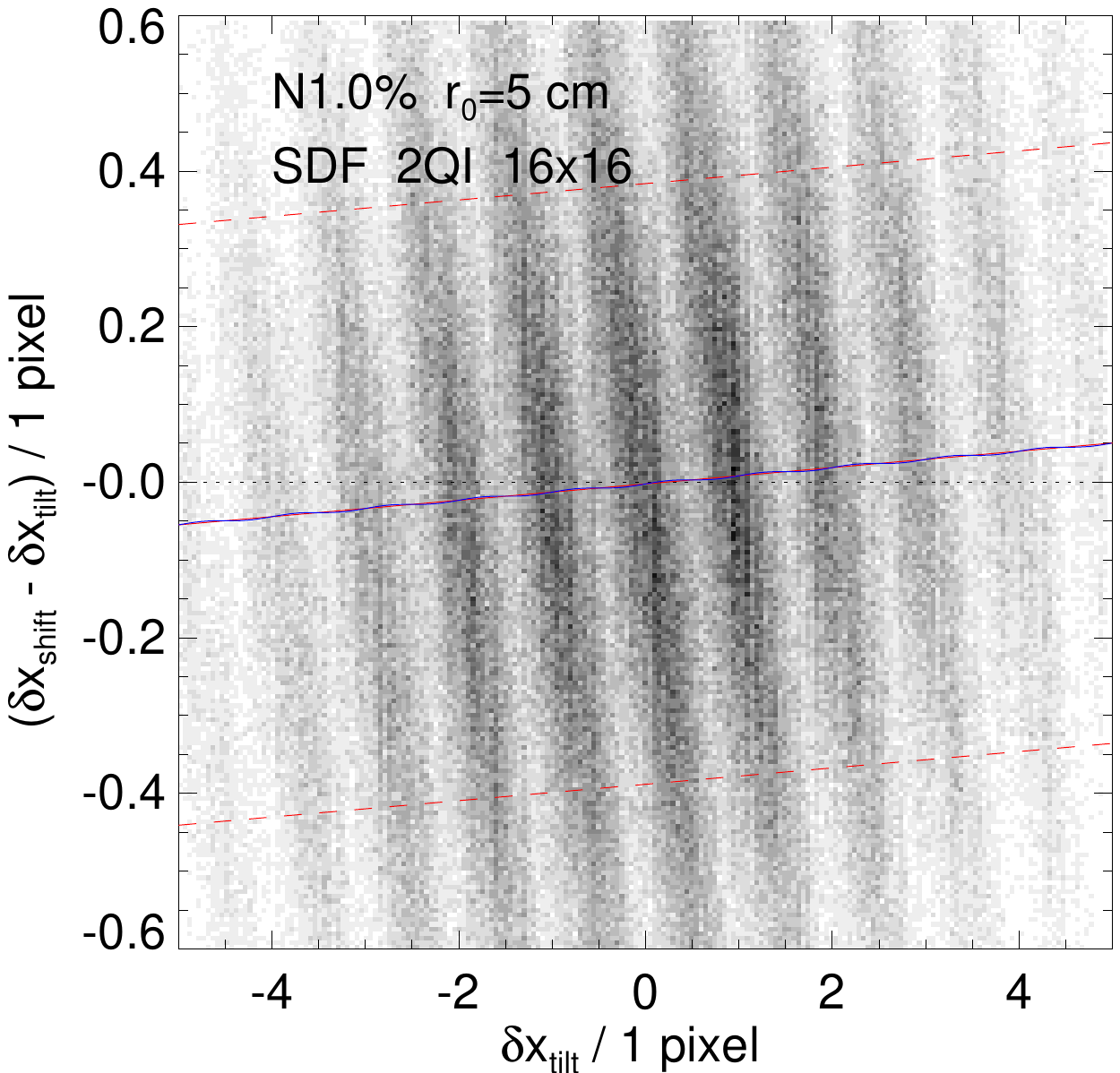}
  \quad
  \includegraphics[bb=58 0 430 345,clip,width=\figwidth]{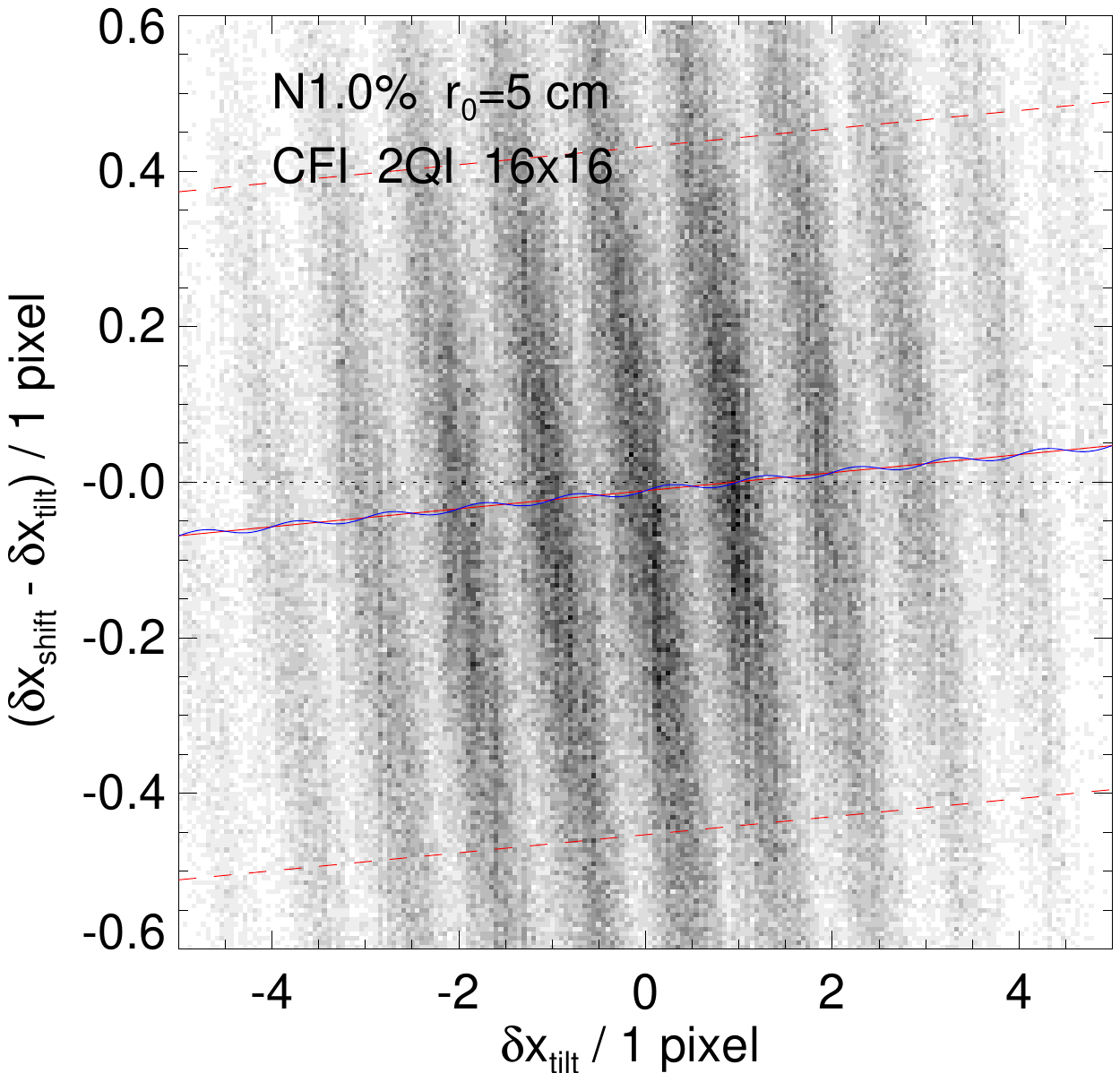}
  \quad
  \includegraphics[bb=58 0 430 345,clip,width=\figwidth]{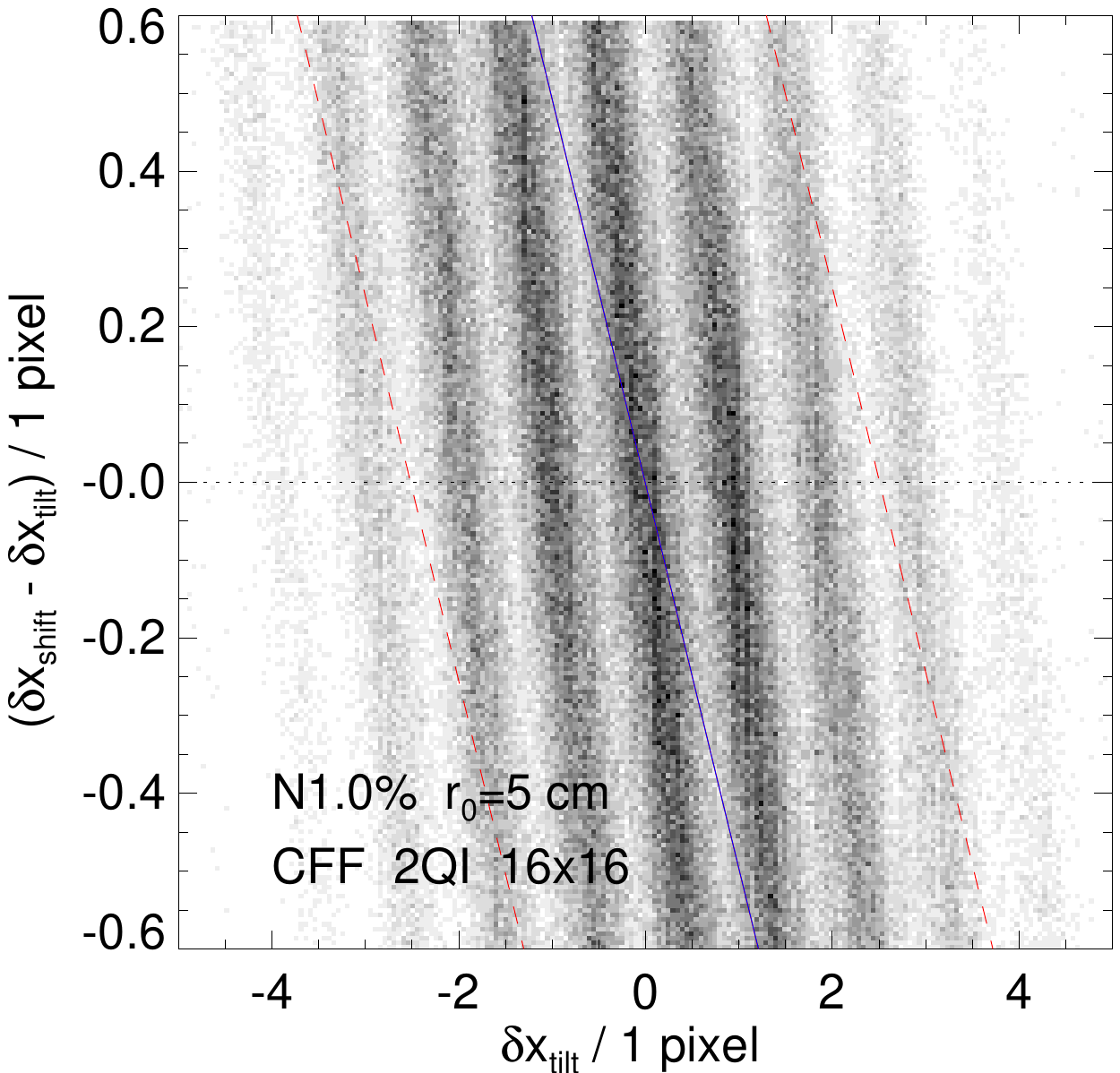}
  \caption{2D histograms of shift measurement errors for simulated
    seeing. 16$\times$16-pixel subfields, 0\farcs41/pixel, $r_0=5$~cm.
    \textbf{Top:} no noise; \textbf{Bottom:} 1\% noise. Blue line: fit
    to Eq.~(\ref{eq:linesine2}); Red lines: linear part of fit
    (dashed: $\pm1\sigma$).}
  \label{fig:scatter-seeing-16x16-r0-5cm}
\end{figure*}

The errors in Fig.~\ref{fig:scatter-seeing-36x36-r0-20cm} are
consistent with the ones in
Fig.~\ref{fig:perfect-scatter-four-methods}. We recognize the
undulations but we also see a linear trend, not only for CFF. SDF and
CFF: $\sigma_\text{err}\approx0.025$ matches undulations plus error
bars (CFF because for 20~cm we are dominated by small shifts). CFI:
$\sigma_\text{err}\approx0.045$ is actually slightly better than in
Fig.~\ref{fig:perfect-scatter-four-methods}. These results (as opposed
to the ones in Fig.~\ref{fig:scatter-seeing-16x16-r0-5cm}) are
comparable because 20~cm is much larger than the subpupil size so the
images are not blurred by local phase curvature. We cannot directly
compare the noisy data, because here we have 1\% noise, in the earlier
experiment we used 0.5\%.

Linear fit offsets are small, $\left|p_0\right|\lesssim 0.002$~pixels
for all CAs, except CFI that has
$\left|p_0\right|\lesssim0.016$~pixels. The undulations fit sines with
an amplitude of $p_2 \approx 0.01$~pixels in almost all cases,
consistent with Figs.~\ref{fig:perfect-scatter-four-methods}
and~\ref{fig:perfect-scatter-SDF-different-versions}, where we used
identical images with known shifts. The exceptions are cases when both
$N$ and $r_0$ are small. In the latter cases the fits to the sine
function sometimes give as small amplitudes as $p_2 \approx
0.002$~pixels or less.

For shifts smaller than 0.2~pixels, the overall overestimation turns
into underestimation. For CFF, the slope and the undulations work in
the same direction, making the underestimation larger for the smaller
shifts.

Figure~\ref{fig:scatter-seeing-16x16-r0-5cm} shows the most demanding
test: the smallest subfields, 16$\times$16 pixels, and the worst
seeing, $r_0=5$~cm, resulting in the largest shifts and the images
most blurred by local phase curvature. Note much larger dispersion and
error in $p_1$ for CFF.

\subsubsection{Z-tilts and G-tilts}
\label{sec:slopes}

The wavefront tilt measured over a subpupil is by necessity an
approximation because in reality the tilts vary over the subpupil. In
the night-time literature, two kinds of tilts are discussed
\citep[e.g., ][]{tokovinin02differential}. G-tilts correspond to
averaging the wavefront Gradient, which is mathematically equivalent
to measuring the center of Gravity of the PSF. However, because of
noise and asymmetrical PSFs this can never be realized in practice.
Windowing and thresholding the PSF gives measurements that are more
related to Z-tilts, corresponding to Zernike tip/tilt and the location
of the PSF peak. When interpreting the measurements, we need to know
what kind of tilts are measured by our methods.

The simulated tilts are implemented as coefficients to the Zernike tip
and tilt polynomials. So if the shifts measure Z-tilts, the expected
$p_1$ is by definition
\begin{equation}
  \label{eq:8}
  E[p_1\mid\text{Z-tilt}]\equiv1 .
\end{equation}
In order to derive the expected $p_1$ for G-tilts, we use formulas
given by \citet{tokovinin02differential}. The variance of the
differential image motion can be written as
\begin{equation}
  \label{eq:4}
  \sigma_\text{d}^2 = K \lambda^2 r_0^{-5/3} D^{-1/3}
\end{equation}
where $D$ is the subpupil diameter and $K$ is a number that depends
on the kind of tilt. The expected $p_1$ should be equal to the ratio
of $\sigma_\text{d}$ for G-tilts and Z-tilts, i.e.,
\begin{equation}
   E[p_1\mid\text{G-tilt}] = \sqrt{K_\text{G}/K_\text{Z}} \approx 0.966,
   \label{eq:7}
\end{equation}
where we used $K=K_\text{G}=0.340$ for G-tilts and
$K=K_\text{Z}=0.364$ for Z-tilts, which is asymptotically correct for
large separations between subpupils. This corresponds to a 3.4\%
difference in tilt measurements with Z-tilts giving the larger
numbers, regardless of wavelength, seeing conditions, and aperture
size. For smaller separations, $E[p_1\mid\text{G-tilt}]$ is even
smaller and depends somewhat on whether the shifts are longitudinal or
transversal (parallel or orthogonal to the line separating the
apertures).

In our results for all the CAs except CFF, $1.007\la p_1 \la 1.010$.
This means they overestimate Z-tilts systematically by $\la1$\% and
G-tilts by $\ga4.6$\%. This result appears to be robust with respect to
noise, subfield size, image scale, and seeing conditions.


\begin{figure}[tb]
  \centering
  \def\tilewidth{\linewidth}
  \includegraphics[bb=15 10 495 344, width=\tilewidth]{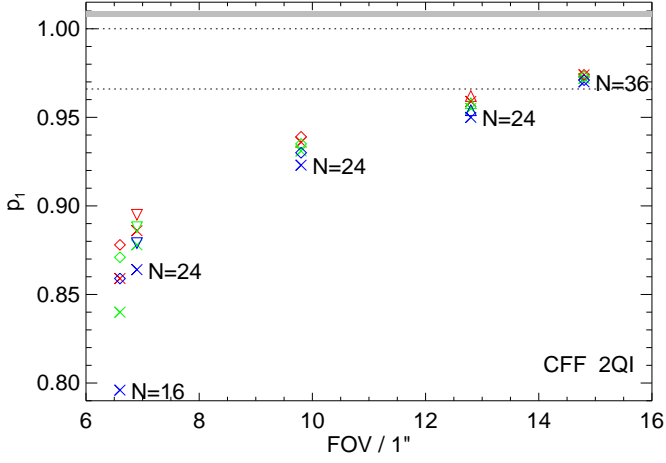}
  \caption{CFF slopes $p_1$ for different FOV sizes, no noise. Red:
    $r_0=20$~cm; Green: $r_0=10$~cm; Blue: $r_0=7$~cm. $N\times
    N$-pixel subfields. Diamonds~($\Diamond$): image scale
    0\farcs41/pixel, varying $N=\{16, 24, 36\}$ as labeled in the
    plots; Down triangles~($\triangledown$): 0\farcs29/pixel, $N=24$;
    Up triangles ($\vartriangle$): 0\farcs53/pixel, $N=24$.  All the
    above without noise. Crosses ($\times$): corresponding FOVs, with
    noise. The dotted lines correspond to Z-tilt ($p_1= 1.0$) and
    G-tilt ($p_1= 0.966$), respectively (see Eq.~(\ref{eq:7})). The
    gray band at the top represents the $1.007 \protect\la p_1
    \protect\la 1.010$ range of the other CA results.}
  \label{fig:slopes}
\end{figure}

The CFF $p_1$ values confirm the underestimation of the image shift
found in Sect.~\ref{sec:perfect-processing}. They depend mostly on the
size of the subfield but also to some extent on $r_0$ and noise. CFF
$p_1$ improves with larger FOV in arcsec, regardless of number of
pixels and $r_0$. For small FOVs, CFF $p_1$ is smaller than the
G-tilts slope of 0.966, grows with FOV size to cross G-tilt at
$\sim$14\arcsec{}. For the other CAs, $p_1$ is independent of FOV
size. CFF slope improved with subfield size. Is it the size in pixels
or the size in arcsec that is important? Figure~\ref{fig:slopes} makes
it clear that the size in arcsec is the important parameter and that
with even larger FOVs we should expect to measure something closer to
Z-tilts. We conjecture that the important parameter is the number of
1\arcsec{} resolution elements that fit within the FOV.

\subsubsection{RMS errors}
\label{sec:rms-errors}

The RMS errors, $\sigma_\text{err}$, are calculated after removing the
outliers and after subtracting the linear fit. In spite of having more
outliers removed from the calculations, the errors are worse for CFF
than for the other methods.

Compared to the noise-free data, adding 1\% image noise significantly
increases $\sigma_\text{err}$. In many cases it also decreases the
number of fails (all cases with $N=24,36$, most cases with $N=16$).
This probably means the noise makes the error distribution more
Gaussian.

$\sigma_\text{err}$ decreases with increasing $r_0$, indicating that
the algorithms perform better in good seeing, as expected. But are the
results worse in bad seeing because of local phase curvature (i.e.,
smearing) or just because the shifts are larger? The relative measure,
$\sigma_\text{err}/\sigma_\text{tilt}$, does not decrease as much with
$r_0$ for zero-noise data and actually tends to \emph{increase} for
noisy data. So the latter should not be the major effect.

\begin{figure}[!tb]
  \centering
  \def\tilewidth{8cm}
  \includegraphics[bb=15 10 495 344, width=\tilewidth]{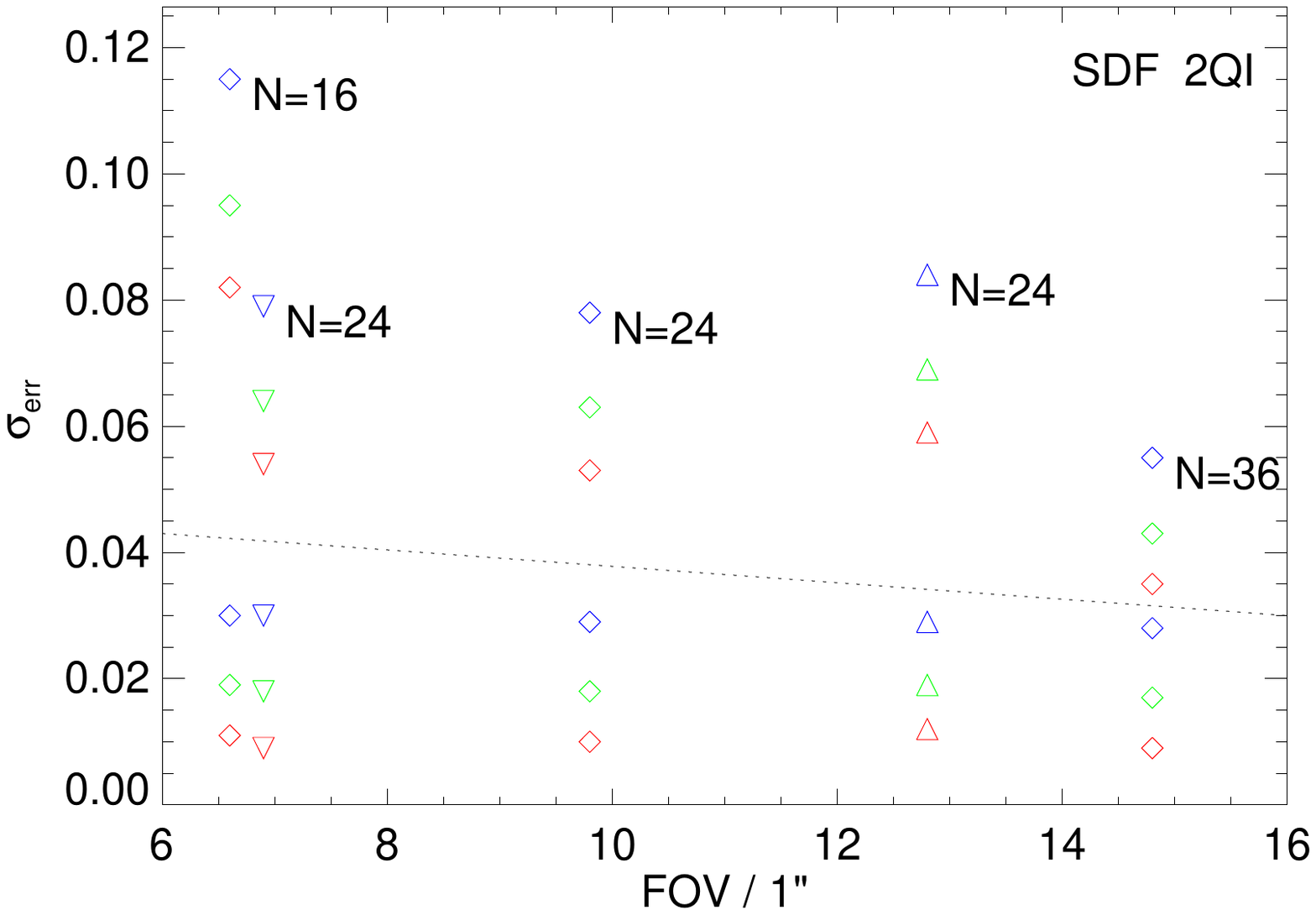}
  \includegraphics[bb=15 10 495 344, width=\tilewidth]{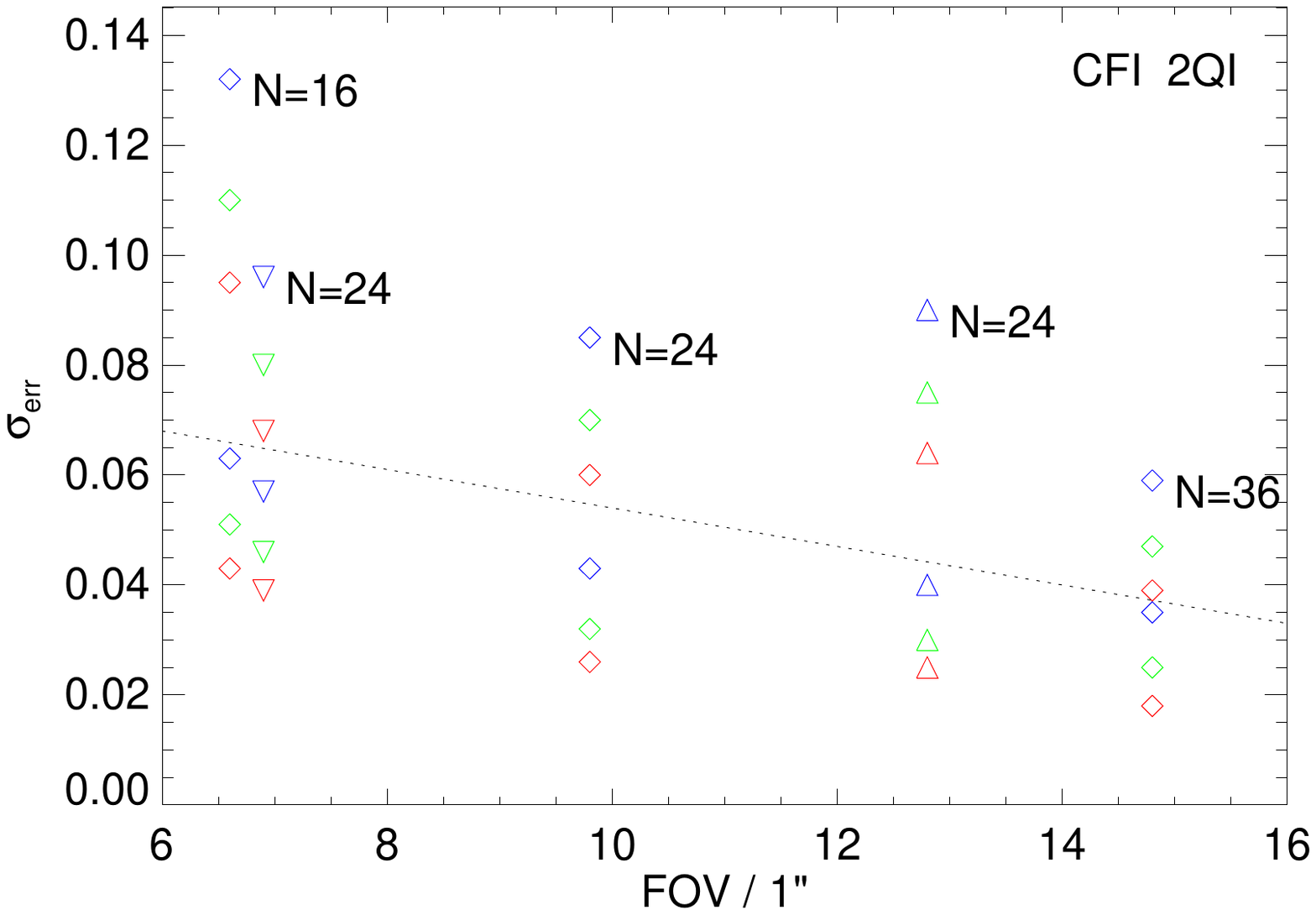}
  \includegraphics[bb=15 10 495 344, width=\tilewidth]{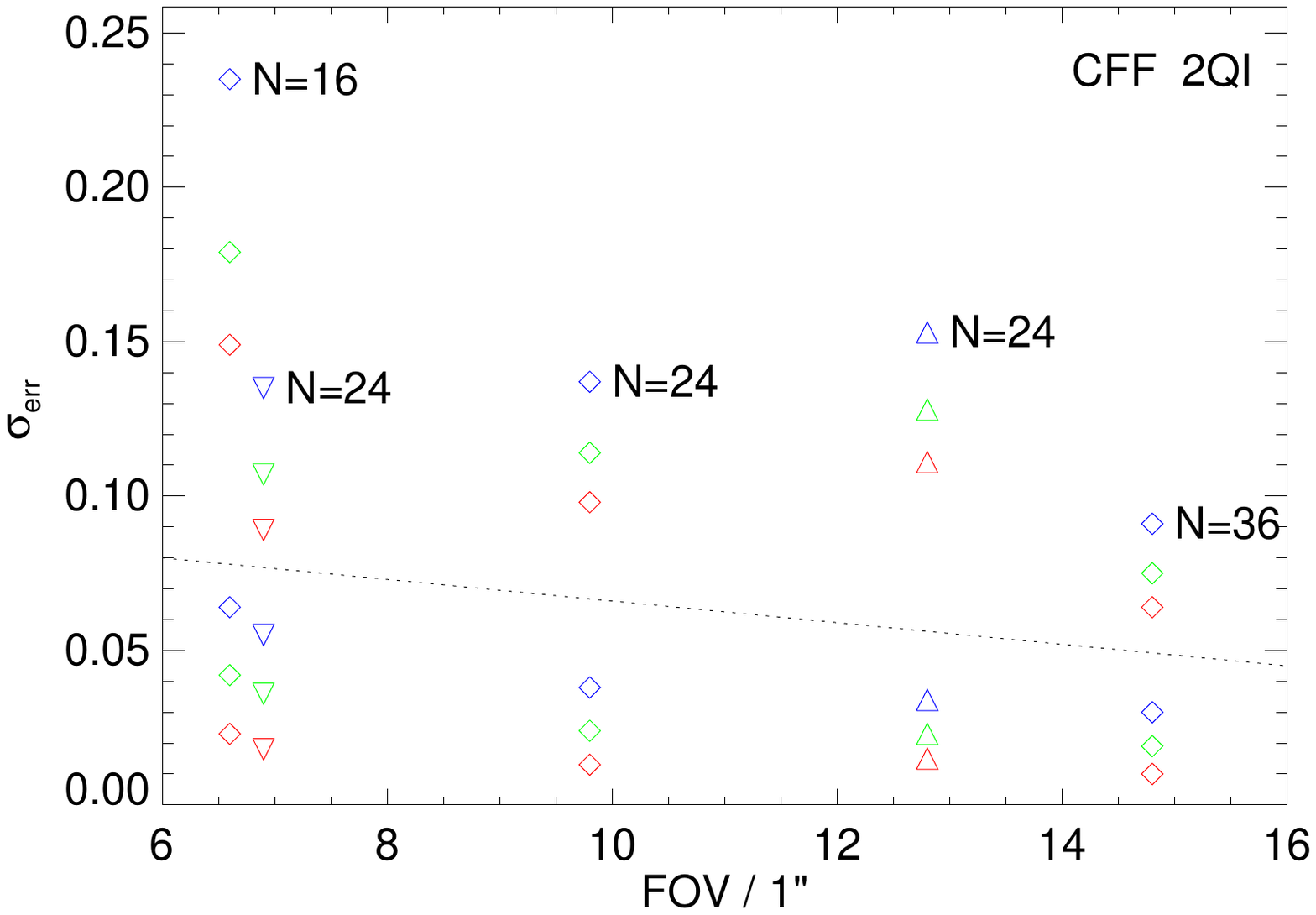}    
  \caption{Standard deviations of errors. $\sigma_\text{err}$ for
    different FOV sizes and CAs. Red:~$r_0=20$~cm; Green:~$r_0=10$~cm;
    Blue:~$r_0=7$~cm. $N\times N$-pixel subfields.
    Diamonds~($\Diamond$): image scale 0\farcs41/pixel, varying
    $N=\{16, 24, 36\}$ as labeled in the plots; Down
    triangles~($\triangledown$): 0\farcs29/pixel, $N=24$; Up triangles
    ($\vartriangle$): 0\farcs53/pixel, $N=24$.  Symbols below the
    dotted line correspond to data without noise, above the line with
    1\% noise. Note different vertical scales.}
  \label{fig:sdevs}
\end{figure}

How do the errors depend on the FOV size in pixels and in seconds of
arc? Figure~\ref{fig:sdevs} visualizes the following results from the
tables. Without noise, the CFI and CFF results improve with $N$ as
well as the FOV size in arcsec with constant $N$ (particularly from
6\farcs9 to 9\farcs8). SDF (and ADF$^2$, the results are so similar
that we show only SDF) errors are more or less independent of FOV
size, whether in pixels or in arcsec. With 1\% noise, however, for all
methods, the errors improve only slightly or \emph{grow} when $N$ is
constant and we increase the FOV size in arcsec (particularly from
7\arcsec{} to 13\arcsec). When $N$ is increased from 16 to 24 with
almost constant 7\arcsec{} FOV, there is a significant improvement.
There is also significant improvement when $N$ is increased from 24 to
36 (although FOV in arcsec is increased too). So for the RMS error,
FOV in pixels is more important than FOV in arc seconds with noisy
data.

\section{Discussion and conclusions}
\label{sec:conclusions}
 
We have evaluated five different correlation algorithms and four
different interpolation algorithms in two experiments with artificial
data. Among these are the algorithms in use for the AO systems
installed at the major high-resolution solar telescopes today. The
experiment in Sect.~\ref{sec:algorithm-accuracy} examined the inherent
performance of the methods with identical images and known shifts, and
introduced also the effects of noise and a mismatch in intensity level
(bias) between images. We used image contrast and noise levels that
resemble the setup at the SST. In Sect.~\ref{sec:image-shift-versus},
we introduced also local wavefront curvature and different seeing
conditions.

Based on the results of both experiments, we recommend SDF and ADF$^2$
(in that order) for calculating the correlation functions. They give
significantly smaller random errors and more predictable systematic
errors than the competing methods. The SDF results are marginally
better so if computational speed is not an issue, use SDF. But ADF$^2$
may, by virtue of speed, be preferable. For subpixel interpolation,
2QI and 2LS perform better than the one-dimensional interpolation
algorithms.

It is clear from Sect.~\ref{sec:algorithm-accuracy} that bias
mismatch has a strong and negative effect on the performance of the
difference based methods. We have not established how much bias
mismatch can be tolerated. Our conclusion is that it should be
compensated for in pre-processing, before the shift measurement
methods are applied. The simulated bias mismatch comes from
multiplication with 1.01 but is compensated for by subtraction for CFI and
CFF. Based on the fact that the latter methods give identical results
with and without the bias mismatch (to the number of digits shown in
the table), we conclude that it does not matter whether the mismatch
is removed by multiplication or by subtraction.

\citet{waldmann07untersuchung} considers only a single correlation
algorithm (CFF) but tries a few different interpolation algorithms:
1LS, 2LS, and a Gaussian fit. He finds a Gaussian fit to give the better
results but the almost as good 1LS is less complicated so he uses
that. On the other hand, \citet{2008SPIE.7015E.154W} try CFF, CFI
(almost: correlation instead of covariance) and ADS, but use only 1LS
for interpolation. They find CFF best and ADF worst. Trying all
combinations of those methods, \citet{johansson10cross-correlation}
confirmed that Gaussian and 1LS perform similarly together with CFF.
She also found a similar performance for SDF and ADF$^2$ with 1LS and
Gaussian, but significantly better performance when interpolated with
2QI and 2LS. Based on this, we did not try a Gaussian fit here because
of its greater computational cost.

We have demonstrated that with the recommended methods, we can measure
shifts in identical, noise-free 24$\times$24-pixel images with an RMS
error of $0.012$~pixels, corresponding to 0\farcs008 in our
setup. With realistic image noise RMS of 0.5\% the errors increase by
40\% to 0.017 pixels. Increasing the noise RMS to 1\% gives
unacceptably high measurement errors in bad seeing, particularly with
small subimages. Increasing the subfield size from 16 to 24 pixels
squared reduced the error by approximately 30\%.

One method is not necessarily the best choice both for open loop and
for closed loop. In open loop, performance for large shifts and
predictable systematic errors are important. However, CFF can compete
with SDF and ADF$^2$ in closed loop, where wavefronts are already
compensated for and the CFF random errors are small. The
underestimation of the shifts can be reduced by using a larger FOV and
can also be compensated for with a different servo gain.
Based on our results, it is difficult to find a reason to ever use the
CFI method.

We have found that SDF, ADF$^2$, and CFI measure Z-tilts with a 1\%
systematic overestimation, rather than G-tilts. For the FOV sizes
relevant to solar SH WFS, the CFF method severely underestimates the
tilts but it is likely that for larger FOVs, also CFF measures
Z-tilts.

\citet{johansson10cross-correlation} processed also images together
with a reference image that was slightly distorted geometrically
(compressed image scale in one direction and expanded in the other)
and found that it could significantly affect the results. We have not
examined this effect here. A relevant further test involving
anisoplanatic effects would require geometrical distortions and
differential blurring whose shape and magnitude can be calculated from
realistic atmospheric turbulence profiles.

We end by emphasizing that small details in processing may have large
effects on the results. Examples from our evaluation include ADF
versus ADF$^2$ and the choice of apodization for CFF. As in so many
other situations, it is important to keep track of implementation
details and other tricks, so they are not lost when upgrading SH
software.

\begin{acknowledgements}

  I acknowledge many discussions with G{\"o}ran Scharmer, Pit
  S\"utterlin and Tim van Werkhoven. Thomas Berkefeld and Torsten
  Waldmann helped me understand some details of their work with the
  VTT adaptive optics system and simulations of wide-field wavefront
  sensors. I am grateful to George (formerly Guang-ming) Dai for
  calculating and sharing the radial Karhunen--L\`oeve functions used
  for the atmospheric turbulence simulations. The Swedish 1-m Solar
  Telescope is operated on the island of La Palma by the Institute for
  Solar Physics of the Royal Swedish Academy of Sciences in the
  Spanish Observatorio del Roque de los Muchachos of the Instituto de
  Astrof\'isica de Canarias.

\end{acknowledgements}


\clearpage
\onecolumn

\begin{table*}[h]
  \caption{Simulation results, no bias mismatch, no noise.}    
  \label{tab:sim-n0b0}
  \centering
  \begin{tabular}{llcD{.}{.}{2.1}@{\,}rcD{.}{.}{2.1}@{\,}rcD{.}{.}{2.1}@{\,}rcD{.}{.}{2.1}@{\,}rcD{.}{.}{2.1}l}
    \hline\hline\noalign{\smallskip}
    \multirow{2}{*}{CA} & 
    \multirow{2}{*}{IA} &
    \multicolumn{2}{c}{All} && 
    \multicolumn{2}{c}{$\delta r<1$} &&
    \multicolumn{2}{c}{$\delta r<2$} &&
    \multicolumn{2}{c}{$3<\delta r<5$} &&
    \multicolumn{2}{c}{$\delta r>5$} &
    \multirow{2}{*}{$p_1$} \\ 
    \cline{3-4}\cline{6-7}\cline{9-10}\cline{12-13}\cline{15-16}\noalign{\smallskip}
    && 
    $\sigma$ (pix) & \multicolumn{1}{c}{\text{Fail (\%)}} && 
    $\sigma$ (pix) & \multicolumn{1}{c}{\text{Fail (\%)}} && 
    $\sigma$ (pix) & \multicolumn{1}{c}{\text{Fail (\%)}} &&
    $\sigma$ (pix) & \multicolumn{1}{c}{\text{Fail (\%)}} && 
    $\sigma$ (pix) & \multicolumn{1}{c}{\text{Fail (\%)}} &  \\ 
    \hline\noalign{\smallskip}
    SDF & 1LS &  0.037 &   1.0 &&  0.042 &   1.5 &&  0.041 &   1.4 &&  0.028 &   0.3 &&  0.027 &   0.3 &  \\
    SDF & 1QI &  0.033 &   1.8 &&  0.039 &   2.5 &&  0.038 &   2.4 &&  0.023 &   0.6 &&  0.022 &   0.6 &  \\
    SDF & 2LS &  0.020 &   0.3 &&  0.021 &   0.3 &&  0.020 &   0.3 &&  0.020 &   0.1 &&  0.020 &   0.3 &  \\
    SDF & 2QI &  0.017 &   0.3 &&  0.017 &   0.2 &&  0.017 &   0.3 &&  0.016 &   0.2 &&  0.016 &   0.3 &  \\
    CFI & 1LS &  0.114 &   0.8 &&  0.119 &   0.5 &&  0.118 &   0.7 &&  0.109 &   0.9 &&  0.106 &   0.9 &  \\
    CFI & 1QI &  0.108 &   0.7 &&  0.112 &   0.5 &&  0.112 &   0.6 &&  0.103 &   0.9 &&  0.100 &   0.8 &  \\
    CFI & 2LS &  0.109 &   0.7 &&  0.109 &   0.4 &&  0.110 &   0.6 &&  0.109 &   1.0 &&  0.105 &   0.9 &  \\
    CFI & 2QI &  0.101 &   0.6 &&  0.102 &   0.4 &&  0.102 &   0.5 &&  0.101 &   0.9 &&  0.099 &   0.9 &  \\
    CFF & 1LS &  0.112 &   5.0 &&  0.060 &  &&  0.078 &  &&  0.187 &   3.0 &&  0.295 &  30.1 &  0.86 \\
    CFF & 1QI &  0.111 &   5.0 &&  0.063 &  &&  0.079 &  &&  0.180 &   3.0 &&  0.280 &  30.1 &  0.86 \\
    CFF & 2LS &  0.091 &   5.1 &&  0.032 &  &&  0.056 &  &&  0.167 &   3.2 &&  0.269 &  30.8 &  0.86 \\
    CFF & 2QI &  0.085 &   5.1 &&  0.031 &  &&  0.054 &  &&  0.156 &   3.2 &&  0.251 &  30.9 &  0.86 \\
    ADF & 1LS &  0.045 &   9.9 &&  0.049 &  14.3 &&  0.049 &  13.7 &&  0.039 &   2.9 &&  0.037 &   2.6 &  \\
    ADF & 1QI &  0.042 &  33.0 &&  0.046 &  39.3 &&  0.047 &  39.1 &&  0.031 &  22.4 &&  0.027 &  20.6 &  \\
    ADF & 2LS &  0.030 &  23.1 &&  0.033 &  25.1 &&  0.033 &  25.1 &&  0.022 &  19.2 &&  0.022 &  17.5 &  \\
    ADF & 2QI &  0.028 &  42.8 &&  0.032 &  49.4 &&  0.032 &  49.7 &&  0.017 &  30.1 &&  0.016 &  27.6 &  \\
    ADF$^2$ & 1LS &  0.038 &   0.9 &&  0.044 &   1.3 &&  0.044 &   1.3 &&  0.029 &   0.3 &&  0.028 &   0.3 &  \\
    ADF$^2$ & 1QI &  0.035 &   1.5 &&  0.041 &   2.2 &&  0.041 &   2.1 &&  0.025 &   0.5 &&  0.023 &   0.5 &  \\
    ADF$^2$ & 2LS &  0.022 &   0.3 &&  0.023 &   0.4 &&  0.023 &   0.4 &&  0.021 &   0.1 &&  0.021 &   0.1 &  \\
    ADF$^2$ & 2QI &  0.019 &   0.1 &&  0.019 &   0.1 &&  0.019 &   0.1 &&  0.018 &  &&  0.018 &  &  \\
    \hline\noalign{\smallskip}
    SDF & 1LS &  0.025 &   1.2 &&  0.029 &   1.7 &&  0.029 &   1.6 &&  0.019 &   0.4 &&  0.019 &   0.5 &  \\
    SDF & 1QI &  0.024 &   1.8 &&  0.028 &   2.6 &&  0.028 &   2.5 &&  0.018 &   0.6 &&  0.017 &   0.7 &  \\
    SDF & 2LS &  0.014 &   0.7 &&  0.014 &   0.9 &&  0.014 &   0.8 &&  0.013 &   0.3 &&  0.013 &   0.2 &  \\
    SDF & 2QI &  0.012 &   0.8 &&  0.012 &   0.9 &&  0.012 &   0.9 &&  0.012 &   0.5 &&  0.011 &   0.4 &  \\
    CFI & 1LS &  0.070 &   0.3 &&  0.073 &   0.2 &&  0.072 &   0.2 &&  0.067 &   0.3 &&  0.064 &   0.4 &  \\
    CFI & 1QI &  0.066 &   0.2 &&  0.070 &   0.2 &&  0.069 &   0.2 &&  0.062 &   0.3 &&  0.060 &   0.4 &  \\
    CFI & 2LS &  0.065 &   0.3 &&  0.067 &   0.1 &&  0.066 &   0.2 &&  0.065 &   0.5 &&  0.062 &   0.6 &  \\
    CFI & 2QI &  0.061 &   0.2 &&  0.062 &  &&  0.062 &   0.1 &&  0.061 &   0.5 &&  0.058 &   0.5 &  \\
    CFF & 1LS &  0.053 &   0.8 &&  0.042 &  &&  0.047 &  &&  0.055 &   0.5 &&  0.083 &   4.5 &  0.94 \\
    CFF & 1QI &  0.053 &   0.8 &&  0.043 &  &&  0.048 &  &&  0.052 &   0.5 &&  0.079 &   4.4 &  0.94 \\
    CFF & 2LS &  0.040 &   0.6 &&  0.017 &  &&  0.025 &  &&  0.062 &   0.4 &&  0.092 &   3.6 &  0.94 \\
    CFF & 2QI &  0.036 &   0.7 &&  0.017 &  &&  0.024 &  &&  0.055 &   0.5 &&  0.082 &   3.8 &  0.94 \\
    ADF & 1LS &  0.029 &  34.9 &&  0.032 &  41.3 &&  0.033 &  41.7 &&  0.021 &  22.7 &&  0.019 &  20.8 &  \\
    ADF & 1QI &  0.027 &  49.6 &&  0.030 &  58.7 &&  0.031 &  59.4 &&  0.015 &  31.6 &&  0.012 &  29.0 &  \\
    ADF & 2LS &  0.016 &  47.9 &&  0.019 &  54.3 &&  0.019 &  55.0 &&  0.009 &  33.8 &&  0.010 &  31.1 &  \\
    ADF & 2QI &  0.014 &  57.3 &&  0.018 &  66.9 &&  0.019 &  67.6 &&  0.009 &  37.0 &&  0.008 &  34.0 &  \\
    ADF$^2$ & 1LS &  0.026 &   1.2 &&  0.029 &   1.8 &&  0.029 &   1.6 &&  0.020 &   0.4 &&  0.019 &   0.5 &  \\
    ADF$^2$ & 1QI &  0.025 &   1.7 &&  0.029 &   2.5 &&  0.029 &   2.3 &&  0.018 &   0.6 &&  0.017 &   0.7 &  \\
    ADF$^2$ & 2LS &  0.015 &   0.6 &&  0.015 &   0.8 &&  0.015 &   0.7 &&  0.014 &   0.1 &&  0.014 &   0.3 &  \\
    ADF$^2$ & 2QI &  0.013 &   0.2 &&  0.014 &   0.4 &&  0.014 &   0.3 &&  0.013 &   0.1 &&  0.012 &   0.1 &  \\
    \hline
 \end{tabular}
 \tablefoot{See
   Tables~\ref{tab:shift_algorithms} and~\ref{tab:subpix_algorithms}
   for the CA and IA acronyms. The ``Fail (\%)'' columns show the
   percentage of $4\sigma$ outliers when not rounded to 0.0. The
   fitted slope, $p_1$, is shown only when not rounded to 1.00. The
     FOV is 16$\times$16 pixels in the upper part and 24$\times$24
     pixels in the bottom part.}  
\end{table*}

\begin{table*}[!t]
  \centering
  \caption{Simulation results, no bias mismatch, 0.5\% noise.}
  \label{tab:sim-n5b0}
  \begin{tabular}{llcD{.}{.}{2.1}@{\,}rcD{.}{.}{2.1}@{\,}rcD{.}{.}{2.1}@{\,}rcD{.}{.}{2.1}@{\,}rcD{.}{.}{2.1}ll}
    \hline\hline\noalign{\smallskip}
    \multirow{2}{*}{CA} & 
    \multirow{2}{*}{IA} &
    \multicolumn{2}{c}{All} && 
    \multicolumn{2}{c}{$\delta r<1$} &&
    \multicolumn{2}{c}{$\delta r<2$} &&
    \multicolumn{2}{c}{$3<\delta r<5$} &&
    \multicolumn{2}{c}{$\delta r>5$} &
    \multirow{2}{*}{$p_1$} \\ 
    \cline{3-4}\cline{6-7}\cline{9-10}\cline{12-13}\cline{15-16}\noalign{\smallskip}
    && 
    $\sigma$ (pix) & \multicolumn{1}{c}{\text{Fail (\%)}} && 
    $\sigma$ (pix) & \multicolumn{1}{c}{\text{Fail (\%)}} && 
    $\sigma$ (pix) & \multicolumn{1}{c}{\text{Fail (\%)}} &&
    $\sigma$ (pix) & \multicolumn{1}{c}{\text{Fail (\%)}} && 
    $\sigma$ (pix) & \multicolumn{1}{c}{\text{Fail (\%)}} &  \\ 
    \hline\noalign{\smallskip}
    SDF & 1LS &  0.041 &   0.6 &&  0.045 &   1.0 &&  0.045 &   0.9 &&  0.033 &   0.2 &&  0.033 &   0.2 &  \\
    SDF & 1QI &  0.038 &   1.0 &&  0.043 &   1.4 &&  0.043 &   1.3 &&  0.030 &   0.3 &&  0.029 &   0.2 &  \\
    SDF & 2LS &  0.026 &   0.1 &&  0.026 &   0.1 &&  0.026 &   0.1 &&  0.026 &   0.1 &&  0.026 &   0.1 &  \\
    SDF & 2QI &  0.024 &   0.1 &&  0.024 &   0.1 &&  0.024 &   0.1 &&  0.023 &   0.1 &&  0.023 &   0.1 &  \\
    CFI & 1LS &  0.117 &   0.7 &&  0.122 &   0.5 &&  0.121 &   0.7 &&  0.111 &   0.9 &&  0.107 &   0.8 &  \\
    CFI & 1QI &  0.110 &   0.7 &&  0.115 &   0.4 &&  0.114 &   0.6 &&  0.105 &   0.8 &&  0.101 &   0.8 &  \\
    CFI & 2LS &  0.111 &   0.7 &&  0.113 &   0.4 &&  0.113 &   0.6 &&  0.111 &   0.9 &&  0.107 &   0.9 &  \\
    CFI & 2QI &  0.104 &   0.6 &&  0.105 &   0.4 &&  0.105 &   0.5 &&  0.103 &   0.9 &&  0.100 &   0.8 &  \\
    CFF & 1LS &  0.118 &   5.0 &&  0.069 &  &&  0.085 &  &&  0.189 &   3.1 &&  0.297 &  30.0 &  0.86 \\
    CFF & 1QI &  0.117 &   5.0 &&  0.072 &  &&  0.086 &  &&  0.182 &   3.1 &&  0.281 &  30.0 &  0.86 \\
    CFF & 2LS &  0.097 &   5.1 &&  0.043 &  &&  0.064 &  &&  0.168 &   3.3 &&  0.271 &  30.7 &  0.86 \\
    CFF & 2QI &  0.092 &   5.2 &&  0.043 &  &&  0.062 &  &&  0.158 &   3.3 &&  0.252 &  30.8 &  0.86 \\
    ADF & 1LS &  0.049 &   4.0 &&  0.053 &   6.2 &&  0.053 &   5.8 &&  0.040 &   0.7 &&  0.038 &   0.7 &  \\
    ADF & 1QI &  0.055 &  11.2 &&  0.059 &  16.1 &&  0.059 &  15.6 &&  0.047 &   3.3 &&  0.044 &   3.1 &  \\
    ADF & 2LS &  0.039 &   5.8 &&  0.041 &   6.5 &&  0.041 &   6.5 &&  0.033 &   4.6 &&  0.032 &   4.3 &  \\
    ADF & 2QI &  0.041 &  20.2 &&  0.044 &  23.5 &&  0.045 &  23.6 &&  0.033 &  14.2 &&  0.032 &  13.1 &  \\
    ADF$^2$ & 1LS &  0.043 &   0.6 &&  0.048 &   0.9 &&  0.048 &   0.8 &&  0.035 &   0.1 &&  0.034 &   0.1 &  \\
    ADF$^2$ & 1QI &  0.041 &   0.8 &&  0.046 &   1.2 &&  0.046 &   1.1 &&  0.032 &   0.2 &&  0.031 &   0.2 &  \\
    ADF$^2$ & 2LS &  0.028 &   0.2 &&  0.029 &   0.2 &&  0.029 &   0.2 &&  0.027 &   0.1 &&  0.028 &   0.1 &  \\
    ADF$^2$ & 2QI &  0.026 &   0.1 &&  0.027 &   0.1 &&  0.027 &   0.1 &&  0.025 &  &&  0.025 &  &  \\
    \hline\noalign{\smallskip}
    SDF & 1LS &  0.028 &   0.7 &&  0.032 &   1.0 &&  0.032 &   0.9 &&  0.022 &   0.2 &&  0.022 &   0.3 &  \\
    SDF & 1QI &  0.027 &   0.9 &&  0.031 &   1.3 &&  0.031 &   1.3 &&  0.021 &   0.3 &&  0.021 &   0.4 &  \\
    SDF & 2LS &  0.018 &   0.1 &&  0.018 &   0.2 &&  0.018 &   0.2 &&  0.017 &   0.1 &&  0.017 &   0.1 &  \\
    SDF & 2QI &  0.017 &   0.1 &&  0.017 &   0.1 &&  0.017 &   0.1 &&  0.016 &  &&  0.016 &  &  \\
    CFI & 1LS &  0.071 &   0.3 &&  0.074 &   0.2 &&  0.073 &   0.2 &&  0.067 &   0.4 &&  0.065 &   0.4 &  \\
    CFI & 1QI &  0.067 &   0.2 &&  0.071 &   0.2 &&  0.070 &   0.2 &&  0.063 &   0.3 &&  0.061 &   0.3 &  \\
    CFI & 2LS &  0.066 &   0.3 &&  0.068 &   0.1 &&  0.067 &   0.2 &&  0.066 &   0.5 &&  0.064 &   0.5 &  \\
    CFI & 2QI &  0.061 &   0.2 &&  0.063 &  &&  0.063 &   0.1 &&  0.061 &   0.5 &&  0.059 &   0.5 &  \\
    CFF & 1LS &  0.057 &   0.8 &&  0.047 &  &&  0.051 &  &&  0.059 &   0.5 &&  0.086 &   4.3 &  0.94 \\
    CFF & 1QI &  0.057 &   0.7 &&  0.049 &  &&  0.053 &  &&  0.057 &   0.5 &&  0.083 &   4.1 &  0.94 \\
    CFF & 2LS &  0.045 &   0.6 &&  0.026 &  &&  0.032 &  &&  0.065 &   0.4 &&  0.093 &   3.5 &  0.94 \\
    CFF & 2QI &  0.042 &   0.6 &&  0.026 &  &&  0.031 &  &&  0.058 &   0.4 &&  0.083 &   3.6 &  0.94 \\
    ADF & 1LS &  0.037 &  13.8 &&  0.040 &  18.0 &&  0.041 &  18.0 &&  0.031 &   6.0 &&  0.029 &   5.8 &  \\
    ADF & 1QI &  0.038 &  29.9 &&  0.042 &  35.4 &&  0.042 &  36.0 &&  0.029 &  19.0 &&  0.027 &  17.5 &  \\
    ADF & 2LS &  0.027 &  22.5 &&  0.029 &  23.6 &&  0.029 &  24.7 &&  0.021 &  18.2 &&  0.020 &  16.7 &  \\
    ADF & 2QI &  0.026 &  39.2 &&  0.029 &  44.1 &&  0.030 &  45.4 &&  0.018 &  27.9 &&  0.017 &  25.5 &  \\
    ADF$^2$ & 1LS &  0.029 &   0.7 &&  0.033 &   1.0 &&  0.033 &   0.9 &&  0.023 &   0.2 &&  0.023 &   0.3 &  \\
    ADF$^2$ & 1QI &  0.029 &   0.8 &&  0.032 &   1.2 &&  0.032 &   1.1 &&  0.022 &   0.2 &&  0.022 &   0.3 &  \\
    ADF$^2$ & 2LS &  0.019 &   0.1 &&  0.020 &   0.2 &&  0.020 &   0.2 &&  0.018 &  &&  0.018 &   0.1 &  \\
    ADF$^2$ & 2QI &  0.018 &  &&  0.018 &  &&  0.018 &  &&  0.017 &  &&  0.017 &  &  \\
    \hline
  \end{tabular}
  \tablefoot{See Tables~\ref{tab:shift_algorithms}
    and~\ref{tab:subpix_algorithms} for the CA and IA acronyms. The
    ``Fail (\%)'' columns show the percentage of $4\sigma$ outliers
    when not rounded to 0.0. The fitted slope, $p_1$, is shown only
    when not rounded to 1.00. The FOV is 16$\times$16 pixels in the
    upper part and 24$\times$24 pixels in the bottom part.}
\end{table*}

\begin{table*}[!t]
  \centering
  \caption{Simulation results, 1\% bias mismatch, no noise.}
  \label{tab:sim-n0b1}
  \begin{tabular}{lllD{.}{.}{2.1}@{\,}rcD{.}{.}{2.1}@{\,}rcD{.}{.}{2.1}@{\,}rcD{.}{.}{2.1}@{\,}rcD{.}{.}{2.1}ll}
    \hline\hline\noalign{\smallskip}
    \multirow{2}{*}{CA} & 
    \multirow{2}{*}{IA} &
    \multicolumn{2}{c}{All} && 
    \multicolumn{2}{c}{$\delta r<1$} &&
    \multicolumn{2}{c}{$\delta r<2$} &&
    \multicolumn{2}{c}{$3<\delta r<5$} &&
    \multicolumn{2}{c}{$\delta r>5$} &
    \multirow{2}{*}{$p_1$} \\ 
    \cline{3-4}\cline{6-7}\cline{9-10}\cline{12-13}\cline{15-16}\noalign{\smallskip}
    && 
    $\sigma$ (pix) & \multicolumn{1}{c}{\text{Fail (\%)}} && 
    $\sigma$ (pix) & \multicolumn{1}{c}{\text{Fail (\%)}} && 
    $\sigma$ (pix) & \multicolumn{1}{c}{\text{Fail (\%)}} &&
    $\sigma$ (pix) & \multicolumn{1}{c}{\text{Fail (\%)}} && 
    $\sigma$ (pix) & \multicolumn{1}{c}{\text{Fail (\%)}} &  \\ 
    \hline\noalign{\smallskip}
    SDF & 1LS &  0.087 &   0.1 &&  0.092 &   0.1 &&  0.090 &   0.1 &&  0.082 &  &&  0.083 &  &  \\
    SDF & 1QI &  0.080 &   0.1 &&  0.084 &   0.2 &&  0.083 &   0.2 &&  0.074 &  &&  0.075 &  &  \\
    SDF & 2LS &  0.079 &   0.1 &&  0.081 &   0.1 &&  0.079 &   0.1 &&  0.079 &   0.1 &&  0.081 &  &  \\
    SDF & 2QI &  0.071 &   0.1 &&  0.072 &   0.1 &&  0.071 &   0.2 &&  0.071 &   0.1 &&  0.072 &  &  \\
    ADF & 1LS &  0.104 &   0.2 &&  0.112 &   0.2 &&  0.109 &   0.2 &&  0.095 &   0.2 &&  0.097 &   0.1 &  \\
    ADF & 1QI &  0.107 &   0.3 &&  0.115 &   0.3 &&  0.112 &   0.3 &&  0.098 &   0.3 &&  0.099 &   0.3 &  \\
    ADF & 2LS &  0.100 &   0.2 &&  0.104 &   0.2 &&  0.102 &   0.2 &&  0.095 &   0.2 &&  0.097 &   0.1 &  \\
    ADF & 2QI &  0.100 &   0.3 &&  0.105 &   0.2 &&  0.103 &   0.3 &&  0.096 &   0.4 &&  0.098 &   0.4 &  \\
    ADF$^2$ & 1LS &  0.108 &   0.1 &&  0.113 &   0.1 &&  0.111 &   0.1 &&  0.102 &  &&  0.104 &  &  \\
    ADF$^2$ & 1QI &  0.108 &   0.1 &&  0.111 &   0.1 &&  0.109 &   0.1 &&  0.106 &   0.1 &&  0.108 &  &  \\
    ADF$^2$ & 2LS &  0.100 &   0.1 &&  0.103 &   0.1 &&  0.100 &   0.2 &&  0.100 &   0.1 &&  0.102 &  &  \\
    ADF$^2$ & 2QI &  0.101 &   0.2 &&  0.101 &   0.2 &&  0.099 &   0.2 &&  0.103 &   0.2 &&  0.106 &   0.1 &  \\
    \hline\noalign{\smallskip}
    SDF & 1LS &  0.055 &   0.1 &&  0.058 &   0.1 &&  0.058 &   0.1 &&  0.050 &  &&  0.049 &  &  \\
    SDF & 1QI &  0.050 &   0.1 &&  0.053 &   0.1 &&  0.053 &   0.1 &&  0.045 &  &&  0.044 &  &  \\
    SDF & 2LS &  0.049 &  &&  0.049 &  &&  0.050 &  &&  0.047 &  &&  0.047 &  &  \\
    SDF & 2QI &  0.043 &  &&  0.044 &  &&  0.044 &  &&  0.042 &  &&  0.041 &  &  \\
    ADF & 1LS &  0.067 &   0.3 &&  0.071 &   0.3 &&  0.071 &   0.4 &&  0.059 &   0.1 &&  0.058 &   0.1 &  \\
    ADF & 1QI &  0.069 &   0.3 &&  0.074 &   0.4 &&  0.074 &   0.3 &&  0.062 &   0.1 &&  0.060 &   0.1 &  \\
    ADF & 2LS &  0.063 &   0.2 &&  0.065 &   0.1 &&  0.065 &   0.2 &&  0.059 &   0.1 &&  0.058 &   0.1 &  \\
    ADF & 2QI &  0.064 &   0.2 &&  0.066 &   0.3 &&  0.066 &   0.3 &&  0.060 &   0.2 &&  0.058 &   0.2 &  \\
    ADF$^2$ & 1LS &  0.068 &  &&  0.070 &  &&  0.070 &  &&  0.063 &  &&  0.062 &  &  \\
    ADF$^2$ & 1QI &  0.069 &  &&  0.070 &  &&  0.070 &  &&  0.066 &  &&  0.065 &  &  \\
    ADF$^2$ & 2LS &  0.062 &  &&  0.063 &  &&  0.063 &  &&  0.061 &  &&  0.060 &  &  \\
    ADF$^2$ & 2QI &  0.064 &  &&  0.064 &  &&  0.064 &  &&  0.064 &  &&  0.063 &  &  \\
    \hline
  \end{tabular}
  \tablefoot{See Tables~\ref{tab:shift_algorithms}
    and~\ref{tab:subpix_algorithms} for the CA and IA acronyms. The
    ``Fail (\%)'' columns show the percentage of $4\sigma$ outliers
    when not rounded to 0.0. The fitted slope, $p_1$, is shown only
    when not rounded to 1.00. The FOV is 16$\times$16 pixels in the
    upper part and 24$\times$24 pixels in the bottom part.}
\end{table*}

\begin{table*}[!t]
  \centering
  \caption{Simulation results, 1\% bias mismatch, 0.5\% noise.}
  \label{tab:sim-n5b1}
  \begin{tabular}{lllD{.}{.}{2.1}@{\,}rcD{.}{.}{2.1}@{\,}rcD{.}{.}{2.1}@{\,}rcD{.}{.}{2.1}@{\,}rcD{.}{.}{2.1}ll}
    \hline\hline\noalign{\smallskip}
    \multirow{2}{*}{CA} & 
    \multirow{2}{*}{IA} &
    \multicolumn{2}{c}{All} && 
    \multicolumn{2}{c}{$\delta r<1$} &&
    \multicolumn{2}{c}{$\delta r<2$} &&
    \multicolumn{2}{c}{$3<\delta r<5$} &&
    \multicolumn{2}{c}{$\delta r>5$} &
    \multirow{2}{*}{$p_1$} \\ 
    \cline{3-4}\cline{6-7}\cline{9-10}\cline{12-13}\cline{15-16}\noalign{\smallskip}
    && 
    $\sigma$ (pix) & \multicolumn{1}{c}{\text{Fail (\%)}} && 
    $\sigma$ (pix) & \multicolumn{1}{c}{\text{Fail (\%)}} && 
    $\sigma$ (pix) & \multicolumn{1}{c}{\text{Fail (\%)}} &&
    $\sigma$ (pix) & \multicolumn{1}{c}{\text{Fail (\%)}} && 
    $\sigma$ (pix) & \multicolumn{1}{c}{\text{Fail (\%)}} &  \\ 
    \hline\noalign{\smallskip}
    SDF & 1LS &  0.089 &   0.1 &&  0.094 &   0.1 &&  0.092 &   0.1 &&  0.084 &  &&  0.086 &  &  \\
    SDF & 1QI &  0.082 &   0.1 &&  0.086 &   0.2 &&  0.085 &   0.1 &&  0.076 &  &&  0.077 &  &  \\
    SDF & 2LS &  0.081 &   0.1 &&  0.083 &   0.1 &&  0.081 &   0.1 &&  0.081 &  &&  0.083 &  &  \\
    SDF & 2QI &  0.073 &   0.1 &&  0.074 &   0.1 &&  0.073 &   0.1 &&  0.073 &  &&  0.074 &  &  \\
    ADF & 1LS &  0.103 &   0.2 &&  0.110 &   0.2 &&  0.108 &   0.2 &&  0.094 &   0.1 &&  0.096 &  &  \\
    ADF & 1QI &  0.103 &   0.3 &&  0.111 &   0.3 &&  0.109 &   0.3 &&  0.093 &   0.2 &&  0.094 &   0.2 &  \\
    ADF & 2LS &  0.098 &   0.2 &&  0.103 &   0.2 &&  0.101 &   0.2 &&  0.094 &   0.1 &&  0.096 &   0.1 &  \\
    ADF & 2QI &  0.096 &   0.3 &&  0.101 &   0.2 &&  0.099 &   0.2 &&  0.091 &   0.3 &&  0.092 &   0.3 &  \\
    ADF$^2$ & 1LS &  0.107 &   0.1 &&  0.112 &   0.1 &&  0.110 &   0.1 &&  0.101 &  &&  0.103 &  &  \\
    ADF$^2$ & 1QI &  0.103 &   0.1 &&  0.107 &   0.2 &&  0.105 &   0.1 &&  0.100 &   0.1 &&  0.102 &  &  \\
    ADF$^2$ & 2LS &  0.099 &   0.1 &&  0.101 &   0.1 &&  0.099 &   0.1 &&  0.098 &  &&  0.101 &  &  \\
    ADF$^2$ & 2QI &  0.096 &   0.1 &&  0.097 &   0.2 &&  0.095 &   0.2 &&  0.097 &   0.1 &&  0.099 &   0.1 &  \\
    \hline\noalign{\smallskip}
    SDF & 1LS &  0.057 &   0.1 &&  0.060 &  &&  0.060 &   0.1 &&  0.052 &  &&  0.051 &  &  \\
    SDF & 1QI &  0.052 &   0.1 &&  0.055 &   0.1 &&  0.055 &   0.1 &&  0.047 &  &&  0.047 &  &  \\
    SDF & 2LS &  0.051 &  &&  0.052 &  &&  0.052 &  &&  0.049 &  &&  0.049 &  &  \\
    SDF & 2QI &  0.046 &  &&  0.046 &  &&  0.046 &  &&  0.044 &  &&  0.044 &  &  \\
    ADF & 1LS &  0.067 &   0.3 &&  0.071 &   0.4 &&  0.072 &   0.4 &&  0.060 &  &&  0.059 &   0.1 &  \\
    ADF & 1QI &  0.068 &   0.3 &&  0.073 &   0.5 &&  0.073 &   0.4 &&  0.059 &   0.1 &&  0.058 &   0.1 &  \\
    ADF & 2LS &  0.064 &   0.2 &&  0.066 &   0.2 &&  0.066 &   0.2 &&  0.059 &   0.1 &&  0.058 &   0.1 &  \\
    ADF & 2QI &  0.063 &   0.3 &&  0.065 &   0.3 &&  0.066 &   0.3 &&  0.058 &   0.2 &&  0.056 &   0.2 &  \\
    ADF$^2$ & 1LS &  0.068 &  &&  0.071 &  &&  0.071 &  &&  0.063 &  &&  0.062 &  &  \\
    ADF$^2$ & 1QI &  0.066 &  &&  0.068 &  &&  0.068 &  &&  0.063 &  &&  0.061 &  &  \\
    ADF$^2$ & 2LS &  0.063 &  &&  0.063 &  &&  0.064 &  &&  0.061 &  &&  0.060 &  &  \\
    ADF$^2$ & 2QI &  0.061 &  &&  0.061 &  &&  0.062 &  &&  0.060 &  &&  0.059 &  &  \\
    \hline
  \end{tabular}
  \tablefoot{See Tables~\ref{tab:shift_algorithms}
    and~\ref{tab:subpix_algorithms} for the CA and IA acronyms. The
    ``Fail (\%)'' columns show the percentage of $4\sigma$ outliers
    when not rounded to 0.0. The fitted slope, $p_1$, is shown only
    when not rounded to 1.00. The FOV is 16$\times$16 pixels in the
    upper part and 24$\times$24 pixels in the bottom part.}
\end{table*}

\begin{table*}[!t]
  \centering
  \caption{Simulation results, no bias mismatch, no noise. Flat top
    window for FFT.}
  \label{tab:sim-n0b0-flat}
  \begin{tabular}{lllD{.}{.}{2.1}@{\,}rcD{.}{.}{2.1}@{\,}rcD{.}{.}{2.1}@{\,}rcD{.}{.}{2.1}@{\,}rcD{.}{.}{2.1}ll}
    \hline\hline\noalign{\smallskip}
    \multirow{2}{*}{CA} & 
    \multirow{2}{*}{IA} &
    \multicolumn{2}{c}{All} && 
    \multicolumn{2}{c}{$\delta r<1$} &&
    \multicolumn{2}{c}{$\delta r<2$} &&
    \multicolumn{2}{c}{$3<\delta r<5$} &&
    \multicolumn{2}{c}{$\delta r>5$} &
    \multirow{2}{*}{$p_1$} \\ 
    \cline{3-4}\cline{6-7}\cline{9-10}\cline{12-13}\cline{15-16}\noalign{\smallskip}
    && 
    $\sigma$ (pix) & \multicolumn{1}{c}{\text{Fail (\%)}} && 
    $\sigma$ (pix) & \multicolumn{1}{c}{\text{Fail (\%)}} && 
    $\sigma$ (pix) & \multicolumn{1}{c}{\text{Fail (\%)}} &&
    $\sigma$ (pix) & \multicolumn{1}{c}{\text{Fail (\%)}} && 
    $\sigma$ (pix) & \multicolumn{1}{c}{\text{Fail (\%)}} &  \\ 
    \hline\noalign{\smallskip}
    CFF & 1LS &  0.146 &   2.0 &&  0.076 &  &&  0.105 &  &&  0.193 &   1.6 &&  0.305 &  11.6 &  0.95 \\
    CFF & 1QI &  0.141 &   1.9 &&  0.076 &  &&  0.103 &  &&  0.183 &   1.5 &&  0.293 &  11.6 &  0.95 \\
    CFF & 2LS &  0.136 &   2.0 &&  0.063 &  &&  0.095 &  &&  0.187 &   1.7 &&  0.295 &  11.8 &  0.95 \\
    CFF & 2QI &  0.131 &   2.0 &&  0.062 &  &&  0.092 &  &&  0.176 &   1.5 &&  0.282 &  11.7 &  0.95 \\
    \hline\noalign{\smallskip}
    CFF & 1LS &  0.077 &   0.4 &&  0.044 &  &&  0.063 &  &&  0.082 &   0.5 &&  0.122 &   1.7 &  0.97 \\
    CFF & 1QI &  0.074 &   0.3 &&  0.044 &  &&  0.061 &  &&  0.077 &   0.5 &&  0.116 &   1.6 &  0.97 \\
    CFF & 2LS &  0.072 &   0.3 &&  0.034 &  &&  0.054 &  &&  0.082 &   0.4 &&  0.123 &   1.5 &  0.97 \\
    CFF & 2QI &  0.068 &   0.3 &&  0.032 &  &&  0.052 &  &&  0.076 &   0.3 &&  0.115 &   1.4 &  0.97 \\
    \hline
  \end{tabular}
  \tablefoot{See Tables~\ref{tab:shift_algorithms}
    and~\ref{tab:subpix_algorithms} for the CA and IA acronyms. The
    ``Fail (\%)'' columns show the percentage of $4\sigma$ outliers
    when not rounded to 0.0. The fitted slope, $p_1$, is shown only
    when not rounded to 1.00. The FOV is 16$\times$16 pixels in the
    upper part and 24$\times$24 pixels in the bottom part.}
  \caption{Simulation results, no bias mismatch, 0.5\% noise. Flat top
    window for FFT.}
  \label{tab:sim-n5b0-flat}
  \begin{tabular}{lllD{.}{.}{2.1}@{\,}rcD{.}{.}{2.1}@{\,}rcD{.}{.}{2.1}@{\,}rcD{.}{.}{2.1}@{\,}rcD{.}{.}{2.1}ll}
    \hline\hline\noalign{\smallskip}
    \multirow{2}{*}{CA} & 
    \multirow{2}{*}{IA} &
    \multicolumn{2}{c}{All} && 
    \multicolumn{2}{c}{$\delta r<1$} &&
    \multicolumn{2}{c}{$\delta r<2$} &&
    \multicolumn{2}{c}{$3<\delta r<5$} &&
    \multicolumn{2}{c}{$\delta r>5$} &
    \multirow{2}{*}{$p_1$} \\ 
    \cline{3-4}\cline{6-7}\cline{9-10}\cline{12-13}\cline{15-16}\noalign{\smallskip}
    && 
    $\sigma$ (pix) & \multicolumn{1}{c}{\text{Fail (\%)}} && 
    $\sigma$ (pix) & \multicolumn{1}{c}{\text{Fail (\%)}} && 
    $\sigma$ (pix) & \multicolumn{1}{c}{\text{Fail (\%)}} &&
    $\sigma$ (pix) & \multicolumn{1}{c}{\text{Fail (\%)}} && 
    $\sigma$ (pix) & \multicolumn{1}{c}{\text{Fail (\%)}} &  \\ 
    \hline\noalign{\smallskip}
    CFF & 1LS &  0.148 &   2.0 &&  0.079 &  &&  0.107 &  &&  0.194 &   1.6 &&  0.307 &  11.8 &  0.95 \\
    CFF & 1QI &  0.144 &   2.0 &&  0.079 &  &&  0.105 &  &&  0.185 &   1.5 &&  0.295 &  11.7 &  0.95 \\
    CFF & 2LS &  0.139 &   2.0 &&  0.066 &  &&  0.098 &  &&  0.188 &   1.6 &&  0.295 &  12.0 &  0.95 \\
    CFF & 2QI &  0.133 &   2.0 &&  0.065 &  &&  0.095 &  &&  0.177 &   1.4 &&  0.283 &  11.9 &  0.95 \\
    \hline
    CFF & 1LS &  0.078 &   0.3 &&  0.046 &  &&  0.064 &  &&  0.083 &   0.5 &&  0.123 &   1.6 &  0.97 \\
    CFF & 1QI &  0.076 &   0.3 &&  0.046 &  &&  0.063 &  &&  0.078 &   0.4 &&  0.117 &   1.5 &  0.97 \\
    CFF & 2LS &  0.073 &   0.3 &&  0.036 &  &&  0.056 &  &&  0.084 &   0.3 &&  0.124 &   1.3 &  0.97 \\
    CFF & 2QI &  0.069 &   0.3 &&  0.035 &  &&  0.054 &  &&  0.078 &   0.3 &&  0.116 &   1.2 &  0.97 \\
    \hline
  \end{tabular}
  \tablefoot{See Tables~\ref{tab:shift_algorithms}
    and~\ref{tab:subpix_algorithms} for the CA and IA acronyms. The
    ``Fail (\%)'' columns show the percentage of $4\sigma$ outliers
    when not rounded to 0.0. The fitted slope, $p_1$, is shown only
    when not rounded to 1.00. The FOV is 16$\times$16 pixels in the
    upper part and 24$\times$24 pixels in the bottom part.}
\end{table*}

\clearpage

\begin{table*}[!t]
  \caption{Results from seeing simulation.}   
  \label{tab:seeing-simulation} 
  \centering
  \begin{tabular}{rl%
      D{.}{.}{1.2}D{.}{.}{1.3}D{.}{.}{1.3}D{.}{.}{1.2}D{.}{.}{1.3}c%
      D{.}{.}{1.2}D{.}{.}{1.3}D{.}{.}{1.3}D{.}{.}{1.2}D{.}{.}{1.3}}
    \hline\hline\noalign{\smallskip}
    \multirow{2}{*}{$r_0$ (cm)}&
    \multirow{2}{*}{CA}&
    \multicolumn{5}{c}{No noise}&
    &
    \multicolumn{5}{c}{1\% noise}\\
    \cline{3-7}\cline{9-13}\noalign{\smallskip}
    & &
    \multicolumn{1}{l}{$\sigma_\text{shift}$ (\arcsec)} &
    \multicolumn{1}{l}{$\sigma_\text{err}$ (\arcsec)} &
    \multicolumn{1}{l}{$\sigma_\text{err}/\sigma_\text{tilt}$} &
    \multicolumn{1}{l}{\text{Fail (\%)}} & 
    \multicolumn{1}{l}{$p_1$} & 
    &
    \multicolumn{1}{l}{$\sigma_\text{shift}$ (\arcsec)} &
    \multicolumn{1}{l}{$\sigma_\text{err}$ (\arcsec)} &
    \multicolumn{1}{l}{$\sigma_\text{err}/\sigma_\text{tilt}$} &
    \multicolumn{1}{l}{\text{Fail (\%)}} &
    \multicolumn{1}{l}{$p_1$} \\
    \hline\noalign{\smallskip}
    5 & SDF & 1.11 & 0.054 & 0.049 &   1.67 &  1.011 &&1.13 & 0.158 & 0.144 &   1.22 &  1.011 \\ 
    5 & ADF$^2$ & 1.11 & 0.055 & 0.050 &   1.65 &  1.010 &&1.13 & 0.169 & 0.154 &   1.27 &  1.010 \\ 
    5 & CFI & 1.12 & 0.090 & 0.082 &   1.64 &  1.012 &&1.16 & 0.181 & 0.165 &   3.44 &  1.012 \\ 
    5 & CFF & 0.84 & 0.111 & 0.101 &  10.43 &  0.832 &&0.81 & 0.509 & 0.464 &   2.06 &  0.505 \\ 
    7 & SDF & 0.84 & 0.030 & 0.037 &   0.55 &  1.009 &&0.85 & 0.115 & 0.138 &   0.52 &  1.009 \\ 
    7 & ADF$^2$ & 0.84 & 0.031 & 0.037 &   0.52 &  1.009 &&0.85 & 0.123 & 0.148 &   0.54 &  1.009 \\ 
    7 & CFI & 0.84 & 0.063 & 0.076 &   0.57 &  1.010 &&0.86 & 0.132 & 0.159 &   1.27 &  1.010 \\ 
    7 & CFF & 0.69 & 0.064 & 0.077 &   4.01 &  0.859 &&0.69 & 0.235 & 0.283 &   5.57 &  0.796 \\ 
    10 & SDF & 0.62 & 0.019 & 0.031 &   0.13 &  1.009 &&0.63 & 0.095 & 0.154 &   0.25 &  1.009 \\ 
    10 & ADF$^2$ & 0.62 & 0.020 & 0.032 &   0.12 &  1.009 &&0.63 & 0.102 & 0.166 &   0.28 &  1.009 \\ 
    10 & CFI & 0.62 & 0.051 & 0.083 &   0.23 &  1.009 &&0.64 & 0.110 & 0.178 &   0.54 &  1.009 \\ 
    10 & CFF & 0.53 & 0.042 & 0.069 &   1.46 &  0.871 &&0.55 & 0.179 & 0.291 &   2.51 &  0.840 \\ 
    15 & SDF & 0.44 & 0.014 & 0.031 &   0.16 &  1.008 &&0.45 & 0.085 & 0.194 &   0.31 &  1.008 \\ 
    15 & ADF$^2$ & 0.44 & 0.014 & 0.033 &   0.14 &  1.008 &&0.45 & 0.091 & 0.208 &   0.33 &  1.008 \\ 
    15 & CFI & 0.45 & 0.046 & 0.104 &   0.24 &  1.008 &&0.46 & 0.099 & 0.225 &   0.48 &  1.009 \\ 
    15 & CFF & 0.38 & 0.029 & 0.066 &   1.03 &  0.876 &&0.41 & 0.157 & 0.357 &   1.06 &  0.855 \\ 
    20 & SDF & 0.35 & 0.011 & 0.033 &   0.04 &  1.008 &&0.36 & 0.082 & 0.236 &   0.20 &  1.008 \\ 
    20 & ADF$^2$ & 0.35 & 0.012 & 0.035 &   0.03 &  1.008 &&0.36 & 0.088 & 0.254 &   0.22 &  1.008 \\ 
    20 & CFI & 0.35 & 0.043 & 0.126 &   0.09 &  1.008 &&0.36 & 0.095 & 0.274 &   0.37 &  1.009 \\ 
    20 & CFF & 0.30 & 0.023 & 0.067 &   0.82 &  0.878 &&0.34 & 0.149 & 0.431 &   0.72 &  0.859 \\ 
    \hline\noalign{\smallskip}
    5 & SDF & 1.11 & 0.052 & 0.047 &   1.66 &  1.010 &&1.12 & 0.111 & 0.102 &   0.97 &  1.011 \\ 
    5 & ADF$^2$ & 1.11 & 0.052 & 0.048 &   1.64 &  1.010 &&1.12 & 0.119 & 0.108 &   0.93 &  1.010 \\ 
    5 & CFI & 1.11 & 0.068 & 0.062 &   1.28 &  1.012 &&1.12 & 0.121 & 0.110 &   1.03 &  1.012 \\ 
    5 & CFF & 1.00 & 0.067 & 0.061 &   2.11 &  0.919 &&0.98 & 0.193 & 0.176 &   3.10 &  0.900 \\ 
    7 & SDF & 0.84 & 0.029 & 0.035 &   0.55 &  1.009 &&0.84 & 0.078 & 0.094 &   0.39 &  1.009 \\ 
    7 & ADF$^2$ & 0.84 & 0.029 & 0.035 &   0.51 &  1.009 &&0.84 & 0.083 & 0.100 &   0.38 &  1.009 \\ 
    7 & CFI & 0.84 & 0.043 & 0.052 &   0.39 &  1.010 &&0.85 & 0.085 & 0.102 &   0.39 &  1.010 \\ 
    7 & CFF & 0.77 & 0.038 & 0.046 &   0.69 &  0.930 &&0.78 & 0.137 & 0.166 &   0.87 &  0.923 \\ 
    10 & SDF & 0.62 & 0.018 & 0.029 &   0.12 &  1.008 &&0.62 & 0.063 & 0.102 &   0.11 &  1.008 \\ 
    10 & ADF$^2$ & 0.62 & 0.018 & 0.030 &   0.12 &  1.008 &&0.63 & 0.068 & 0.110 &   0.13 &  1.008 \\ 
    10 & CFI & 0.62 & 0.032 & 0.053 &   0.09 &  1.009 &&0.63 & 0.070 & 0.113 &   0.10 &  1.009 \\ 
    10 & CFF & 0.58 & 0.024 & 0.040 &   0.36 &  0.935 &&0.59 & 0.114 & 0.185 &   0.34 &  0.931 \\ 
    15 & SDF & 0.44 & 0.012 & 0.028 &   0.16 &  1.008 &&0.45 & 0.056 & 0.127 &   0.19 &  1.008 \\ 
    15 & ADF$^2$ & 0.44 & 0.012 & 0.028 &   0.14 &  1.008 &&0.45 & 0.060 & 0.137 &   0.19 &  1.008 \\ 
    15 & CFI & 0.44 & 0.028 & 0.063 &   0.18 &  1.009 &&0.45 & 0.062 & 0.142 &   0.18 &  1.009 \\ 
    15 & CFF & 0.41 & 0.017 & 0.038 &   0.39 &  0.938 &&0.42 & 0.102 & 0.232 &   0.36 &  0.935 \\ 
    20 & SDF & 0.35 & 0.010 & 0.029 &   0.02 &  1.008 &&0.35 & 0.053 & 0.155 &   0.07 &  1.008 \\ 
    20 & ADF$^2$ & 0.35 & 0.010 & 0.029 &   0.03 &  1.008 &&0.35 & 0.057 & 0.166 &   0.09 &  1.008 \\ 
    20 & CFI & 0.35 & 0.026 & 0.075 &   0.05 &  1.009 &&0.35 & 0.060 & 0.173 &   0.06 &  1.009 \\ 
    20 & CFF & 0.32 & 0.013 & 0.039 &   0.25 &  0.939 &&0.34 & 0.098 & 0.282 &   0.27 &  0.935 \\ 
    \hline\noalign{\smallskip}
    5 & SDF & 1.11 & 0.051 & 0.047 &   1.66 &  1.010 &&1.11 & 0.083 & 0.075 &   1.08 &  1.011 \\ 
    5 & ADF$^2$ & 1.11 & 0.051 & 0.047 &   1.64 &  1.010 &&1.11 & 0.087 & 0.079 &   1.05 &  1.011 \\ 
    5 & CFI & 1.11 & 0.058 & 0.053 &   1.41 &  1.011 &&1.11 & 0.087 & 0.079 &   1.06 &  1.011 \\ 
    5 & CFF & 1.06 & 0.054 & 0.049 &   1.47 &  0.965 &&1.06 & 0.129 & 0.117 &   0.88 &  0.962 \\ 
    7 & SDF & 0.84 & 0.028 & 0.034 &   0.58 &  1.009 &&0.84 & 0.055 & 0.066 &   0.38 &  1.009 \\ 
    7 & ADF$^2$ & 0.84 & 0.028 & 0.034 &   0.52 &  1.009 &&0.84 & 0.058 & 0.070 &   0.34 &  1.009 \\ 
    7 & CFI & 0.84 & 0.035 & 0.042 &   0.46 &  1.009 &&0.84 & 0.059 & 0.071 &   0.36 &  1.009 \\ 
    7 & CFF & 0.81 & 0.030 & 0.036 &   0.51 &  0.971 &&0.81 & 0.091 & 0.110 &   0.41 &  0.970 \\ 
    10 & SDF & 0.62 & 0.017 & 0.028 &   0.11 &  1.008 &&0.62 & 0.043 & 0.070 &   0.05 &  1.008 \\ 
    10 & ADF$^2$ & 0.62 & 0.018 & 0.028 &   0.11 &  1.008 &&0.62 & 0.046 & 0.074 &   0.06 &  1.008 \\ 
    10 & CFI & 0.62 & 0.025 & 0.040 &   0.12 &  1.008 &&0.62 & 0.047 & 0.076 &   0.05 &  1.008 \\ 
    10 & CFF & 0.60 & 0.019 & 0.031 &   0.12 &  0.973 &&0.60 & 0.075 & 0.122 &   0.15 &  0.972 \\ 
    15 & SDF & 0.44 & 0.012 & 0.027 &   0.17 &  1.008 &&0.44 & 0.037 & 0.085 &   0.16 &  1.008 \\ 
    15 & ADF$^2$ & 0.44 & 0.012 & 0.027 &   0.14 &  1.008 &&0.44 & 0.040 & 0.091 &   0.13 &  1.008 \\ 
    15 & CFI & 0.44 & 0.020 & 0.045 &   0.26 &  1.007 &&0.44 & 0.041 & 0.093 &   0.15 &  1.007 \\ 
    15 & CFF & 0.43 & 0.013 & 0.029 &   0.18 &  0.974 &&0.43 & 0.067 & 0.152 &   0.23 &  0.974 \\ 
    20 & SDF & 0.35 & 0.009 & 0.027 &   0.02 &  1.008 &&0.35 & 0.035 & 0.102 &   0.03 &  1.008 \\ 
    20 & ADF$^2$ & 0.35 & 0.009 & 0.027 &   0.02 &  1.008 &&0.35 & 0.038 & 0.110 &   0.03 &  1.008 \\ 
    20 & CFI & 0.35 & 0.018 & 0.053 &   0.15 &  1.007 &&0.35 & 0.039 & 0.113 &   0.03 &  1.007 \\ 
    20 & CFF & 0.34 & 0.010 & 0.029 &   0.04 &  0.974 &&0.34 & 0.064 & 0.185 &   0.11 &  0.974 \\ 
    \hline
  \end{tabular}
  \tablefoot{Different CAs, IA always 2QI. The ``Fail (\%)'' columns
    show the percentage of $4\sigma$ outliers. $p_1$ is the linear
    coefficient fitted to Eq.~(\ref{eq:linesine2}). The image scale is 
    0\farcs41/pixel. In the top part, the FOV is 16$\times$16
    pixels or 6\farcs6$\times$6\farcs6. In the center part, the FOV
    is 24$\times$24 pixels or 9\farcs8$\times$9\farcs8. In the
    bottom part, the FOV is 36$\times$36 pixels or
    14\farcs8$\times$14\farcs8.}
\end{table*}

\begin{table*}[!t]
  \caption{Results from seeing simulation.}   
  \label{tab:seeing-simulation-comp} 
  \centering
  \begin{tabular}{rl%
      D{.}{.}{1.2}D{.}{.}{1.3}D{.}{.}{1.3}D{.}{.}{1.2}D{.}{.}{1.3}c%
      D{.}{.}{1.2}D{.}{.}{1.3}D{.}{.}{1.3}D{.}{.}{1.2}D{.}{.}{1.3}}
    \hline\hline\noalign{\smallskip}
    \multirow{2}{*}{$r_0$ (cm)}&
    \multirow{2}{*}{CA}&
    \multicolumn{5}{c}{No noise}&
    &
    \multicolumn{5}{c}{1\% noise}\\
    \cline{3-7}\cline{9-13}\noalign{\smallskip}
    & &
    \multicolumn{1}{l}{$\sigma_\text{shift}$ (\arcsec)} &
    \multicolumn{1}{l}{$\sigma_\text{err}$ (\arcsec)} &
    \multicolumn{1}{l}{$\sigma_\text{err}/\sigma_\text{tilt}$} &
    \multicolumn{1}{l}{\text{Fail (\%)}} & 
    \multicolumn{1}{l}{$p_1$} & 
    &
    \multicolumn{1}{l}{$\sigma_\text{shift}$ (\arcsec)} &
    \multicolumn{1}{l}{$\sigma_\text{err}$ (\arcsec)} &
    \multicolumn{1}{l}{$\sigma_\text{err}/\sigma_\text{tilt}$} &
    \multicolumn{1}{l}{\text{Fail (\%)}} &
    \multicolumn{1}{l}{$p_1$} \\
    \hline\noalign{\smallskip}
    5 & SDF & 1.09 & 0.054 & 0.049 &   4.09 &  1.011 &&1.09 & 0.114 & 0.104 &   3.50 &  1.010 \\ 
    5 & ADF$^2$ & 1.09 & 0.055 & 0.050 &   4.08 &  1.011 &&1.09 & 0.122 & 0.111 &   3.47 &  1.009 \\ 
    5 & CFI & 1.09 & 0.084 & 0.077 &   3.91 &  1.011 &&1.10 & 0.135 & 0.123 &   3.88 &  1.010 \\ 
    5 & CFF & 0.88 & 0.095 & 0.087 &   8.39 &  0.859 &&0.85 & 0.212 & 0.193 &   8.50 &  0.814 \\ 
    7 & SDF & 0.84 & 0.030 & 0.036 &   0.64 &  1.009 &&0.84 & 0.079 & 0.095 &   0.53 &  1.009 \\ 
    7 & ADF$^2$ & 0.84 & 0.030 & 0.036 &   0.64 &  1.009 &&0.84 & 0.085 & 0.103 &   0.53 &  1.009 \\ 
    7 & CFI & 0.84 & 0.057 & 0.069 &   0.61 &  1.010 &&0.84 & 0.096 & 0.116 &   0.57 &  1.010 \\ 
    7 & CFF & 0.71 & 0.055 & 0.066 &   2.69 &  0.879 &&0.71 & 0.135 & 0.163 &   3.20 &  0.864 \\ 
    10 & SDF & 0.62 & 0.018 & 0.030 &   0.17 &  1.009 &&0.63 & 0.064 & 0.104 &   0.21 &  1.009 \\ 
    10 & ADF$^2$ & 0.62 & 0.019 & 0.030 &   0.16 &  1.009 &&0.63 & 0.069 & 0.113 &   0.22 &  1.009 \\ 
    10 & CFI & 0.62 & 0.046 & 0.075 &   0.24 &  1.009 &&0.63 & 0.080 & 0.129 &   0.19 &  1.009 \\ 
    10 & CFF & 0.54 & 0.036 & 0.058 &   1.11 &  0.888 &&0.55 & 0.107 & 0.174 &   0.82 &  0.878 \\ 
    15 & SDF & 0.44 & 0.012 & 0.027 &   0.08 &  1.008 &&0.45 & 0.057 & 0.129 &   0.17 &  1.008 \\ 
    15 & ADF$^2$ & 0.44 & 0.012 & 0.028 &   0.07 &  1.008 &&0.45 & 0.062 & 0.140 &   0.18 &  1.008 \\ 
    15 & CFI & 0.45 & 0.041 & 0.093 &   0.12 &  1.008 &&0.45 & 0.071 & 0.163 &   0.12 &  1.008 \\ 
    15 & CFF & 0.39 & 0.024 & 0.054 &   0.89 &  0.893 &&0.40 & 0.094 & 0.213 &   0.37 &  0.884 \\ 
    20 & SDF & 0.35 & 0.009 & 0.027 &   0.06 &  1.008 &&0.35 & 0.054 & 0.156 &   0.16 &  1.008 \\ 
    20 & ADF$^2$ & 0.35 & 0.010 & 0.028 &   0.05 &  1.008 &&0.35 & 0.059 & 0.169 &   0.18 &  1.008 \\ 
    20 & CFI & 0.35 & 0.039 & 0.112 &   0.07 &  1.008 &&0.36 & 0.068 & 0.197 &   0.11 &  1.008 \\ 
    20 & CFF & 0.31 & 0.018 & 0.053 &   0.88 &  0.895 &&0.32 & 0.089 & 0.256 &   0.33 &  0.886 \\ 
    \hline\noalign{\smallskip}
    5 & SDF & 1.11 & 0.051 & 0.047 &   1.50 &  1.010 &&1.12 & 0.119 & 0.109 &   0.80 &  1.010 \\ 
    5 & ADF$^2$ & 1.11 & 0.052 & 0.047 &   1.48 &  1.010 &&1.12 & 0.129 & 0.118 &   0.78 &  1.010 \\ 
    5 & CFI & 1.11 & 0.064 & 0.058 &   1.16 &  1.012 &&1.12 & 0.127 & 0.115 &   0.84 &  1.012 \\ 
    5 & CFF & 1.04 & 0.059 & 0.053 &   1.33 &  0.946 &&1.04 & 0.209 & 0.190 &   1.79 &  0.933 \\ 
    7 & SDF & 0.84 & 0.029 & 0.035 &   0.40 &  1.009 &&0.84 & 0.084 & 0.102 &   0.20 &  1.009 \\ 
    7 & ADF$^2$ & 0.84 & 0.029 & 0.035 &   0.39 &  1.009 &&0.84 & 0.092 & 0.111 &   0.23 &  1.009 \\ 
    7 & CFI & 0.84 & 0.040 & 0.048 &   0.30 &  1.010 &&0.84 & 0.090 & 0.109 &   0.21 &  1.010 \\ 
    7 & CFF & 0.79 & 0.034 & 0.041 &   0.34 &  0.954 &&0.80 & 0.153 & 0.184 &   0.56 &  0.950 \\ 
    10 & SDF & 0.62 & 0.019 & 0.031 &   0.08 &  1.008 &&0.62 & 0.069 & 0.112 &   0.07 &  1.008 \\ 
    10 & ADF$^2$ & 0.62 & 0.019 & 0.031 &   0.08 &  1.008 &&0.63 & 0.075 & 0.122 &   0.08 &  1.008 \\ 
    10 & CFI & 0.62 & 0.030 & 0.049 &   0.24 &  1.009 &&0.63 & 0.075 & 0.121 &   0.07 &  1.009 \\ 
    10 & CFF & 0.59 & 0.023 & 0.037 &   0.11 &  0.958 &&0.60 & 0.128 & 0.208 &   0.31 &  0.955 \\ 
    15 & SDF & 0.44 & 0.014 & 0.032 &   0.02 &  1.008 &&0.45 & 0.062 & 0.141 &   0.04 &  1.007 \\ 
    15 & ADF$^2$ & 0.44 & 0.014 & 0.033 &   0.02 &  1.007 &&0.45 & 0.067 & 0.153 &   0.05 &  1.007 \\ 
    15 & CFI & 0.44 & 0.026 & 0.059 &   0.38 &  1.008 &&0.45 & 0.067 & 0.152 &   0.06 &  1.008 \\ 
    15 & CFF & 0.42 & 0.017 & 0.038 &   0.04 &  0.960 &&0.44 & 0.116 & 0.263 &   0.25 &  0.958 \\ 
    20 & SDF & 0.35 & 0.012 & 0.036 &   0.01 &  1.008 &&0.35 & 0.059 & 0.171 &   0.04 &  1.008 \\ 
    20 & ADF$^2$ & 0.35 & 0.013 & 0.036 &   0.01 &  1.008 &&0.35 & 0.064 & 0.186 &   0.05 &  1.008 \\ 
    20 & CFI & 0.35 & 0.025 & 0.071 &   0.44 &  1.008 &&0.35 & 0.064 & 0.186 &   0.05 &  1.009 \\ 
    20 & CFF & 0.33 & 0.015 & 0.043 &   0.02 &  0.962 &&0.35 & 0.111 & 0.321 &   0.22 &  0.959 \\ 
   \hline
  \end{tabular}
  \tablefoot{Different CAs, IA always 2QI. The ``Fail (\%)''
    columns show the percentage of $4\sigma$ outliers. $p_1$  is the linear coefficient fitted to
    Eq.~(\ref{eq:linesine2}). The FOV is 24$\times$24 pixels. In the
    top part, with an image scale of 0\farcs29/pixel,  this corresponds to
    6\farcs9$\times$6\farcs9. In the bottom part the image scale 
    is 0\farcs53/pixel and the FOV is 12\farcs8$\times$12\farcs8.}
\end{table*}

\end{document}